\documentclass[a4paper,11pt]{article}
\usepackage{preamble}
\usepackage{comment}
\usepackage{relsize}
\usepackage{graphbox}
\usepackage{array}
\usepackage[shortlabels]{enumitem}
\usepackage{CJKutf8}

\usepackage[sorting=none,style=numeric-comp,maxbibnames=10]{biblatex}
\newcommand{\betaJ}{\beta\bJ}
\newcounter{yjcc}

\newcommand{\vev}[1]{\left\langle #1\right\rangle}
\date{\today}

\renewcommand{\baselinestretch}{1.2}
\makeatletter
\def\blfootnote{\gdef\@thefnmark{}\@footnotetext}
\makeatother

\begin{document}

\thispagestyle{empty}
\vspace{13mm}  %\vspace{10mm}
\begin{center}
{\huge A Path Integral for Chord Diagrams and}
\\[5mm]
{\huge Chaotic-Integrable Transitions in Double Scaled SYK}
\\[13mm]
{\large Micha Berkooz$^{1}$, Nadav Brukner$^{1}$, Yiyang Jia\begin{CJK*}{UTF8}{gbsn}
		(贾抑扬)
\end{CJK*}$^1$, Ohad Mamroud$^{2,3}$}
 
\bigskip
{\it
$^1$ Department of Particle Physics and     Astrophysics, \\[.0em]
Weizmann Institute of Science, Rehovot 7610001, Israel \\[.2em]
$^2$ SISSA, via Bonomea 265, 34136 Trieste, Italy \\[.2em]
$^3$ INFN, Sezione di Trieste, Via Valerio 2, 34127 Trieste, Italy \\[.2em]
}

\bigskip
\bigskip

{\bf Abstract}\\[8mm]
{\parbox{16cm}{\hspace{5mm}

% Insert abstract here

We study transitions from chaotic to integrable Hamiltonians in the double scaled SYK and $p$-spin systems. The dynamics of our models is described by chord diagrams with two species. We begin by developing a path integral formalism of coarse graining chord diagrams with a single species of chords, which has the same equations of motion as the bi-local ($G\Sigma$) Liouville action, yet appears otherwise to be different and in particular well defined. We then develop a similar formalism for two types of chords, allowing us to study different types of deformations of double scaled SYK and in particular a deformation by an integrable Hamiltonian. The system has two distinct thermodynamic phases: one is continuously connected to the chaotic SYK Hamiltonian, the other is continuously connected to the integrable Hamiltonian, separated at low temperature by a first order phase transition. We also analyze the phase diagram for generic deformations, which in some cases includes a zero-temperature phase transition.

}}
\end{center}

\newpage
\pagenumbering{arabic}
\setcounter{page}{1}
\setcounter{footnote}{0}
\renewcommand{\thefootnote}{\arabic{footnote}}

{\renewcommand{\baselinestretch}{.88} \parskip=0pt
\setcounter{tocdepth}{2}
\tableofcontents}

%%%%%%%%%%%%%%%%%%%%%%%%%%%%%%%%%%%%%%%%%%%%%%%%%%%%%%
%%%%%%%%%%%%%%%%%%%%%%%%%%%%%%%%%%%%%%%%%%%%%%%%%%%%%%
\newpage

\section{Introduction}
\label{sec:intro}
The Sachdev-Ye-Kitaev (SYK) model \cite{kitaev2015simple, Maldacena:2016hyu,  sachdev1993,sachdev2010} is a quantum mechanical model of $N$ Majorana fermions that exhibits many interesting properties of quantum chaos, ranging from level repulsion \cite{you2017sachdev, garcia2016,Cotler:2016fpe} to a maximal chaos exponent \cite{kitaev2015simple, Maldacena:2016hyu, Maldacena:2015waa}. The SYK model starts its life in nuclear physics \cite{french1970,bohigas1971} and in condensed matter physics \cite{sachdev1993,sachdev2010}, and the maximal chaos property in particular made it important as a controllable model of holography \cite{kitaev2015simple, Maldacena:2016upp, Cotler:2016fpe, Saad:2018bqo, Maldacena:2018lmt, Goel:2018ubv, Jensen:2016pah, Polchinski:2016xgd, sachdev2010}. 

Our goal is to study the phase diagram that arises from deformations of the SYK model, and in particular from deformation of the chaotic SYK model (and other models in its universality class) by an integrable Hamiltonian on the same set of degrees of freedom. This problem is interesting for the study of quantum chaos (as an analogue of the KAM theorem for this case) and it also promises to be an interesting case for holography, where the nature of any putative dual for the integrable case changes significantly compared to the chaotic one \cite{Gao:2023gta, AlmehiriPaper}. These types of models can even be studied experimentally \cite{Pikulin:2017mhj} or by quantum simulations \cite{Jafferis:2022crx}. Previous works have already observed that certain generalizations of the model quell its chaotic nature, but these examples involve the introduction of additional species of fermions 
\cite{banerjee2017solvable,jian2017model,Jian:2017unn,Peng:2017kro}. Moreover, numerical investigations suggest that a similar transition might occur by certain deformations of the SYK model \cite{Garcia-Garcia:2017bkg}. In this work, we discuss analytically controlled deformations of the Hamiltonian that also induce a transition, without introducing new degrees of freedom, and map their phase diagram. On the way we will develop some new techniques to control the double scaled SYK model.

To do so, we focus on a simple model that allows us to interpolate between an integrable and a chaotic SYK-like system. The specific model that we will discuss is of the form
\begin{equation}\label{eq:Cha_integ_Ham}
    H = \nu H_{\text{Chaotic}} + \kappa H_{\text{Integrable}} \,,\qquad \nu^2 + \kappa^2 = 1\,,\qquad \nu,\kappa \in [0,1] \,,
\end{equation}
where $H_{\text{Chaotic}}$ ($H_{\text{Integrable}}$) are random $p$-local chaotic (integrable) Hamiltonians, and the weights interpolate between the purely integrable and the fully chaotic system\footnote{The detailed normalization conditions are given in the next section.}. We will then generalize our approach to similar interpolations between two generic Hamiltonians.

\begin{figure}[t]
    \centering
    \includegraphics[width=0.5\textwidth]{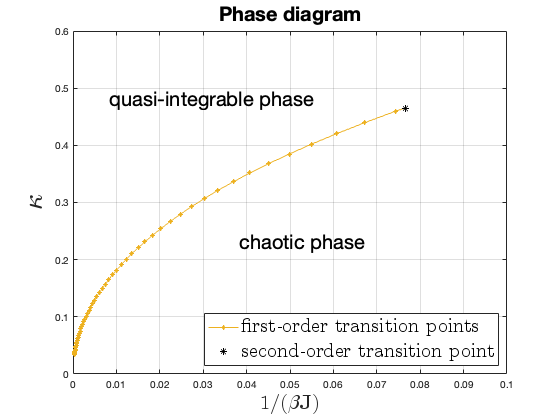}
    \caption{Phase diagrams in the $\kappa-1/\beta \mathbb{J}$ plane. The orange dots denote the numerically obtained first order transition points, and the orange line is the first order transition line. The black dot is where the first-order transition line terminates. }
    \label{fig:phase transition line}
\end{figure}

In the examples we consider below the Hamiltonians act, via a $p$-local interaction, either on systems of $N$ fermions or on spin systems of $N$ qubits.  Our main requirement is that the Hamiltonians have a double scaling limit in which the dynamics are governed by chord diagrams, and we will work in this limit \cite{erdHos2014phase,Cotler:2016fpe,Berkooz:2018qkz}. Many $p$-local Hamiltonians\footnote{In fact, some systems are not $p$-local and are still governed by chord diagrams, such as the Parisi's hypercube model \cite{Berkooz:2023scv, Berkooz:2023cqc, Jia:2020rfn}. See Appendix \ref{app:parisiChordRules}. Some of our results apply to them as well.} on $N$ qubits have such a limit, which is 
\begin{equation}
    N,p\rightarrow\infty,\qquad  p^2/N=\text{const} \,.
\end{equation}
Hence, our construction is quite general. We give two concrete examples -- a system of $N$ fermions with a Hamiltonian that interpolates between the commuting SYK model \cite{Gao:2023gta} and the SYK model, or a system of $N$ qubits with an interpolation between the chaotic $p$-spin model \cite{erdHos2014phase,Berkooz:2018qkz} and the integrable $p$-spin model \cite{derridaPRL, derridaPRB}. We will often use the notation $\lambda \sim p^2/N$. In all such cases we can write down an exact partition function for the model, but it becomes particularly tractable if one takes a further limit $\lambda \rightarrow 0$ afterwards, which is what we will do. The end result of the analysis is Figure~\ref{fig:phase transition line} which exhibits an abrupt change in the free energy of the system, and implies a very clean first-order phase transition line between two phases as we vary $\kappa$ or the temperature. The first-order line  terminates at a (presumably) second-order phase transition point. Much of the phase diagram can be controlled analytically. One of the phases is continuously connected to the purely chaotic system at $\kappa = 0$, and we call it the \emph{chaotic phase}. The other phase is continuously connected to the purely integrable system, and we call it, with a slight abuse of terminology, the \emph{quasi-integrable phase}.  We stress that in this work we do not establish the existence of a transition of out-of-time-ordered four-point functions or a transition of level statistics.  We hope to address these questions in future works.

The main technical novelty in this paper is a new approach to handle chord diagrams. We first describe it for the case of a single Hamiltonian, i.e., a single type of chords in the diagrams, in Section~\ref{sec:path integral single chord}. We introduce an (exact) coarse graining procedure, where we divide the thermal circle into $s$ segments, and introduce a new variable---the number of chords that stretch between any two segments, denoted by $n_{ij}$. Both the partition function and correlation functions are written as a sum over all possible configurations of the $n_{ij}$'s, weighted by all diagrams with the same configuration. The limit $s\to\infty$ should be thought of as a continuum limit, when taken in parallel to the semi-classical $\lambda \to 0$ limit. Under these circumstances, the $n_{ij}$'s become functions of the Euclidean times $n(\tau_i,\tau_j)$, and we arrive at a simple path integral expression for the partition function \eqref{eq:partition function path int 1 chord}. Schematically,
\begin{equation}
    Z_{\text{1-chord}} = \sum_{\{n_{ij}\}}\left[\cdots\right] \xrightarrow{s \to \infty,\; \lambda \to 0} \int \cD n \, e^{-\frac{1}{\lambda} S[n(\tau_i,\tau_j)]} \,,
\end{equation}
and the expression can then be analyzed via a saddle point approximation. 
The equations of motion are equivalent to the Liouville equation obtained from the $G\Sigma$ approach of \cite{Maldacena:2016hyu, Cotler:2016fpe, Sachdev:2015efa, Stanford-talk-kitp}, yet beyond the saddle point the actions are different,\footnote{At the very least, we are not yet able to change variables in the path integral to transform one to the other.} and in particular the new one gives a well defined path integral. We will further elaborate on the similarities and differences in a future work \cite{single-chord-future-paper}. 

In our model \eqref{eq:Cha_integ_Ham}, the two types of Hamiltonians give rise to two types of chords. While in some special systems \cite{Berkooz:2020xne,Berkooz:2020uly}  the two-chord case is solvable, generically it is not. Despite that, we find the generic case to be solvable in the semi-classical limit. In Section~\ref{sec:path integral two chords} we repeat the coarse graining trick, this time with $n_{ij}$ denoting the chords associated with the chaotic Hamiltonian, and $z_{ij}$ denoting those associated with the integrable one. We again find an action formulation, \eqref{eq:2 chord continuum action}, in the continuum and semi-classical limit,
\begin{equation}
    Z_{\text{2-chords}} = \sum_{\{n_{ij}, z_{ij}\}}\left[\cdots\right] \xrightarrow{s \to \infty,\; \lambda \to 0} \int \cD n \cD z\, e^{-\frac{1}{\lambda} S[n(\tau_i,\tau_j), z(\tau_i,\tau_j)]} \,,
\end{equation}
The equations of motion are simply expressed in terms of a couple of new fields, $g_n(\tau_1,\tau_2)$ and $g_z(\tau_1,\tau_2)$, that capture the number of $n$- and $z$-chords that stretch ``across" $\tau_1$ and $\tau_2$, i.e., the chords that would intersect a chord stretching between these two times. The saddle point equations are
\begin{align}
    \partial_{\tau_1}\partial_{\tau_2}g_n(\tau_1,\tau_2) &= -2\bJ^2\nu^2 e^{g_n(\tau_1,\tau_2) + g_z(\tau_1,\tau_2)} \,, \\ 
    \partial_{\tau_1}\partial_{\tau_2}g_z(\tau_1,\tau_2) &= -2\bJ^2\kappa^2 e^{g_n(\tau_1,\tau_2)} \,.
\end{align}
In the purely chaotic model where $\kappa$ and $g_z$ vanish, they reproduce the Liouville equation discussed above. As we will see in Section~\ref{sec:phase transition}, the numerical solutions to these equations exhibit a clear first-order phase transition, and using analytic approximations we can explain some of the features of the phase diagram at low temperatures. For any $\kappa > 0$ the low-temperature phase is the quasi-integrable phase. The numerics suggest that the transition ends at some critical temperature and $\kappa$, see Figure~\ref{fig:phase transition line}.

The technique developed here allows us to study an interpolation between any two Hamiltonians that have a double scaling limit, not just those describing chaotic and integrable systems. One particular family of such interpolations describes renormalization-group (RG) flows between SYK models, which were studied in \cite{jiang2019,Anninos:2022qgy}. We comment on the phase diagram for general type of chord rules in Section~\ref{sec:generic 2 chords}. In some cases, such as the case of the SYK RG-flows, there is no phase transition, while in other cases there could even be a quantum phase transition at zero temperature.

A companion paper \cite{Berkooz:2024evs} summarizes the case of the integrable-to-chaotic transition, while here we present a more thorough derivation of the path integral technique, and the generalization to other systems.

This paper is organized as follows: in Section~\ref{sec:model-defs} we define the different microscopic models and review how, in the double scaling limit, their dynamics are entirely captured by chord diagrams. In Section~\ref{sec:path integral single chord} we present the coarse graining process and a path integral for the simpler case of a single type of chords, and review its semi-classical limit. In Section~\ref{sec:path integral two chords} we present a path integral formulation for the semi-classical limit for Hamiltonians whose dynamics are captured by two types of chords for the interpolating Hamiltonian \eqref{eq:Cha_integ_Ham}. In Section~\ref{sec:phase transition} we analyze the chaotic and integrable phases of the interpolating Hamiltonian and describe the phase transition. In Section~\ref{sec:generic 2 chords} we repeat the analysis for a system that interpolates between any two Hamiltonians with a chord description, and find its phase diagram. The work also contains several appendices. In Appendix~\ref{app:p-spin model} we briefly review our integrable system. In Appendix~\ref{app:CrshCrdDiag} we give a more thorough introduction to chord diagrams, explaining how they arise from the microscopic Hamiltonians. In Appendix~\ref{app:SpecialFunctions} we define some of the special functions used throughout the paper. In Appendix~\ref{app:matrix_element} we give some more details regarding our coarse graining procedure.

\section{The microscopic Hamiltonians}
\label{sec:model-defs}
As mentioned in the introduction, we will be studying a simple model which interpolates between an integrable and a chaotic system,
\begin{equation}\label{eq:Cha_integ_Ham_2}
    H = \nu H_{\text{Chaotic}} + \kappa H_{\text{Integrable}} \,,\qquad \nu^2 + \kappa^2 = 1\,,\qquad \nu,\kappa \in [0,1] \,,
\end{equation}
where $H_{\text{Chaotic}}$ ($H_{\text{Integrable}}$) is a random chaotic (integrable) Hamiltonian. The two Hamiltonians are $p$-local Hamiltonians whose random couplings are drawn independently. For convenience we will choose (the trace is normalized such that $\Tr(\1)=1$) 
\begin{equation}
    \braket{\Tr(H_{\text{Integrable}}^2)} = \braket{ \Tr(H_{\text{Chaotic}}^2)} = 1, \quad \braket{ \Tr(H_{\text{Integrable}}H_{\text{Chaotic}})} = 0 \quad \implies \quad \Tr(H^2)=1
\end{equation}
where $\langle\cdots\rangle$ means the average over the ensemble of couplings. In subsection \ref{sec:ChaosHExmp} we will list two examples of chaotic Hamiltonians and in section \ref{sec:IntegHexmp} we will list the corresponding integrable Hamiltonians. It will be clear there that the construction can be generalized to many other $p$-local systems. Finally in subsection \ref{sec:SoltnRev} we will review how to formally write the partition function of these models using chord diagrams (as well as some other pieces of lore).

% \subsection{The combined Hamiltonian}
% We would like to study the Hamiltonian
% \begin{equation}
% \label{eq:combined_Hamiltonian_spins}
%     H = \nu H_{\text{sSYK}} + \kappa H_{p\text{-spin}} \,,\qquad \nu^2 + \kappa^2 = 1\,,\qquad \kappa \in [0,1] \,,
% \end{equation}
% where we keep the normalization $\Tr(H^2) = \cJ^2$. An equivalent realization in our limit would be a system of (even) $N$ Majorana fermions with an all-to-all interaction,
% \begin{equation}
% \label{eq:combined_Hamiltonian_fermions}
%     H = \nu H_{\text{SYK}} + \kappa H_{\text{cSYK}}^{(N,p)} \,, \qquad \nu^2 + \kappa^2 = 1\,,\qquad \kappa \in [0,1]\,.
% \end{equation}
% The averages over the two sets of couplings, $J$ and $B$, are taken independently of one another, i.e. 
% \begin{equation}
% \braket{J_{I_1}\cdots J_{I_k} B_{J_1}\cdots B_{J_\ell}} = \braket{J_{I_1}\cdots J_{I_k}} \braket{B_{J_1}\cdots B_{J_\ell}}\,.
% \end{equation}

\subsection{The chaotic Hamiltonian $H_{\text{Chaotic}}$ }\label{sec:ChaosHExmp}

We would like to demonstrate two chaotic Hamiltonians for which our technique can be applied. One is the SYK system and the other is a $p$-spin model. In the double scaling limit the two models are equivalent\footnote{At least in leading order in  $1/N$. At finite $p$, the fermion model and the spin model can have qualitative differences \cite{Baldwin_2020}, but they disappear in the double scaling limit where the dependence on $p$ and $N$ appears through $\lambda$.} for all values of $\lambda$, as both are described purely by chord diagrams. This will be explained below.

\paragraph{1. The double scaled SYK (DS-SYK)}
The SYK system is a quantum mechanical system of $N$ Majorana fermions $\psi_i$, $\left\{ \psi_{i},\psi_{j}\right\} =2\delta_{ij}$, with the Hamiltonian 
\begin{equation}
\label{eq:H_SYK_def}
    H_{\text{SYK}} = i^{p/2}\sum_{1\leq i_{1}<i_{2}<\dots<i_{p}\leq N}J_{i_{1}\cdots i_{p}}\psi_{i_{1}}\cdots\psi_{i_{p}}\;.
\end{equation}
The couplings are random Gaussian variables that satisfy
\begin{equation}
    \label{eq:cJ norm}
    \braket{J_{I}} = 0 \;, \qquad \braket{J_{I}J_{J}} = \binom{N}{p}^{-1}\cJ^{2}\delta_{IJ} \;, 
\end{equation}
and capital indices $I,J$ denote sets of $I=\{i_1,..,i_p\}$ indices of size $p$, $1\leq i_1<i_2<\cdots<i_p\leq N$. The aforementioned double scaling limit amounts to taking the limit $N \to \infty$ and $p \to \infty$ while keeping fixed the ratio
\begin{equation}
    \lambda \equiv \frac{2p^2}{N}\,, \quad q \equiv e^{-\lambda} \;.
\end{equation}
\sloppy This normalization is convenient in the chord diagram language described below, and is chosen such that ${\left\langle\Tr\left(H_{\text{SYK}}^2\right)\right\rangle = \cJ^2}$ with the convention that $\Tr (\1) = 1$. A different normalization used in the literature, e.g. \cite{Maldacena:2016hyu, Lin:2022rbf, Goel:2023svz}, reads
\begin{equation}\label{eq:bJnormalization}
    \braket{J_I J_J} = \frac{1}{\lambda}\binom{N}{p}^{-1} \bJ^2 \delta_{IJ} \,,
\end{equation}
the two normalizations are related by
\begin{equation}
\label{eq:def bJ}
    \bJ^2 = \lambda \cJ^2 \;.
\end{equation}
We will find it convenient to work with the $\cJ$ normalization for evaluating chord diagrams, and we will even set $\cJ = 1$ for convenience. Factors of $\cJ$ can always be restored by dimensional analysis. After the coarse graining procedure our effective action would have a nicer form using the $\bJ$ variable, and we will re-introduce it at that point.

The SYK model was extensively studied in the past decade. It is usually studied for finite $p$ and at the large $N$ limit, where it was shown to be maximally chaotic \cite{kitaev2015simple,Maldacena:2016hyu} in the sense that its Lyapunov exponent saturates the universal bound \cite{Maldacena:2015waa}, and its spectrum exhibits (numerically) random-matrix level statistics \cite{you2017sachdev,garcia2016, Cotler:2016fpe}. These features gave rise to a holographic interpretation for the model \cite{sachdev2010, Maldacena:2016hyu, Jensen:2016pah}, where its low energy sector is dual to JT gravity on near-AdS$_2$ spacetime.

The double scaling limit of the model was also studied in \cite{Cotler:2016fpe, erdHos2014phase,Berkooz:2018jqr,Berkooz:2018qkz, garcia2017,garcia2018c} and subsequent work. In this limit the entire model is solvable at all energies, in the sense that the partition function and the correlation functions can be computed for any $\lambda$ \cite{Berkooz:2018jqr}. The semi-classical limit, $\lambda \to 0$, agrees with that of the large $p$ expansion of \cite{Maldacena:2016hyu}, see \cite{Goel:2023svz,Mukhametzhanov:2023tcg,Okuyama:2023bch}. In the semi-classical and low temperature limit, $\beta \bJ \sim \lambda^{-1}$, the model exhibits a Schwarzian density of states \cite{Berkooz:2018jqr, Berkooz:2018qkz}. Moreover, by using the chord diagram techniques described below, one can define an auxiliary Hilbert space which can be associated with the Hilbert space of the bulk dual \cite{Lin:2022rbf,Berkooz:2018qkz}. 

At finite temperature, when $\beta \bJ \sim \lambda^{0}$, the situation is less clear. At finite temperature the chaos exponent is no longer maximal, but rather depends on the dimensionless temperature $\beta \bJ$ \cite{Maldacena:2016hyu,Berkooz:2018jqr,Berkooz:2018qkz,Streicher:2019wek,Choi:2019bmd,Goel:2023svz,Okuyama:2023bch}. A suggested dual geometry, the fake disk, was proposed in \cite{Lin:2023trc}. At finite $q$, spacetime becomes even stranger. The dual theory is conjectured to be given by a particle moving on a non-commutative deformation of AdS$_3$ \cite{Berkooz:2022mfk}, or in another variant, a BF theory whose boundary dynamics is given by the $q$-Schwarzian theory, defined by a particle travelling on the quantum group $SU_q(1,1)$ \cite{Blommaert:2023opb, Blommaert:2023wad}.

\paragraph{2. The chaotic $p$-spin model (C-Spin)} In the double scaling limit the SYK model is just one microscopic realization of a broad universality class. Another realization, studied in \cite{erdHos2014phase, Berkooz:2018qkz} (recently, outside of the double scaled limit, also in \cite{Swingle:2023nvv}), is a system of $N$ sites with a spin-$\frac{1}{2}$ variable (a qubit) at each site. The Hamiltonian is given by an all-to-all random Hamiltonian where random Pauli matrices $\sigma^{a}_i$, $a=1, 2, 3$ (or $x, y, z$), act on the $i$th qubit:
\begin{equation} \label{eq:def spin-SYK}
    H_{\text{C-Spin}} = \sum_{\substack{1\leq j_1<\cdots<j_p\leq N \\ a_{1},\cdots,a_{p}=\{1,2,3\}}}J_{j_1\cdots j_p}^{a_{1}\cdots a_{p}}\sigma_{j_1}^{a_{j_1}}\cdots \sigma_{j_p}^{a_{j_p}}.
\end{equation}
The random couplings are again independent, Gaussian, and 
\begin{equation} \label{eq:Gaussian_dist}
    \braket{J_I^A} = 0 \,, \qquad \braket{J_I^{A} J_J^{A'}} = 3^{-p}\cJ^2\binom{N}{p}^{-1} \delta_{IJ}\delta^{AA'} \,,
\end{equation}
where $I$ is a multi-index of the sites, and $A$ is a vector of length $p$ in which each entry takes one of the values $\{1,2,3\}$. The normalization here again corresponds to $\Tr\left(H_{\text{C-Spin}}^2\right) = \cJ^2$ in the normalization where $\Tr(\1) = 1$. The double scaling limit of this model is taken by keeping $\lambda = \frac{4p^2}{3N}$ fixed, and as before $q=e^{-\lambda}$.

\subsection{The integrable Hamiltonian $H_{\text{Integrable}}$}
\label{sec:IntegHexmp}

The Hamiltonians in the previous subsection will play the role of $H_{\text{Chaotic}}$ in equation \eqref{eq:Cha_integ_Ham_2}. Next we define the $H_{\text{Integrable}}$ parts in each of the systems. These two integrable Hamiltonians are equivalent to each other, where the one for $N$ Majorana fermions exactly maps to the one for $N/2$ qubits, and are briefly reviewed in Appendix~\ref{app:p-spin model}.  

In particular, it is known that at low enough temperatures these models have a spin-glass phase  \cite{derridaPRL,derridaPRB,gross1984, gardener1985}. However with our normalization the critical temperature is suppressed by $1/N$ as discussed in Appendix~\ref{app:p-spin model}. Therefore we will always be above the glassy phase, where the computations using annealed averages, which we will do, are justified.

\paragraph{1. The integrable SYK model}
The isometry of the Clifford algebra of the Majorana fermions is $SO(N)$ generated by $\psi^{[i}\psi^{j]}$. We will work with even $N$ here. To obtain an integrable model we can take the product of any generators within a Cartan subalgebra. For simplicity we will choose the Cartan generators to be
$i Q_k = \psi_{2k-1}\psi_{2k}$,
and the Hamiltionian becomes
\begin{equation}
\label{eq:commuting-SYK}
H_{\text{I-SYK}} = i^{p/2} \sum_{1\leq i_1<\dots<i_{p/2}=N/2} B_{i_1,\dots,i_p} \big(\psi_{2{i_1}-1}\psi_{2{i_1}}\big)\cdots \big(\psi_{2i_{p/2}-1}\psi_{2i_{p/2}}\big) \,,
\end{equation}
where the $B$'s are drawn from a random Gaussian distribution, 
\begin{equation}
    \braket{B_I} = 0 \,,\qquad \braket{B_I B_J} = \cJ^2 \binom{N/2}{p/2}^{-1} \delta_{IJ} \,.
\end{equation}
This model was studied by \cite{Gao:2023gta} under the name ``commuting SYK'', and its double scaled limit is considered in \cite{AlmehiriPaper}.

\paragraph{2. The integrable $p$-spin model (I-Spin)}
The integrable model that we will study is the so-called $p$-spin model \cite{derridaPRL,derridaPRB,gross1984,gardener1985}. We take the same system of $N$ qubits as before with a random, all-to-all interaction where products of $\sigma^{z}$ act on different sites together,
\begin{equation}
\label{eq:p-spin-Hamiltonian}
    H_{\text{I-Spin}} = \sum_{1\leq i_1<\dots<i_p\leq N} B_{i_1,\dots,i_p} \sigma^{z}_{i_1}\cdots \sigma^{z}_{i_p} \,.
\end{equation}
The $B$'s are drawn from a random Gaussian distribution, 
\begin{equation}
    \braket{B_I} = 0 \,,\qquad \braket{B_I B_J} = \cJ^2 \binom{N}{p}^{-1} \delta_{IJ} \,.
\end{equation}
It is actually the same as the integrable SYK for $2N$ fermions when we go from one description to the other via a Jordan-Wigner transformation, but once we couple it to the chaotic SYK this isomorphism is not helpful. 

\subsubsection{The combined system}
Our interpolating Hamiltonian \eqref{eq:Cha_integ_Ham_2} can now be written for (even) $N$ Majorana fermions as 
\begin{equation} \label{eq:Interpolating_Hamiltonian_Majorana}
    H = \nu H_{\text{SYK}} + \kappa H_{\text{I-SYK}} \,, \qquad \kappa^2 + \nu^2 = 1 \,, \qquad \kappa,\nu \in [0,1] \,, 
\end{equation}
or for $N$ qubits as
\begin{equation} \label{eq:def Pauli_combined_system}
    H = \nu H_{\text{C-Spin}} + \kappa H_{\text{I-Spin}} \,, \qquad \kappa^2 + \nu^2 = 1 \,, \qquad \kappa,\nu \in [0,1] \,.
\end{equation}
When $\kappa = 0$ we are left with the chaotic model, while when $\kappa = 1$ we have the integrable one. 

Generally, the interaction lengths (denoted by $p$ above) for the two Hamiltonians can differ, but we will specialize to the case where they are the same from here on. The generic case will result in slightly different chord intersection rules than the ones explained below. Another generalization is described in Section~\ref{sec:generic 2 chords}, where we consider Hamiltonians that give rise to generic chord intersections.

 \subsection{The solution of the double scaled models}\label{sec:SoltnRev}

\paragraph{The partition function for a single species of chords:} In the double scaling limit, systems such as these can be solved using chord diagrams \cite{erdHos2014phase,Berkooz:2018jqr,Berkooz:2018qkz}. We briefly review this approach, and elaborate it in Appendix~\ref{app:CrshCrdDiag}. 

\begin{figure}[t]
    % \begin{subfigure}[t]{0.45\textwidth}
    %     \centering
    %     \includegraphics[width=0.6\textwidth]{Figures/Illustrations/chord diagram contractions.pdf}
    % \end{subfigure}
    % \qquad
    % \begin{subfigure}[t]{0.45\textwidth}
        \centering
        \includegraphics[width=0.35\textwidth]{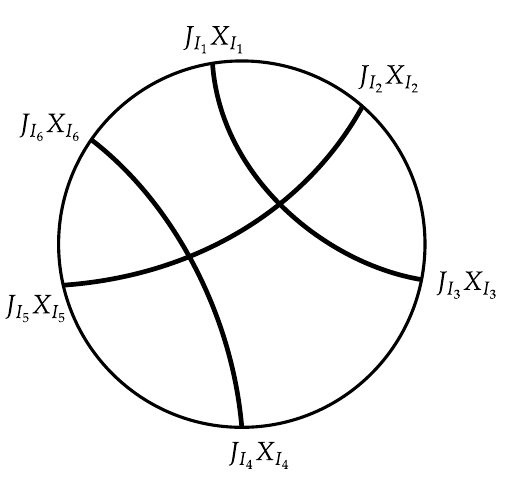}
    % \end{subfigure}
    \caption{%Illustration of chord diagrams. Left: The chord diagram that represents the contractions in $\langle J_{I_1}J_{I_3} \rangle \langle J_{I_2} J_{I_4} \rangle \Tr\left(X_{I_1} \cdots X_{I_2} \cdots X_{I_3} \cdots X_{I_4} \cdots\right)$. Right: 
    A chord diagram that contributes to $m_6$, representing the Wick contractions of $\langle J_{I_1} J_{I_3}\rangle \langle J_{I_2} J_{I_5}\rangle \langle J_{I_4} J_{I_6}\rangle \Tr(X_{I_1}\cdots X_{I_6})$. It contributes as $q^2$ to the sum as it has two chord intersections.}
    \label{fig:chord diagram}
\end{figure}

The different chaotic and integrable Hamiltonians described above are of the form 
\begin{equation}
    H = \sum_I J_I X_I \,,
\end{equation}
where the $J$'s are random couplings with Gaussian distribution and the $X$'s are operators. Their averaged\footnote{To leading order in $N$ the model self averages and there is no difference between annealed and quenched averaging \cite{Sachdev:2015efa}.} partition function can be computed by a series expansion. In each term in the series we have to compute the average moment, $\langle \Tr(H^{k})\rangle = \sum_{I_1,\cdots,I_k} \langle J_{I_1} \cdots J_{I_k}\rangle \Tr(X_{I_1} \cdots X_{I_k})$. Since our couplings have a Gaussian distribution, we can use Wick's theorem to express the moment as a sum over all pairwise contractions of the couplings times the appropriate trace. These contractions can then be represented diagrammatically via a \emph{chord diagram}, where there are $k$ nodes on a circle, each representing an element in the trace, and $k/2$ chords connecting them, representing the Wick contractions between the appropriate couplings, see Figure~\ref{fig:chord diagram}. In the double scaling limit, the contribution of a single chord diagram is given by the number of chord intersections in the diagram, each giving a factor of $q$. For the integrable models, $q=1$. The partition function then takes the form
\begin{equation} \label{eq:CrdPrtFnctn}
 \langle \Tr (e^{-\beta H})\rangle = \sum_k \frac{(-\beta)^{2k}}{(2k)!} m_{2k},\ \ \ m_{2k} = \langle \Tr(H^{2k})\rangle = \sum_{\text{CD}(2k)}  q^{\text{\# intersections}},
\end{equation}
where $\text{CD}(2k)$ are chord diagrams with $2k$ nodes, i.e., $k$ chords.

\paragraph{The transfer matrix}
Next, a systematic evaluation of these diagrams is possible via a transfer matrix approach \cite{Berkooz:2018qkz, Berkooz:2018jqr}. Imagine a diagram with $k$ chords, and choose some point on the boundary of the diagram. By convention, at this point there are no open chords. Then as one moves along the diagram there are $2k$ instances in which either a chord open or closes. We choose chords to intersect only when they close, to count each intersection exactly once. After each instance, the ``state'' is given by the number of open chords. The space of all states is the chord Hilbert space,
\begin{equation}
\cH_{\text{chord}} = \left\{\ket{n} | n\in\bZ_+\right\} \,.
\end{equation}
We will find it useful to also equip it with the inner product 
\begin{equation}
    \braket{n|m} = \delta_{nm} \,.
\end{equation}
At each step in our convention there is exactly one way of opening a chord, but any of the open chords may close, with an overall weight $1 + \cdots + q^{n-1}$ for a state with $n$ open chords. A transfer matrix $T$ is then constructed out of chord creation and annihilation operators\footnote{Note that this is not the standard inner product on this space \cite{Speicher:1993kt, pluma2022dynamical}. In particular, $a^+$ is not the conjugate of $a$ in the inner product above, but this will not play a role for us.},
\begin{equation}
\label{eq:single chord T def}
    T = a + a^+ \,, \quad a \ket{n} = \frac{1-q^n}{1-q} \ket{n-1} \,, \quad a^+\ket{n} = \ket{n+1} \,,
\end{equation}
such that the contribution of all diagrams with $k$ chords to the moment $m_{2k}$ amount to acting with the transfer matrix $2k$ times and finishing with no open chords. The partition function is finally given by
\begin{equation}
    Z = \braket{0|e^{-\beta T}|0} \,.
\end{equation}
The creation and annihilation operators satisfy the algebraic relation
\begin{equation}
    [a,a^+]_q \equiv aa^+ - q a^+a = 1 \,.
\end{equation}

\begin{figure}[t]
    \centering
    \includegraphics[width=0.35\textwidth]{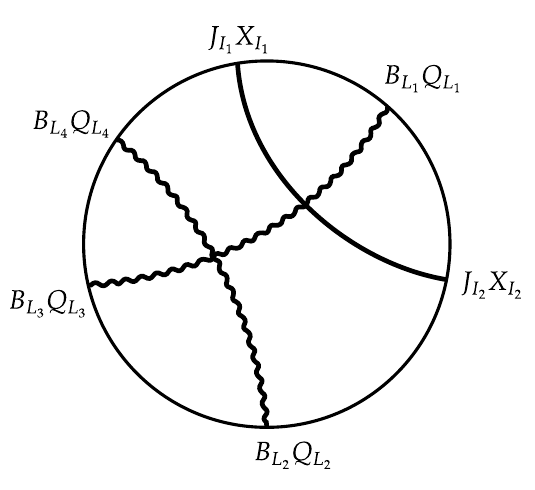}
    \caption{A chord diagram that contributes to $\left\langle\Tr\left(H^6\right)\right\rangle$, representing the Wick contractions of $\langle J_{I_1} J_{I_2}\rangle \langle B_{L_1} B_{L_3}\rangle \langle B_{L_2} B_{L_4}\rangle \Tr(X_{I_1} Q_{L_1} X_{I_2} Q_{L_2} Q_{L_3} Q_{L_4})$. It has one $n$-chord and two $z$-chords, and contributes $\nu^2\kappa^4 q$ to the sum over diagrams.}
    \label{fig:chord diagram 2 types}
\end{figure}

\paragraph{Two types of chords}
In this work we would like to study systems of the sort \eqref{eq:Cha_integ_Ham}, which have two types of random couplings, schematically
\begin{equation}
    H = \sum_I J_I X_I + \sum_L B_L Q_L \,,
\end{equation}
where $J_I$ and $B_L$ denote random couplings, while $X_I$ and $Q_L$ are operators, which, in the case of \eqref{eq:Cha_integ_Ham}, correspond to the chaotic and integrable Hamiltonians, respectively. As in the single chord case, the ensemble averaged partition function can be computed by a series expansion followed by ensemble average for each term\footnote{Odd terms in the sum vanish due to the ensemble average.}
\begin{equation}
Z = \left\langle\Tr\left(e^{-\beta H}\right)\right\rangle = \sum_{k=0}^\infty \frac{(-\beta)^{2k}}{(2k)!} \braket{\Tr(H^{2k})} \,.
\end{equation}
This sum can be diagrammatically represented as a sum over all chord diagrams with two types of chords  (see Figure~\ref{fig:chord diagram 2 types} for illustration):
\begin{itemize}
    \item $n$-chords: represent the Wick contractions between the $J$ couplings of the chaotic Hamiltonians.
    \item $z$-chords: represent the Wick contractions between the $B$ couplings of the integrable Hamiltonians.
\end{itemize}

The weight of each chord diagram depends on the number of chord intersections, as explained in Appendix~\ref{app:CrshCrdDiag}. Each intersection between two $n$-chords contributes a factor of $q$, each intersection between a $n$-chord and a $z$-chord contributes a factor of $q$ as well, and each intersection between two $z$-chords contributes a factor of $1$. Additionally, for a diagram with $n$ $n$-chords and $z$ $z$-chords there is a factor of $\nu^{2n}\kappa^{2z}$ coming from the pre-factors of the Hamiltonian,
\begin{equation}\label{eqn:twoChordRules}
    \left\langle\Tr\left(H^{2k}\right)\right\rangle = \sum_{\substack{\text{chord diagrams with} \\ \text{$n+z=k$ chords}}} \nu^{2n}\kappa^{2z} q^{\text{\#$n$-$n$ intersections}}q^{\text{\#$n$-$z$ intersections}} \,.
\end{equation}

\paragraph{2-chord transfer operator:} Just like in the single chord case, the sum over chord diagrams can be analyzed by a transfer matrix approach. Given a chord diagram with $k$ chords, one chooses an arbitrary point on the boundary of the diagram and declares that there are no open chords at this point. As one then encircles the diagram, there are $2k$ points of interest where either a chord open or closes. Unlike the case of a single type of chords, here the ``state'' of a diagram at each point is not just described by the overall number of open chords but also by their ordering,
\begin{equation}
\label{eq:2 chord Hilbert space}
\cH_{\text{2-types}} = \bigoplus_{k=0}^\infty \cH_k \,,\qquad \cH_k = \Span\left\{ \ket{n,z;\vec{r}} \; \big| \quad \vec{r} \in \{0,1\}^k,\quad \sum_{i=1}^k r_i = n, \quad n+z = k\right\} \,,
\end{equation}
where $\cH_k$ is the Hilbert space with exactly $k$ chords. $n$ ($z$) denotes how many $n$-chords ($z$-chords) there are, and the $i^{\text th}$ component of $\vec r$ represents whether there is an $n$- or a $z$-chord in the $i^{\text{th}}$ place. We choose a convention where chords intersect when a chord closes. There are now two types of chord creation and annihilation operators, $a_{n,z}$ and $a_{n,z}^+$, which satisfy the algebra 
\begin{equation}
    [a_n, a_n^+]_q = 1\,,\qquad [a_n, a_z^+]_q = 0\,, \qquad [a_z, a_n^+]_q = 0\,,\qquad [a_z, a_z^+] = 1 \,.
\end{equation}
The transfer matrix is
\begin{equation}
    T = \nu \left(a_n + a_n^+\right) + \kappa \left(a_z + a_z^+\right) \,,
\end{equation}
and the partition function is 
\begin{equation}
    Z = \braket{0|e^{-\beta T}|0} \,.
\end{equation}

Later on, we will need to sum over the weight of all sub-diagrams where $n$ $n$-chords and $z$ $z$-chords with ordering $\vec r_i$ come out of a segment of length $x$, without allowing for intersections between the outgoing chords, but only with chords that start and end inside the segment. This weight is exactly the coefficient of $|n,z;{\vec r} \rangle$ when spanning $e^{-x T} | 0 \rangle$ over the basis in \eqref{eq:2 chord Hilbert space}. Therefore, given some vector $v \in \cH_{\text{2-types}}$, we are going to define (by some mild abuse of notation) 
\begin{equation}
 \llangle n, z;\vec r | v \rangle = \text{coefficient of $| n, z;\vec r \rangle$ in the expansion of $v$.}
\end{equation}
i.e., 
\begin{equation}
\label{eq:not inner product}
    \llangle n, z;\vec r | n^\prime, z^\prime; \vec{s}\rangle = \delta_{n n'} \delta_{z z'} \delta_{\vec r \vec{s}} \,.
\end{equation}
but we will stop short of declaring this to be an inner product, as 
the Hilbert space $\cH_{\text{2-types}}$ carries a canonical positive definite inner product \cite{Speicher:1993kt,pluma2022dynamical}. The weight of all the sub-diagrams described above is therefore $\left\llangle n, z;\vec r | e^{-x T} |0 \right\rangle$.

While the chord Hilbert space is huge, $\dim(\cH_k) = 2^k$, we will later see that we can set up the computation such that the ordering can be partially neglected. In fact, this is one of the motivations for setting up the computation as we do later on.

% While the chord Hilbert space is huge, $\dim(\cH_k) = 2^k$, we will later see that we can set up the computation  

% in the semi-classical limit the ordering $\vec r$ does not play a role, and the only relevant quantities are the overall number of chords of each type. 

\subsubsection{The Liouville description}
\label{sec:Liouville}

A standard approach for solving the SYK model at fixed $p$ (length of interaction) is the $G\Sigma$ approach \cite{Jevicki:2016bwu, Maldacena:2016hyu}, also called the collective field approach---an effective action (and Schwinger-Dyson equations) for two bilocal fields, $G$ that represents the two-point function for the fundamental fermions of the model, and $\Sigma$ which is the Lagrange multiplier that enforces this. The action is found after integrating out the fermions and taking the large $N$ limit.

In the double scaling limit this action simplifies considerably, as explained in \cite{Cotler:2016fpe,Lin:2023trc,Goel:2023svz}. After introducing the re-scaled variables $g$ and $\sigma$,
\begin{equation}
    \Sigma(\tau_1,\tau_2) = \frac{\sign(\tau_1-\tau_2)\sigma(\tau_1,\tau_2)}{p},\qquad G(\tau_1,\tau_2) = \sign(\tau_1-\tau_2)\left(1+\frac{g(\tau_1,\tau_2)}{p}\right)\,,
\end{equation}
where note that we have taken $\sigma$ to be symmetric and real, to highlight the similarity to some formulas we later develop, and as opposed to the literature. The partition function can be written in terms of the action 
\begin{multline}
\label{eq:action g sigma}
    Z = \int D\sigma Dg\,\exp\Bigg\{-\frac{1}{\lambda}\Bigg[-\int_{0}^{\beta}d\tau_{1}\int_{0}^{\beta}d\tau_{2}\int_{\tau_{1}}^{\tau_{2}}d\tau_{3}\int_{\tau_{2}}^{\tau_{1}}d\tau_{4}\,\sigma\left(\tau_{1},\tau_{2}\right)\sigma\left(\tau_{3},\tau_{4}\right) \\
    + \int_0^\beta d\tau_{1}d\tau_{2}\left[i\sigma\left(\tau_{1},\tau_{2}\right)g\left(\tau_{1},\tau_{2}\right) - \frac{\bJ^{2}}{2}e^{g\left(\tau_{1},\tau_{2}\right)} \right]\Bigg]\Bigg\} \,.
\end{multline}
where the bounds of the integral are written for $\tau$'s that live on a circle, and flipping the limits of the integration should be understood as integrating over the other side of the circle. At this point $\sigma$ can be integrated out, leaving us with a ``lightlike-Liouville'' action
\begin{equation}
\label{eq:Liouville g action}
    Z = \int Dg \,\exp\left[-\frac{1}{2\lambda}\int_0^\beta d\tau_1 \int_0^\beta d\tau_2\, \left[\frac{1}{4} \partial_1 g(\tau_1,\tau_2) \partial_2 g(\tau_1,\tau_2) - \bJ^2 e^{g(\tau_1,\tau_2)}  + O(1/p)\right]\right] \,.
\end{equation}
Unfortunately, this action is not bounded from below\footnote{At least for the naive integration contour for $g$ and $\sigma$.}, and neither is \eqref{eq:action g sigma}. The action does have several alluring properties, though---a perturbative expansion in $\bJ$ produces the chord diagrams discussed above \cite{Lin:2023trc}, and one can use semi-classical methods to compute correlation functions in the model, with the equations of motion being
\begin{equation}
\label{eq:Liouville_eq_GSigma}
    \partial_1\partial_2 g =  -2\bJ^2 e^{g} \,.
\end{equation}

Below we start from the chord diagram approach, then develop a coarse graining technique that allows us to write the partition function (and correlation functions). For the single chord case, the partition function is written in terms of the number of chords stretching between two boundary points on the diagram, termed $n(\tau_1,\tau_2)$. In this language, $g(\tau_1,\tau_2)$ is the number of chords that cross a chord stretching between the points $\tau_{1}$ and $\tau_{2}$. The action we get, \eqref{eq:Continuum action}, looks like a version of \eqref{eq:action g sigma} after integrating out $g$, where $n$ takes the role of $\sigma$. One can reproduce the same equation of motion \eqref{eq:Liouville_eq_GSigma} from this new action, but unlike the Liouville-like action above it is bounded from below. Away from the saddle point the action is different, and we regard it as a well-posed version of the bi-local Liouville theory \eqref{eq:Liouville g action}.

\section{A coarse grained approach to chord diagrams}
\label{sec:path integral single chord}

The transfer matrix method allows us to compute $n$-point functions relatively efficiently, but it is not useful for discussing the 2-chord case which is needed for the integrability-to-chaos transition. Furthermore, it contains some arbitrariness which clouds some issues. For example, one can choose a different convention of how chords are arranged when opened/closed. This gives a new transfer matrix which is the transpose of the one in Section~\ref{sec:SoltnRev} but fortunately, it is related to it by a conjugation so none of the physical quantities change. In fact there are other conventions as well. This seems like a technicality but, in some moral sense, we are assigning the same chord intersection to different regions of the dual spacetime which muddies the water when trying to obtain a bulk interpretation\footnote{It is tempting to think of this as different gauges of the same object but it is not clear how to make this precise.}. So we would like to find a more invariant approach to the computation.

The final result of this section will be such a path integral formulation which will be akin to the Liouville description discussed above. It will not be the same though---its saddle point equation will be the same as the Liouville description, but the full form of the action, \eqref{eq:Continuum action}, will not be the same\footnote{At least, we have not been able to find a map from one to the other.}. Unlike the Liouville description, it will also be completely well defined as a converging sum over positive quantities. In the semi-classical limit, this will result in an action which is bounded from below. Additional details and comparisons will be made in \cite{single-chord-future-paper}.

Suppose we want to compute the partition function. In order to do so, divide the thermal circle into $s$ segments, each of length $\beta_i = \beta / s$, where $s$ is an arbitrary number. Consider now
\begin{equation}
n_{ij}\text{ = the number of chords stretching between the $i$th and the $j$th segments,}\quad i,j=1,\cdots,s \,.
\end{equation}
We use the conventions that $n_{ii}=0$ and that $n_i \equiv \sum_j n_{ij}$ is the total number of chords leaving a segment. For a given collection of $n_{ij}$ there are many different chord diagrams\footnote{As long we do not push $s$ to be so large such that $n_i=0,1$ for all $i$.}. Our approach is to evaluate the weight of all the diagrams with a specific set of values of $n_{ij}$'s, and then sum over all these possible values. 

The motivation for this construction is the following. We will be interested in the case in which there are many chords in the diagrams, so $s$ will be large but such that the $n_{ij}$ are also typically large. We can think about it as if we are interested in some physical scale which is not close to the ``Planck scale"---the analogue of a single chord---but a much larger distance scale. Hence we want to coarse grain the microscopic UV data. On the other hand, we want to discuss distances much smaller then the thermal circle (which is the largest length scale if we are in Euclidean space), hence we need to formulate the theory in a way which still coarse grains over many chords but is still much finer than the thermal circle---the $s$ intervals above do precisely that. 

\begin{figure}[t]
\begin{subfigure}[t]{0.3\textwidth}
    \centering
    \includegraphics[width=1\textwidth]{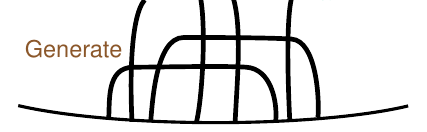}
    \subcaption{Generate.}
\end{subfigure}
\hfill
\begin{subfigure}[t]{0.3\textwidth}
    \centering
    \includegraphics[width=1\textwidth]{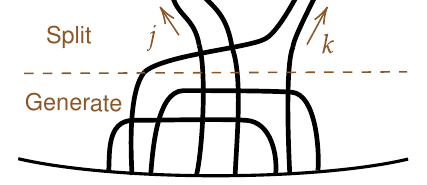}
    \subcaption{Split.}
\end{subfigure}
\hfill
\begin{subfigure}[t]{0.3\textwidth}
    \centering
    \includegraphics[width=1\textwidth]{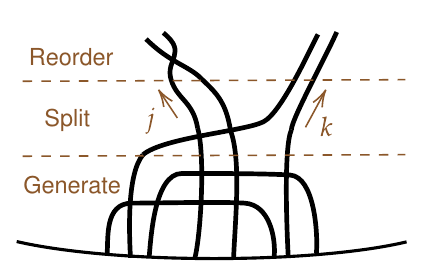}
    \subcaption{Reorder.}
\end{subfigure}

\centering
\begin{subfigure}[t]{0.3\textwidth}
    \includegraphics[width=1\textwidth]{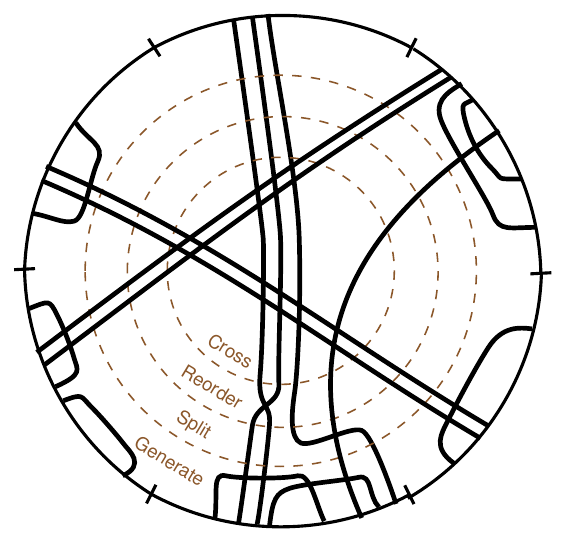}
    \subcaption{Cross.}
\end{subfigure}
\qquad\quad
\begin{subfigure}[t]{0.3\textwidth}
    \includegraphics[width=1\textwidth]{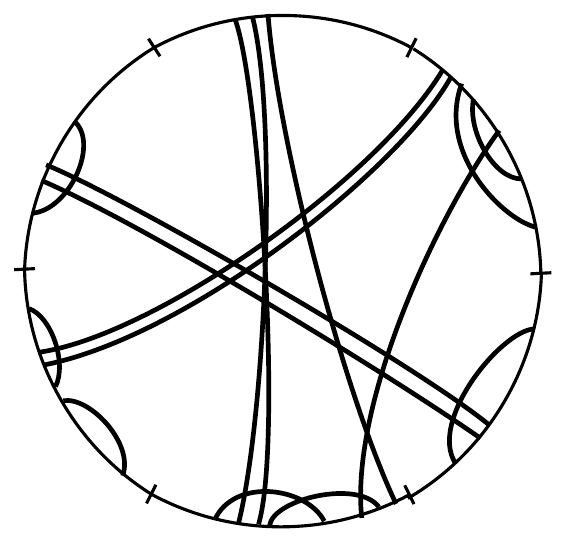}
    \subcaption{Without splitting into stages.}
\end{subfigure}
\caption{Illustration of the different steps in the coarse graining scheme.}
\label{fig:coarse_grain_single_chord}
\end{figure}

So we are left with evaluating the weights for a specific choice of $n_{ij}$'s, and then summing over the $n_{ij}$. The weights of the diagrams come from four sources, as illustrated in Figure~\ref{fig:coarse_grain_single_chord}. We annotate by bold face the names of the steps:
\begin{enumerate}[(a)]

    \item The amplitudes of {\bf generating} $n_i$ outgoing chords from the $i$th segment without counting the intersections of these chords with themselves, $\prod_i \braket{n_i | e^{-\beta_i T}| 0}$. 
    
    \item {\bf Splitting} the chords going from a single segment, $n_i$, into groups that will attach to the other segments, $n_{ij}$. The overall weight\footnote{The $q$-factorial $[n]_q!$ counts permutations keeping track of the number of inversions, and so counts the reordering of $n$ chords giving weight $q$ to any intersections.} is $\prod_i\binom{n_i}{n_{i1}\cdots n_{is}}_q$, defined in \eqref{eq:q multinomial def}. We note that chords within each group $n_{ij}$ do not intersect at this stage. 
    
    \item {\bf Reordering} the chords that stretch between two segments, and counting the weights associated with the intersections. Each group of chords $n_{ij}$ connecting two segments gives an additional overall weight of $[n_{ij}]_q!$, due to their possible intersections, so overall we have an additional factor of $\prod_i \prod_{j>i} [n_{ij}]_q!$.
    
    \item The {\bf crossings} of chords from different segments, which gives a factor of $q^{\sum_{i<k<j<\ell}n_{ij}n_{k\ell}}$.
\end{enumerate}
Combining them results in
\begin{equation}
\label{eq:exact_Z_coarse_grained}
    Z = \mathlarger{\sum}_{\{n_{ij}\}} \left[q^{\sum_{i<k<j<\ell}n_{ij}n_{k\ell}} \mathlarger{\prod}_{i=1}^s \left(\frac{[n_i]_q!}{\sqrt{\prod_{j\neq i}[n_{ij}]_q!}} \braket{n_i|e^{-\beta_i T}|0}\right)\right] \,,
\end{equation}
where the sum is over all non-negative integer $n_{ij}$'s.
We stress that no approximation has been made---this is the exact partition function of the model at any $q$ (and any number of segments, $s$). A similar expression which applies to special cases appears in (C.7), (C.10) of \cite{Berkooz:2018jqr} and in \cite{ismail1987combinatorics}.

\subsection{The semiclassical $q \to 1$ limit and a regulated Liouville} 

While this is an exact expression for any value of $q$, it is quite cumbersome to work with.  However, it simplifies considerably when we take the limit $q\to 1$ (equivalently, $\lambda \to 0$) and becomes semi-classical, in the sense that it is controlled by a saddle point approximation to a path integral. 

Let us split the partition function into the terms that depend explicitly on $n_{ij}$ and to the part that depends only on $n_i$, and write the partition function as 
\begin{align}
    Z &= \sum_{\{n_{ij}\}}\left[\prod_{i<k<j<\ell=1}^{s}q^{n_{ij}n_{k\ell}}\times\prod_{\substack{i,j=1\\ i\neq j}}^{s}\sqrt{\frac{\left(q^{n_{ij}};q\right)_{\infty}}{1-q^{n_{ij}}}}\times\prod_{i=1}^{s} C\left(n_{i}\right)\right]\,,\\
    C(n_{i}) &\equiv \sqrt{\left(q;q\right)_{\infty}^{1-s}\cdot \left(1-q\right)^{n_{i}}}\cdot [n_i]_q! \cdot \braket{n_{i}|e^{-\beta_{i}T}|0} \,.
    \label{eq:def C}
\end{align}
where we have used \eqref{eq:q factorial def} $[n_{ij}]_q!= {(q;q)_\infty \over (q^{n_{ij}};q)_\infty}{1-q^{n_{ij}}\over (1-q)^{n_{ij}}}$. We now take the semi-classical limit $q\to 1$, where the expansion \eqref{eq:q-Pochhammer approx} applies and the partition function becomes
\begin{equation}
\label{eq:action in semiclassical limit}
    Z = \sum_{\{n_{ij}\}}\left[e^{-\frac{1}{\lambda}\left[\sum_{i<k<j<\ell=1}^{s}\tilde{n}_{ij}\tilde{n}_{k\ell}+\frac{1}{2}\sum_{i\neq j=1}^{s}\Li_{2}\left(e^{-\tilde{n}_{ij}}\right)\right]+O\left(\lambda^{0}\right)} \times \prod_{i=1}^{s} C\left(n_{i}\right) \right] \;, \quad\qquad \tilde{n}_{ij}\equiv\lambda n_{ij}\;.
\end{equation}
Our next step is to take a continuum limit, where the number of segments is large, $s \gg 1$, such that the size of each segment is small, $\tilde \beta_i \equiv \sqrt{\lambda}\beta/s  \ll 1$. Since the segments are small, the (rescaled) number of chords that connect any two segments is also small\footnote{But still the non-rescaled number of chords is large, $n_{ij}\gg 1$.}, $\tilde{n}_{ij} \ll 1$. We will later check that our final result is consistent with this assumption. In this regime $e^{-\tilde n_{ij}}$ is close to 1 and we can expand the dilogarithm using \eqref{eq:Dilogarithm expansions} to obtain
\begin{equation}
    \label{eq:action before continuum limit}
    Z \equiv \sum_{\{n_{ij}\}} e^{-\frac{1}{\lambda} S[n_{ij}]} \,, 
\end{equation}
where our ``action'' $S$ is
\begin{equation}
\label{eq:discrete action}
    S = \sum_{i<k<j<\ell=1}^{s}\tilde{n}_{ij}\tilde{n}_{k\ell}+\frac{\pi^{2}s(s-1)}{12} + \frac{1}{2}\sum_{\substack{i,j=1\\i\neq j}}^{s}\tilde{n}_{ij}\left(\log\tilde{n}_{ij}-1\right) - \lambda\sum_{i=1}^{s} \log C(n_i) + O(\lambda, \tilde n_{ij}^2)\,.
\end{equation}
We still need to evaluate $C$, and in particular $\braket{n_i | e^{-\beta_i T}| 0}$. Let us demonstrate that in the limit\footnote{We do not take the limit of the denominator simply because it exactly cancels when computing $C$ in \eqref{eq:def C}.} $\lambda \to 0$, $\tilde\beta_i \to 0$ we have
\begin{equation}
    \label{eq:0-n amplitude high temp}
    \braket{n_i|e^{-\beta_i T}|0} = e^{\frac{\beta_i^2}{2}}\frac{(-\beta_i)^{n_i}}{[n_i]_q!}\left(1 + O(\tilde\beta_i, \lambda, \lambda n_i^2)\right) \,,
\end{equation}
We show this by first introducing the $q$-exponential, $e_{q}^{-\beta_iT}\equiv \sum_{n=0}^{\infty}\frac{(-\beta_iT)^{n}}{\left[n\right]_{q}!}$, and then studying the matrix element $\braket{n_i|e^{-\beta_i T}_q|0}$. By the $q$-Zassenhaus formula \cite{katriel1996q}, which truncates here since $[a,a^+]_q = 1$, the matrix element takes the simple form\footnote{We remind the reader that the inner product is $\braket{n|m} = \delta_{nm}$.}
\begin{equation} 
\braket{n_i|e^{-\beta_i T}_q|0} = \braket{n_i|e^{-\beta_i (a+a^+)}_q|0} = \braket{n_i|e_{q}^{-\beta_i a^{+}}e_{q}^{-\beta_i a}e_{q}^{\frac{\beta_i^{2}}{\left[2\right]_{q}}}|0} = e_q^{\frac{\beta_i^2}{[2]_q}}\frac{(-\beta_i)^{n_i}}{[n_i]_q!} \,.
\end{equation}
Let us now argue that both of the matrix elements are equal in our limit (that of the usual exponential and of the $q$-exponential). The $q$-exponential admits the plethystic expansion \eqref{eq:q exp plethystic}
\begin{equation}
\label{eq:def_q_exp}
e_{q}^{-\beta_iT} = \exp\left[\sum_{k=1}^{\infty}\frac{(-\beta_iT)^{k}\left(1-q\right)^{k}}{k\left(1-q^{k}\right)}\right] = \exp\left[ \sum_{k=1}^{\infty}\frac{1}{\lambda}\frac{\left(-\beta_i \lambda T\right)^k}{k^2}\left(1 + O(\lambda)\right) \right]\,.
\end{equation}
Insert a complete set of eigenstates \cite{Berkooz:2018jqr} of $T$ with eigenvalues $\frac{2\cos\theta}{\sqrt{1-q}}$, and in the limit $\lambda,\tilde \beta_i \to 0$ we get
\begin{multline}
    \braket{n_i|e^{-\beta_i T}_q|0} = \int d\theta \braket{n_i|\theta} \exp\left[ \sum_{k=1}^{\infty}\frac{1}{\lambda}\frac{\left(-2\tilde\beta_i \cos\theta\right)^k}{k^2}\left(1 + O(\lambda)\right) \right] \braket{\theta|0} \\ 
    = \int d\theta \braket{n_i|\theta} \exp\left[ \frac{1}{\lambda}(-2\tilde\beta_i \cos\theta)(1+O(\tilde\beta_i^2,\lambda))\right] \braket{\theta|0} = \braket{n_i|e^{-\beta_i T}|0} \left(1 + O(\tilde \beta_i^2, \lambda)\right) \,.
\end{multline}
Hence the matrix elements are equal, and \eqref{eq:0-n amplitude high temp} follows\footnote{Actually, there is a correction of order $O(\lambda n_i^2)$ to this formula from taking into account higher order corrections in this derivation. These terms, however, are subleading in the subsequent continuum limit.}. We present another derivation in Appendix~\ref{app:matrix_element}, which evaluates the explicit form of the matrix element using a saddle point technique. 

After substituting \eqref{eq:0-n amplitude high temp} into \eqref{eq:def C} and using \eqref{eq:q-Pochhammer approx} we find the explicit form of $C$,
\begin{multline}
    C(n_i) = \lambda^{\frac{s-1}{4}+\frac{n_i}{2}} e^{\frac{(s-1)}{\lambda}\frac{\pi^2}{12}} e^{-\frac{\beta_i^2}{2}}(-\beta_i)^{n_i} 
    \\ 
    = (-1)^{n_i}\lambda^{\frac{s-1}{4}} \exp\left[-\frac{1}{\lambda}\left(-\frac{(s-1)\pi^2}{12} - \frac{\tilde\beta_i^{2}}{2}-\tilde{n}_{i}\log(\sqrt{\lambda}\beta_i) \right)\right] \,.
\end{multline}
The factors of $(-1)^{n_i}$ cancel in the product $\prod_i C(n_i)$, as each chord exits a segment and enters another and so $\sum_i n_i$ is even, and we can ignore the $(-1)$ factors from now on. The second term, $\tilde\beta_i^2$, is subleading in the continuum limit in \eqref{eq:discrete action}. By dimensional analysis we can restore the dimensionful parameter $\cJ$ which we set to 1 at the beginning of the computation by simply replacing $\beta_i$ by $\beta_i \cJ$. Moreover, from here on we find it more convenient to switch to the normalization \eqref{eq:bJnormalization}, which ultimately gives
\begin{equation}
\label{eq:tilde C high temp}
    C(n_i) = \lambda^{\frac{s-1}{4}} \exp\left[-\frac{1}{\lambda}\left(-\frac{(s-1)\pi^2}{12} -\tilde{n}_{i}\log(\beta_i\bJ) \right)\right] \,.
\end{equation}

Next we take the continuum limit, $s\rightarrow \infty$, in the action \eqref{eq:discrete action}. The index of a segment, $i$, now becomes a Euclidean time, $\tau_i$, where $i/s = \tau_i / \beta$. We introduce a rescaled continuum variable, and by slight abuse of notation denote it by $n$,
\begin{equation} \label{eq:Continuum_n}
\tilde{n}_{ij}\equiv\frac{\beta^{2}}{s^{2}} n\left(\tau_{i},\tau_{j}\right) \,,
\end{equation}
such that sums over $\tilde n$ become integrals over $n$, i.e.,  $\sum_{i,j=1}^{s}\tilde{n}_{ij}\to\int_{0}^{\beta}d\tau_{1}\int_{0}^{\beta}d\tau_{2}\, n\left(\tau_{1},\tau_{2}\right)$. The summation over all possible values of $n_{ij}$ becomes a path integral over the non-negative $n$ with a uniform measure. 

The first term in the action \eqref{eq:discrete action} becomes 
\begin{equation}
    \frac{1}{4}\int_{0}^{\beta}d\tau_{1}\int_{0}^{\beta}d\tau_{2}\int_{\tau_{1}}^{\tau_{2}}d\tau_{3}\int_{\tau_{2}}^{\tau_{1}}d\tau_{4}\, n(\tau_1,\tau_2)n(\tau_3,\tau_4) \,,
\end{equation}
where the bounds of the integral are written for $\tau$'s that live on a circle, and flipping the limits of the integration should be understood as integrating over the other side of the circle. The pre-factor corrects the over-counting when switching $\tau_1$ and $\tau_2$, and when switching the pairs $\tau_{1,2}$ and $\tau_{3,4}$. As for the continuum limit of the other term, the constant term cancels between \eqref{eq:discrete action} and \eqref{eq:tilde C high temp}. For the log term\footnote{Note that $n$ is dimensionful, so one should combine the arguments of both logs to $n/\beta_i^2$, but separating them makes the various cancellations more obvious.}
\begin{equation}
    \frac{1}{2}\sum_{i,j} \tilde n_{ij} \left(\log \tilde n_{ij} - 1\right) \longrightarrow \frac{1}{2}\int_0^\beta d\tau_1 \int_0^\beta d\tau_2 \, n(\tau_1,\tau_2) \left[\log n(\tau_1,\tau_2) - 1 - 2\log(\beta_i)\right] \,
\end{equation}
and the last term cancels against \eqref{eq:tilde C high temp}. Finally, we are left with a simple path integral expression for the partition function,
\begin{equation}
    \label{eq:partition function path int 1 chord}
    Z = \int Dn \, \exp\left(-\frac{1}{\lambda} S[n]\right) \,,
\end{equation}
with a flat measure over positive and symmetric $n(\tau_i,\tau_j)$ and $n(\tau,\tau) = 0$. We discuss the measure a bit more below and show that the final result is independent of the unphysical number of segments, $s$. Our action is 
\begin{multline}
\label{eq:Continuum action}
    S = \frac{1}{4}\int_{0}^{\beta}d\tau_{1}\int_{0}^{\beta}d\tau_{2}\int_{\tau_{1}}^{\tau_{2}}d\tau_{3}\int_{\tau_{2}}^{\tau_{1}}d\tau_{4}\, n(\tau_1,\tau_2)n(\tau_3,\tau_4) \\ + \frac{1}{2} \int_{0}^{\beta}d\tau_1 \int_{0}^{\beta}d\tau_2\, n(\tau_1,\tau_2)\left[\log\left(\frac{n(\tau_1,\tau_2)}{\bJ^2}\right) - 1\right] \,.
\end{multline}
Having originated from the finite sum over chord diagrams, this is a completely well defined path integral, and indeed the action is bounded from below. While the action looks a priori very different than \eqref{eq:Liouville g action}, we will see momentarily that that it reproduces the equations of motion \eqref{eq:Liouville_eq_GSigma} and has the same on-shell action. We therefore regard it as a well-posed version of that action.

\subsection{The saddle point}
Since we took $\lambda \to 0$ the action is dominated by its saddle point, which turns out to be equivalent to that of the Liouville approach of Section~\ref{sec:Liouville}. Further comments about the relation between the two approaches will be made in \cite{single-chord-future-paper}. The saddle point equation is
\begin{equation}
    \label{eq:n_hat_eom} \int_{\tau_{1}}^{\tau_{2}}d\tau_{3}\int_{\tau_{2}}^{\tau_{1}}d\tau_{4}\, n\left(\tau_{3},\tau_{4}\right) + \log \left(\frac{n\left(\tau_{1},\tau_{2}\right)}{\bJ^2}\right) = 0 \;.
\end{equation}
To solve this integral equation, we define\footnote{We are a little sloppy with \eqref{eq:def g}. As the integral is really defined on a circle, the proper definition on a real line should periodically extend $n(\tau_1,\tau_2)$ and define
\begin{equation}  
g\left(\tau_{1},\tau_{2}\right) \equiv - \theta(\tau_1-\tau_2)\int_{\tau_{1}}^{\tau_{2}+\beta}d\tau_{3}\int_{\tau_{2}}^{\tau_{1}}d\tau_{4}\, n\left(\tau_{3},\tau_{4}\right)- \theta(\tau_2-\tau_1)\int_{\tau_{1}}^{\tau_{2}}d\tau_{3}\int_{\tau_{2}}^{\tau_{1}+\beta}d\tau_{4}\, n\left(\tau_{3},\tau_{4}\right) \end{equation}
for $\tau_1,\tau_2 \in [0,\beta]$ and then periodically extend $g$. This gives the modified relation 
\begin{equation}
-\frac{1}{2}\partial_{\tau_1}\partial_{\tau_1}g(\tau_1,\tau_2)=n(\tau_1,\tau_2) - \delta(\tau_1-\tau_2) \int_{0}^\beta d\tau_3 n(\tau_3, \tau_1).\end{equation}
The addition of the contact term is needed to make the boundary conditions \eqref{eqn:gBdyConditions} consistent with the positive-definiteness of $n(\tau_1,\tau_2)$, but otherwise does not affect our saddle-point analysis. \label{foo:counterterms}} % end of footnote
another function, $g\left(\tau_{1},\tau_{2}\right)$, that counts (up to a sign) all the chords going across $\tau_1$ and $\tau_2$, i.e., those that intersect the chords connecting $\tau_{1}$ and $\tau_{2}$, 
\begin{equation}
\label{eq:def g}
    g\left(\tau_{1},\tau_{2}\right) \equiv -\int_{\tau_{1}}^{\tau_{2}}d\tau_{3}\int_{\tau_{2}}^{\tau_{1}}d\tau_{4}\, n\left(\tau_{3},\tau_{4}\right) \quad\implies\quad n\left(\tau_{1},\tau_{2}\right) = -\frac{1}{2}\partial_{\tau_{1}}\partial_{\tau_{2}}g\left(\tau_{1},\tau_{2}\right) \;, \qquad \text{for }\tau_1 \neq \tau_2 \,.
\end{equation}
In terms of $g$ and after slight rearrangement, the equation of motion is recognized as the Liouville equation of motion,
\begin{equation} \label{eq:Liouville_eq}
    \partial_{\tau_{1}}\partial_{\tau_{2}}g\left(\tau_{1},\tau_{2}\right) + 2\bJ^2 e^{g\left(\tau_{1},\tau_{2}\right)} = 0\,,
\end{equation}
with the boundary conditions coming from \eqref{eq:def g}
\begin{equation}\label{eqn:gBdyConditions}
    g(0,0) = g(0,\beta) = g(\beta,0) = g(\beta,\beta) = 0\,.
\end{equation}
As we will see momentarily, this is intimately related to the 2-point functions, and agrees with the result\footnote{Our $\bJ$ is denoted there as $\cJ$, and we compare to their large $q$ limit. We use here the notation of \cite{Goel:2023svz}.} of \cite{Maldacena:2016hyu}. 
One solution to the saddle point equations is 
\begin{equation} \label{eq:g_saddle_value}
g(\tau_1, \tau_2) = 2\log\left[\frac{\cos\left(\frac{\pi v}{2}\right)}{\cos\left(\frac{\pi v}{2}\left(1-\frac{2|\tau_2 - \tau_1|}{\beta}\right)\right)}\right] \,, \qquad \beta\bJ = \frac{\pi v}{\cos\frac{\pi v}{2}} \,.
\end{equation}
The saddle for the variable $g$ is exactly the same as the one for the similarly named variable in the collective field approach of \cite{Maldacena:2016hyu, Cotler:2016fpe, Lin:2023trc}.  The saddle point value for the number of chords going between two segments is
\begin{equation}
\label{eq:n_hat_saddle_value}
    n(\tau_2,\tau_1) = \bJ^2 \frac{\cos^2\left(\frac{\pi v}{2}\right)}{\cos^2\left[\frac{\pi v}{2}\left(1 - \frac{2\left|\tau_2-\tau_1\right|}{\beta}\right)\right]} \,,
\end{equation}
and in terms of the original discrete $n_{ij}$ variable,
\begin{equation}
    n_{ij} = \frac{\beta^2}{s^2\lambda}  n(\tau_i, \tau_j) = \frac{1}{\lambda}\frac{(\beta\bJ)^2}{s^2}\left[\frac{\cos\left(\frac{\pi v}{2}\right)}{\cos\left(\frac{\pi v}{2}\left(1-\frac{2|i - j|}{s}\right)\right)}\right]^2 \;.
\end{equation} 
Since our approach relies on having many chords going out of each segment, we can only go from the continuum variables back to the discrete ones when $n_{ij} \gg 1$, i.e., $\frac{\beta \bJ}{s} \equiv \tilde\beta_i \gg \sqrt{\lambda}$. Moreover, we neglected terms of order $O(\tilde n_{ij}^2)$ along the way, which at the saddle point are of order $O(\tilde\beta_i^4)$. The overall range of validity of our approximation is thus $1 \gg \tilde\beta_i \gg \sqrt{\lambda}$, which is consistent with the order of limits we took -- first $\lambda \to 0$, then the continuum limit.

\paragraph{The continuum limit of the measure} 
Let us argue that the result of the path integral expression is independent of the arbitrary number of segments, $s$. When we changed the summation over all possible values of the discrete $n_{ij}$ to a path integral over the continuous $\tilde n_{ij} \equiv \lambda n_{ij}$, we had to compensate by
\begin{equation}
\label{eq:Transition to tilde n path integral}
    \sum_{\{n_{ij}\}} \to \lambda^{-\frac{s(s-1)}{2}} \int \cD \tilde n_{ij} \;,
\end{equation}
where the power comes from the fact that the $n_{ij}$'s are symmetric and the diagonal terms vanish, and we have $s$ segments. Additionally, the one-loop determinant for the saddle point in $\tilde n$ will contribute a $\sqrt{\lambda}$ for each mode, or $\lambda^{\frac{s(s-1)}{4}}$ overall. We also have a contribution from the $C(n_i)$ due to \eqref{eq:tilde C high temp}. There are $s$ such contributions, and so overall there is another factor of $\lambda^{\frac{s(s-1)} {4}}$. These last two factors cancel against the one coming from \eqref{eq:Transition to tilde n path integral}, such that the $s$ dependence cancels and none of the terms to this order depends on the number of segments. Similarly, the transition to the path integral over $n_{ij} \equiv \frac{\beta^2}{s^2}\tilde n_{ij}$ will require a factor of $\left(\beta/s\right)^{s(s-1)}$ which will cancel exactly against the factor coming from the one-loop determinants over these bilocal variables.

\paragraph{The partition function} 
As our saddle point equations are the same as those in \cite{Maldacena:2016hyu, Cotler:2016fpe} whereas the off-shell action is different, let us verify that they agree on-shell and that we can reproduce their results. At the saddle point, to leading order in $\lambda$, the partition function is determined by evaluating the action \eqref{eq:Continuum action} at the saddle point. Using the equations of motion \eqref{eq:n_hat_eom} we re-write
\begin{equation}
\begin{aligned}
    S &= \frac{1}{4}\iint_0^\beta\left[n(\tau_1,\tau_2)\left(\log\left(\frac{n(\tau_1,\tau_2)}{\bJ^2}\right) - 2\right) \right]\,d\tau_1\,d\tau_2\,,\qquad\qquad \text{(at the saddle point)} 
\end{aligned}
\end{equation}
and then substitute the value for $n_{ij}$ at the saddle, \eqref{eq:n_hat_saddle_value}. One eventually finds that the partition function at the saddle point agrees with the results of \cite{Maldacena:2016hyu},
\begin{equation}
    Z = \exp\left[-\frac{1}{\lambda}\frac{\pi^2 v^2}{2} + \frac{2}{\lambda}\pi v \tan\left(\frac{\pi v}{2}\right)\right] \,.
\end{equation}

\paragraph{The 2-point function}
Let us compute the two point function of a random operator $M$ \cite{Berkooz:2018jqr, Berkooz:2018qkz}, see Appendix~\ref{app:CrshCrdDiag} for its definition. The two-point function is 
$G(\tau_1,\tau_2) = \frac{1}{Z}\vev{\Tr \left(e^{-\beta H}M(\tau_1)M(\tau_2)\right)}$. It can again be expanded into moments, and each moment us computed by a weighted sum over all chord diagrams, where there is a single matter chord stretching between $\tau_1$ and $\tau_2$. The weight for an intersection of a matter chord with a Hamiltonian chord is $\tilde q = e^{-\tilde \lambda}$. Intersections between Hamiltonian chords and themselves are of weight $q = e^{-\lambda}$, as before. We will be interested in the limit $\lambda\to 0$, where we take $\Delta \equiv \frac{\tilde\lambda}{\lambda}$ to be finite.

This can be computed using our method. We have a single matter chord that stretches between $\tau_1$ and $\tau_2$. We again divide the thermal circle into $s$ segments\footnote{Before we assumed the segments to be equal. Here, the segments that end on the matter chords might be of different lengths. In the continuum limit, this doesn't matter, but it might slightly change the subleading terms in $1/s$.} with the matter chord stemming out of the ends of segments. The action now gets an additional contribution due to Hamiltonian chords that intersect the matter chord, i.e., those who have ends on opposite sides of the thermal circle with respect to $\tau_{1,2}$. We denote the sum over such chords as $\sum^\prime_{i,j}$. For every chord diagram, this additional weight is $e^{-\tilde \lambda \sum^\prime_{i,j} n_{ij}} = e^{-\Delta \sum^\prime_{i,j} \tilde n_{ij}}$. This has a natural continuum limit,
\begin{equation}
    G(\tau_1,\tau_2) = \braket{e^{\Delta \cdot g(\tau_1,\tau_2)}} = \frac{1}{Z} \int \cD  n\, e^{-\frac{1}{\lambda}S - \Delta \int_{\tau_1}^{\tau_2}d\tau_3\int_{\tau_2}^{\tau_1}d\tau_{4}\,n(\tau_3,\tau_4)} \,.
\end{equation}
When $\Delta$ is finite the insertion of the additional operator does not affect the saddle point, and therefore to leading order in $\lambda$ the two point function agrees with \cite{Maldacena:2016hyu, Goel:2023svz}, and is given by
\begin{equation}
    G(\tau_1,\tau_2) = \left[\frac{\cos^2\left(\frac{\pi v}{2}\right)}{\cos^2\left[\frac{\pi v}{2}\left(1-\frac{2|\tau_{1}-\tau_{2}|}{\beta}\right)\right]}\right]^{\tilde{\lambda}/\lambda} \,.
\end{equation}

\paragraph{Relation to kinematic space} In the semi-classical limit of the holographic dual, the two point function of some boundary operator dual to a field of mass $m$ is related to $\ell(\tau_1,\tau_2)$, the (renormalized) length of a geodesic connecting the boundary at the points $\tau_1, \tau_2$,
\begin{equation}
    G(\tau_1, \tau_2) = e^{-m \ell(\tau_1,\tau_2)} \;.
\end{equation}
On the other hand, from our chord computation we see that
\begin{equation}
    G(\tau_1,\tau_2) = e^{\frac{\tilde\lambda}{\lambda} g(\tau_1,\tau_2)} \;,
\end{equation}
and so we can identify $g(\tau_1,\tau_2) \equiv -\ell(\tau_1,\tau_2)$ and $\tilde\lambda/\lambda$ with the mass. Since we found earlier that the chord density is related to $g$ by $n(\tau_1,\tau_2) = -\frac{1}{2}\partial_{\tau_1}\partial_{\tau_2} g(\tau_1,\tau_2)$, we find that the chord density is given by the second derivative of the length of a boundary geodesic with respect to its endpoints (up to a constant factor), which is exactly the definition of the Crofton form \cite{Czech:2016xec}. Therefore the chord density $n(\tau_1,\tau_2)$ is the natural measure on the space of boundary geodesics of the holographic dual space in the semi-classical limit.

\section{A path integral for two types of chords}
\label{sec:path integral two chords}

Here we generalize the coarse graining technique of the previous section to the case of systems with multiple chord species. One particularly interesting example is that of the chaos to integrability transition of \eqref{eq:Cha_integ_Ham}, but our techniques apply also to cases with more general weights for chord intersections, and they are considered in Section~\ref{sec:generic 2 chords}.

The outline of this section is the following. In Section~\ref{sec:MltiPthInteg} we write the formal path integral expression and discuss its $q\rightarrow 1$ limit. This section relies a lot on the finer details of chord constructions, so readers interested in the final equations for the integrability-chaos transition can go to the following section, \ref{sec:MltiSddl}. In that section we compute the saddle point equations \eqref{eq:eom 2 chords} of the $q\rightarrow 1$ action. In the next section we analyze these equations and the resulting transition phase structure.

\subsection{The multi chord path integral}\label{sec:MltiPthInteg}

\subsubsection{The exact path integral}
Consider a system with two types of chords, whose microscopic origin comes from averaging over two types of random couplings, as discussed in Section~\ref{sec:model-defs}. We consider the system $\eqref{eq:Cha_integ_Ham}$, which has two types of chords---the $n$-chords, associated with the chaotic Hamiltonian, and the $z$-chords, associated with the integrable one. Intersections of $n$-chords with themselves or with $z$-chords are weighted\footnote{Recall that we are discussing the case where the lengths of $H_{\text{chaotic}}$ and $H_{\text{Integrable}}$ are the same.} by a factor of $q = e^{-\lambda}$, and intersections of $z$-chords with themselves are weighted by $1$.

As in the single chord case, we would like to to write the sum over all chord diagrams as a path integral. We will divide the boundary of the chord diagram into $s$ segments, each of length $\beta_i = \beta/s$, and denote by $n_{ij}$ and $z_{ij}$ the number of $n$-chords and $z$-chords connecting the $i$th and $j$th segments. We will also denote the overall number of $n$-chords stemming out of the $i$th segment by $n_{i} = \sum_{j\neq i} n_{ij}$, and the overall number of $z$-chords by $z_{i} = \sum_{j\neq i} z_{ij}$.

Before taking any limit the ordering of the two types of chords within each segment---which is an $n$-chord and which is a $z$-chord---also matters, as different orderings are associated to different diagrams and carry different weights. We called this ordering $\vec r_i$ in Section~\ref{sec:model-defs}. Similarly, we also need to keep track of the order of the chords that exit the $i$th segment towards\footnote{Our notation is such that $\vec r_{ij}$ is not necessarily symmetric. In the limit that concerns us, this will have no effect.} the $j$th segment, and we will denote it by $\vec r_{ij}$.

\begin{figure}[t]
\begin{subfigure}[t]{0.3\textwidth}
    \centering
    \includegraphics[width=1\textwidth]{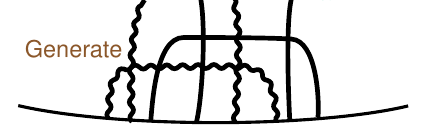}
    \subcaption{Generate.}
\end{subfigure}
\hfill
\begin{subfigure}[t]{0.3\textwidth}
    \centering
    \includegraphics[width=1\textwidth]{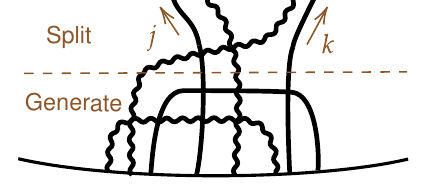}
    \subcaption{Split.}
\end{subfigure}
\hfill
\begin{subfigure}[t]{0.3\textwidth}
    \centering
    \includegraphics[width=1\textwidth]{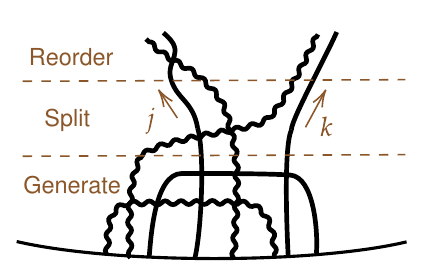}
    \subcaption{Reorder.}
\end{subfigure}

\centering
\begin{subfigure}[t]{0.3\textwidth}
    \includegraphics[width=1\textwidth]{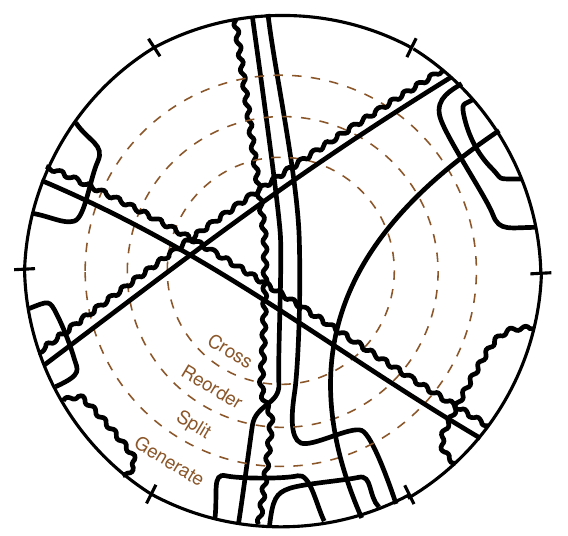}
    \subcaption{Cross.}
\end{subfigure}
\caption{Illustration of the different steps in the coarse graining scheme for two types of chords.}
\label{fig:coarse_grain_two_chords}
\end{figure}

The weight of the chord diagrams is found by multiplying four factors, as illustrated in Figure~\ref{fig:coarse_grain_two_chords}:
\begin{enumerate}[(a)]
    \item \sloppy The amplitudes of generating $n_i$, $z_i$ outgoing chords from the $i$th segment in the order $\vec r_i$ without counting the intersections of these chords with themselves, $\prod_{i=1}^s \llangle n_i, z_i; \vec r_i| e^{-\beta_i T}|0 \rangle$, see the discussion around \eqref{eq:not inner product}.
    \item Splitting the chords going from a single segment, $n_i$, $z_i$ into groups that will attach to the other segments, $n_{ij}$, $z_{ij}$. This term depends only on the chords emanating from each interval, i.e., it is of the form $\prod_{i=1}^s Z_{\text{split}}\left(\{n_{ij},z_{ij};\vec r_{ij}\}_{j=1}^s\right)$. We do not present the closed form for $Z_{\text{split}}$ for general $q$.
    \item Reordering the chords that stretch between the same two segments. The weight associated with this part is denoted by $\prod_{j>i} Z_{\text{reorder}}\left(n_{ij},z_{ij};\vec r_{ij}\right)$. 
    \item \sloppy The crossing of chords that connect different segments, whose weight is $q^{\sum_{i<k<j<\ell}\left[n_{ij}n_{k\ell}+n_{ij}z_{k\ell}+z_{ij}n_{k\ell}\right]}$.
\end{enumerate}
The overall expression for the partition function is then
\begin{multline}
    Z = \sum_{\{n_{ij},z_{ij},\vec r_{ij}\}} \Bigg[q^{\sum_{i<k<j<\ell}\left[n_{ij}n_{k\ell}+n_{ij}z_{k\ell}+z_{ij}n_{k\ell}\right]} \prod_{i=1}^s \Big[\llangle n_i, z_i; \vec r_i| e^{-\beta_i T}|0 \rangle \\
    \times  Z_{\text{split}}\left(\{n_{ij},z_{ij};\vec r_{ij}\}\right) \times \prod_{j>i} Z_{\text{reorder}}\left(n_{ij},z_{ij};\vec r_{ij}\right)\Big]  \Bigg] \,.
\end{multline}

\subsubsection{The $q\rightarrow 1$ limit}

We would like to study this expression in the semi-classical limit $q \to 1$, and in a continuum limit where there are many segments of small length, $\tilde\beta_i \equiv \sqrt{\lambda}\beta_i  \ll 1$, where as in the previous section we work with the normalization $\cJ = 1$ \eqref{eq:cJ norm} for now, and reinstate that of \eqref{eq:bJnormalization} when writing the final path integral expression. As in the previous section, in this limit we can also take $\tilde z_{ij}, \tilde n_{ij} \ll 1$, where $\tilde z_{ij} = \lambda z_{ij}$ and $\tilde n_{ij} = \lambda n_{ij}$. Luckily, at this limit the various factors considerably simplify and become independent of the ordering. The computation is rather cumbersome, and readers who prefer can jump directly to the final expression, \eqref{eq:2-chord partition function before limit}: 
\begin{enumerate}[(a)]
    \item Let us denote the quantity of interest by
    $M^{(i)}_{q,q,1} \equiv \llangle n_i, z_i; \vec r_i| e^{-\beta_i T}|0 \rangle$, where the subscripts denote the weights for intersections between two $n$-chords, $n$-chord and $z$-chord, and two $z$-chords, respectively. Since $M^{(i)}_{q,q,1}$ is composed of a sum over positive quantities, we have
    \begin{equation}
    M_{q,q,q}^{(i)} \leq M_{q,q,1}^{(i)} \leq M_{1,1,1}^{(i)} \,.
    \end{equation}
    \begin{itemize}
        \item For the upper bound, $M_{1,1,1}^{(i)}$, we use the Zassenhaus formula which truncates when $[a_i,a_j^+] = \delta_{ij}$, 
    \begin{multline}
    \label{eq:2 chord upper bound amplitude}
        \llangle n_i,z_i;\vec r_i|e^{-\beta_i(\nu a_n + \kappa a_z + \nu a_n^+ + \kappa a_z^+)}|0 \rangle \\ = \llangle n_i,z_i;\vec r_i|e^{-\beta_i(\nu a_n^+ + \kappa a_z^+)} e^{-\beta_i(\nu a_n + \kappa a_z)} e^{\frac{\beta_i^2}{2}}|0 \rangle = e^{\frac{\beta_i^2}{2}} \frac{(-\nu\beta_i)^{n_i}(-\kappa\beta_i)^{z_i}}{(n_i + z_i)!} \,.
    \end{multline}
    
    \item For the lower bound, $M_{q,q,q}^{(i)}$, consider the case where $[a_i,a_j^+]_q = \delta_{ij}$, and define
    \begin{equation}
    \label{eq:2-chord to 1-chord creation operator}
    a = \nu a_n + \kappa a_z \,, \qquad a^+ = \nu a_n^+ + \kappa a_z^+ \,, \qquad [a,a^+]_q = 1\,.
    \end{equation}
    As in the single chord case, we can first study a related amplitude given by the $q$-exponential, $\llangle n_i,z_i;\vec r_i|e^{-\beta_i (a+a^+)}_q|0 \rangle$. By virtue of the $q$-Zassenhaus formula \cite{katriel1996q}, which truncates here, we find
    \begin{equation} 
    %\llangle n_i,z_i;\vec r_i|e^{-\beta_i T}_q|0 \rangle = 
    \llangle n_i,z_i;\vec r_i|e^{-\beta_i (a+a^+)}_q|0 \rangle = \llangle n_i,z_i; \vec r_i|e_{q}^{-\beta_i a^{+}}e_{q}^{-\beta_i a}e_{q}^{\frac{\beta_i^{2}}{\left[2\right]_{q}}}|0 \rangle
    = e_q^{\frac{\beta_i^2}{[2]_q}}\frac{(-\nu\beta_i)^{n_i} (-\kappa\beta)^{z_i}}{[n_i+z_i]_q!} \,.
    \end{equation}
    In order to argue that in the limit $\lambda \to 0$, $\tilde \beta_i \to 0$ this amplitude is the same as $M_{q,q,q}^{(i)}$, note that the transfer matrix acting on the vacuum $\ket{0}$ only moves us within the sub-Hilbert space generated by $a^+$ from the vacuum, ${\cH_a = ~\{(a^+)^k\ket{0} | k\in \bZ_+ \}}$, and therefore we can restrict ourselves to it. Within this subspace we can diagonalize the transfer matrix using the same states $\ket{\theta}$ and with the same eigenvalues $\frac{2\cos\theta}{\sqrt{1-q}}$ as in the single chord case \cite{Berkooz:2018qkz}. We repeat the argument from below \eqref{eq:q exp plethystic} to find that in this limit the two amplitudes are the same, and we find
    \begin{equation}
        \label{eq:2 chord lower bound amplitude}
        M_{q,q,q}^{(i)} = e^{\frac{\beta_i^2}{2}}\frac{(-\nu\beta_i)^{n_i} (-\kappa\beta_i)^{z_i}}{(n_i+z_i)!}\left(1 + O(\tilde\beta_i, \lambda)\right) \,.
    \end{equation}
    \end{itemize}
    
    We see that in the semi-classical and continuum limits the upper bound \eqref{eq:2 chord upper bound amplitude} and lower bound \eqref{eq:2 chord lower bound amplitude} are the same, and so our matrix element is 
    \begin{equation}
        \label{eq:2 chord amplitude}
        \llangle n_i, z_i; \vec r_i| e^{-\beta_i T}|0 \rangle = e^{\frac{\beta_i^2}{2}}\frac{(-\nu\beta_i)^{n_i} (-\kappa\beta_i)^{z_i}}{(n_i+z_i)!}\left(1 + O(\tilde\beta_i, \lambda)\right) \,.
    \end{equation}
    \item Splitting. As we work in the $q\to 1$ limit, the leading order of the combinatorial factor comes just from the number of ways of splitting the $n_i$ and $z_i$ chords into subsets $n_{ij}$, $z_{ij}$, which is $Z_{\text{split}}\left(\{n_{ij},z_{ij};\vec r_{ij}\}_{j=1}^s\right) = \binom{n_i}{n_{i1}\cdots n_{is}}\binom{z_i}{z_{i1}\cdots z_{is}}$. 
    \item Reordering. Each group of $n$-chords $n_{ij}$ connecting two segments gives an additional overall weight of $n_{ij}!$ due to their possible intersections. A similar factor of $z_{ij}!$ comes from the intersections of the $z$-chords, $Z_{\text{reorder}}\left(n_{ij},z_{ij};\vec r_{ij}\right) = n_{ij}!z_{ij}!$.
    \item The crossing of chords from different segments gives a factor of $q^{\sum_{i<k<j<\ell}\left[n_{ij}n_{k\ell}+n_{ij}z_{k\ell}+z_{ij}n_{k\ell}\right]}$.
\end{enumerate}
Note that we have taken the $\lambda \to 0$ limit in slightly different ways in the different terms. In the first three factors, (a)--(c), we have effectively taken a strict $\lambda \to 0$ limit, while in the "bulk" crossing term (d) we have kept the leading $\lambda$ dependence. This is because we will shortly take the continuum limit $s\rightarrow \infty$ and use the fact that the first three terms are separate for each small interval, and therefore these factors depend only on the number of chords that leave the segment. In that limit, this becomes a negligible contribution, and the ambiguities in weights associated with crossing and ordering within each short interval disappear (in fact they are shifted into the last ``bulk'' term). Hence we can strictly take $\lambda\rightarrow 0$ with impunity. In the bulk crossing term, (d), the contribution is in some sense macroscopic, and in order to capture any non-trivial contribution to it we need to account for its $\lambda$ dependence. Our approach below is essentially to balance between the factors (a)--(c) and (d).

In the semi-classical ($q \to 1$) and continuum ($\tilde\beta_i \to 0$) limits none of these factors depend explicitly on the ordering $\vec r_i$, and so the sum over the different orderings simply gives another combinatorial factor, $\binom{n_i+z_i}{n_i}$. Overall, the partition function becomes
\begin{equation}
\label{eq:2-chord partition function before limit}
Z = \sum_{\{n_{ij},z_{ij}\}} \left[q^{\sum_{i<k<j<\ell}\left[n_{ij}n_{k\ell}+n_{ij}z_{k\ell}+z_{ij}n_{k\ell}\right]} \prod_{i=1}^s \left[e^{\frac{\beta_i^2}{2}} \frac{(-\nu\beta_i)^{n_i}(-\kappa\beta_i)^{z_i}}{\sqrt{\prod_{j\neq i} n_{ij}! z_{ij}!}}\right]\right] \,.
\end{equation}
Taking the limit\footnote{Technically, the matrix element was computed in the limit where $\lambda \to 0$ while keeping $\tilde \beta = \sqrt{\lambda}\beta_i$ and $\tilde n_{ij} = \lambda n_{ij}$ fixed, and only then taking the $\tilde \beta_i, \tilde n_{ij} \to 0$ limit, while the combinatorial factor was computed when $\lambda\to 0$ first with $n_{ij}$ fixed, then taking the $n_{ij} \gg 1$ limit. In the single chord case the leading order for both limits agrees, and therefore we expect these limits to commute also in the two chord case.} where the number of chords is large, $z_{ij}, n_{ij} \gg 1$, we can write the partition function as
\begin{equation}
    Z = \sum_{\{n_{ij}\},\{z_{ij}\}} e^{-\frac{1}{\lambda}S} \,,
\end{equation}
with the action\footnote{The factors of $(-1)$ cancel as each chord is counted twice, and so the overall power is always even.} written in terms of the new variables $\tilde n_{ij} \equiv \lambda n_{ij}$ and $\tilde z_{ij} \equiv \lambda z_{ij}$, and restoring the dimensionful coupling $\bJ$ \eqref{eq:def bJ} via dimensional analysis as in the previous section,
\begin{multline}
S = \sum_{i<k<j<\ell}\left[\tilde n_{ij}\tilde n_{k\ell}+\tilde n_{ij} \tilde z_{k\ell}+ \tilde z_{ij} \tilde n_{k\ell}\right] + \frac{1}{2}\sum_{i,j=1}^s \Bigg[\tilde n_{ij}\left(\log\left[\frac{\tilde n_{ij}}{(\nu\beta_i\bJ)^2}\right]-1\right) \\ 
+ \tilde z_{ij}\left(\log\left[\frac{\tilde z_{ij}}{(\kappa\beta_i\bJ)^2}\right]-1\right)\Bigg] \,.
\end{multline}
As in the previous section, in the continuum limit the $i$th segment corresponds to a specific Euclidean time $\tau_i$, such that $i/s = \tau_i/\beta$. It is then convenient to work with rescaled variables $n(\tau_i,\tau_j)$, $z(\tau_i,\tau_j)$,
\begin{equation} \label{eq:Continuum_n_z}
\tilde{n}_{ij}\equiv\frac{\beta^{2}}{s^{2}} n\left(\tau_{i},\tau_{j}\right) \,,  \qquad \tilde{z}_{ij}\equiv\frac{\beta^{2}}{s^{2}} z\left(\tau_{i},\tau_{j}\right)\,,
\end{equation}
so when $s\to \infty$ sums over the discrete quantities become integrals over the continuous ones, i.e.,  $\sum_{i,j=1}^{s}\tilde{n}_{ij}\to\int_{0}^{\beta}d\tau_{1}\int_{0}^{\beta}d\tau_{2}\, n\left(\tau_{1},\tau_{2}\right)$, and similarly for sums over $\tilde z$. Our action transforms into
\begin{multline}
\label{eq:2 chord continuum action}
    S = \frac{1}{4}\int_{0}^{\beta}d\tau_{1}\int_{0}^{\beta}d\tau_{2}\int_{\tau_{1}}^{\tau_{2}}d\tau_{3}\int_{\tau_{2}}^{\tau_{1}}d\tau_{4}\,\left[n(\tau_1,\tau_2)n(\tau_3,\tau_4) + 2n(\tau_1,\tau_2)z(\tau_3,\tau_4)\right] \\ + \frac{1}{2} \int_{0}^{\beta}d\tau_1 \int_{0}^{\beta}d\tau_2\,\left[n(\tau_1,\tau_2)\left[\log\left(\frac{n(\tau_1,\tau_2)}{\nu^2\bJ^2}\right) - 1\right]
    + z(\tau_1,\tau_2)\left[\log\left(\frac{z(\tau_1,\tau_2)}{\kappa^2\bJ^2}\right) - 1\right]\right] \,,
\end{multline}
As in the case with a single type of chords, we remember in the first term that the $\tau$'s live on a circle, and flipping the limits of the integration should be understood as integrating over the other side of the circle. The factor for the first term corrects the overcounting when switching $\tau_1$ and $\tau_2$, and when switching the pairs $\tau_{1,2}$ and $\tau_{3,4}$. The summation over all possible values of $n_{ij}$, $z_{ij}$ becomes a path integral over the non-negative, periodic, symmetric, bi-local functions $n$, $z$ and the partition function becomes
\begin{equation}
    Z = \int \cD n \cD z \,e^{-\frac{1}{\lambda}S[n, z]} \,.
\end{equation}

\subsection{The saddle point}\label{sec:MltiSddl}

Since we work in the semi-classical limit $\lambda \to 0$, we can use a saddle point approximation to study the partition function. The equations of motion are
\begin{equation}
\begin{aligned}
    \log\left[\frac{n(\tau_1,\tau_2)}{\nu^2\bJ^2}\right] &= -\int_{\tau_{1}}^{\tau_{2}}d\tau_{3}\int_{\tau_{2}}^{\tau_{1}}d\tau_{4}\,\left[n(\tau_3,\tau_4) + z(\tau_3,\tau_4)\right], \,\\
    \log\left[\frac{z(\tau_1,\tau_2)}{\kappa^2\bJ^2}\right] &= -\int_{\tau_{1}}^{\tau_{2}}d\tau_{3}\int_{\tau_{2}}^{\tau_{1}}d\tau_{4}\,n(\tau_3,\tau_4) \,.
\end{aligned}
\end{equation}
The equations become more familiar after defining the quantities\footnote{The derivative relations only apply when $\tau_i \neq \tau_j$, see footnote~\ref{foo:counterterms}.}
\begin{equation}
\label{eq:def gn gz}
    \begin{aligned}
        g_n(\tau_1, \tau_2) &= -\int_{\tau_{1}}^{\tau_{2}}d\tau_{3}\int_{\tau_{2}}^{\tau_{1}}d\tau_{4}\, n(\tau_3,\tau_4) \,, & n(\tau_1,\tau_2) &= -\frac{1}{2}\partial_{\tau_1}\partial_{\tau_2} g_n(\tau_1,\tau_2) \,, \\
        g_z(\tau_1, \tau_2) &= -\int_{\tau_{1}}^{\tau_{2}}d\tau_{3}\int_{\tau_{2}}^{\tau_{1}}d\tau_{4}\,z(\tau_3,\tau_4) \,, & z(\tau_1,\tau_2) &= -\frac{1}{2}\partial_{\tau_1}\partial_{\tau_2} g_z(\tau_1,\tau_2) \,,
    \end{aligned}
\end{equation}
as they take the form
\begin{equation}
\label{eq:eom 2 chords}
\begin{aligned}
    \partial_{\tau_1}\partial_{\tau_2}g_n(\tau_1,\tau_2) &= -2\bJ^2\nu^2 e^{g_n(\tau_1,\tau_2) + g_z(\tau_1,\tau_2)} \,, \\ 
    \partial_{\tau_1}\partial_{\tau_2}g_z(\tau_1,\tau_2) &= -2\bJ^2\kappa^2 e^{g_n(\tau_1,\tau_2)} \,.
\end{aligned}
\end{equation}
The boundary conditions for these differential equations are
\begin{equation}
    g_{n,z}(0,0) = g_{n,z}(0,\beta) = g_{n,z}(\beta,0) = g_{n,z}(\beta,\beta) = 0\,.
\end{equation}
We remind the reader that $\nu^2 + \kappa^2 = 1$. When $\kappa = 0$ the system is simply the chaotic model discussed in the previous section. Indeed, the equations of motion imply $g_z = 0$, and reduce to the equations of motion \eqref{eq:Liouville_eq}. When $\kappa = 1$ the system is the purely integrable model.

\section{From chaos to integrability}
\label{sec:phase transition}
So far we have seen that in the semi-classical limit the dynamics of the system \eqref{eq:Cha_integ_Ham} is described by the saddle point equations \eqref{eq:eom 2 chords}. We will now see that there is one solution which is continuously connected to the purely chaotic system, and one which is continuously connected to the pure integrable one. We will then argue, both numerically and analytically, that there is a first-order phase transition between these two phases. In Section~\ref{sec:generic 2 chords} we will discuss more complicated Hamiltonians and their phase diagram.

We will find it convenient to solve the equations of motion by introducing a new variable, $\ell=g_{n}+g_{z}$, which is related to the two point function $G_\Delta$ of a random operator of the SYK-type via $G_\Delta = \braket{e^{\Delta\ell}}$, as this two point function is related to the number of chords of both types that cross the matter chords. We will also assume that, at the saddle, the fields depend only on $\tau = |\tau_1 - \tau_2|$. We will write the equations of motion in terms of the $\ell$ variable below. In order to determine what  the dominant phase is and to estimate the phase transition point, we will also need the on-shell action on the solution,
% \begin{equation}
% \label{eq:saddle_point_action_ell}
%     S_{\text{on-shell}} = -\frac{1}{4}\int_{0}^{\beta}d\tau_{1}\int_{0}^{\beta}d\tau_{2}\,\Bigg[\frac{1}{2}g_{n}\partial_{1}\partial_{2}g_{n}+g_{n}\partial_{1}\partial_{2}g_{z}+2\left(\nu\bJ\right)^{2}e^{g_{n}+g_{z}}+2\left(\kappa\bJ\right)^{2}e^{g_{n}}\Bigg]  \,.
% \end{equation} 
\begin{equation}
\label{eq:saddle_point_action_ell}
    S_{\text{on-shell}} = \frac{\beta}{4}\int_{0}^{\beta}d\tau\,\Bigg[\frac{1}{2}\ell\partial_\tau^2\ell - g_{z}\partial_\tau^2g_{z} - 2\left(\nu\bJ\right)^{2}e^{\ell} - 2\left(\kappa\bJ\right)^{2}e^{\ell - g_z}\Bigg]  \,.
\end{equation} 

\begin{figure}[t]
    \centering
    \begin{subfigure}[t]{0.4\textwidth}
        \centering
        \includegraphics[width=0.7\textwidth]{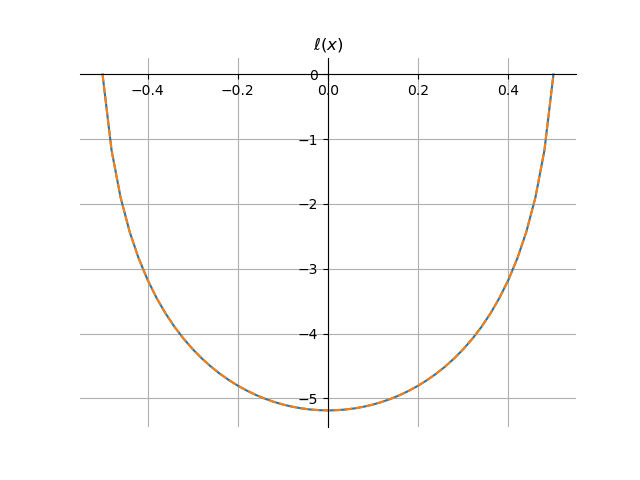}
        \subcaption{$\ell$ for $\kappa = 0.1$.}
    \end{subfigure}
    \qquad
    \begin{subfigure}[t]{0.4\textwidth}
        \centering
        \includegraphics[width=0.7\textwidth]{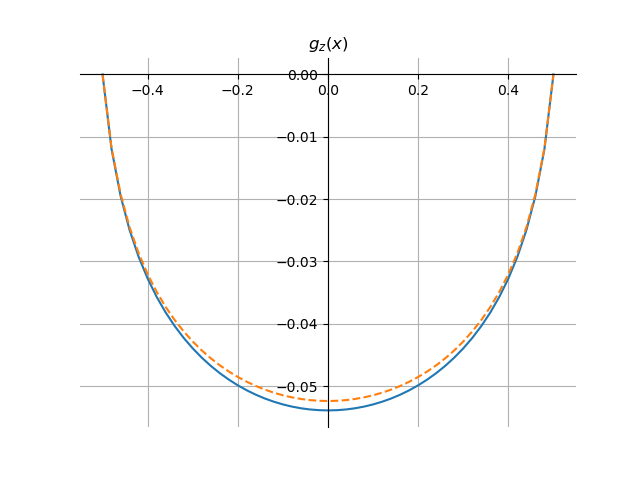}
        \subcaption{$g_z$ for $\kappa = 0.1$.}
    \end{subfigure}

    \centering
    \begin{subfigure}[t]{0.4\textwidth}
        \centering
        \includegraphics[width=0.7\textwidth]{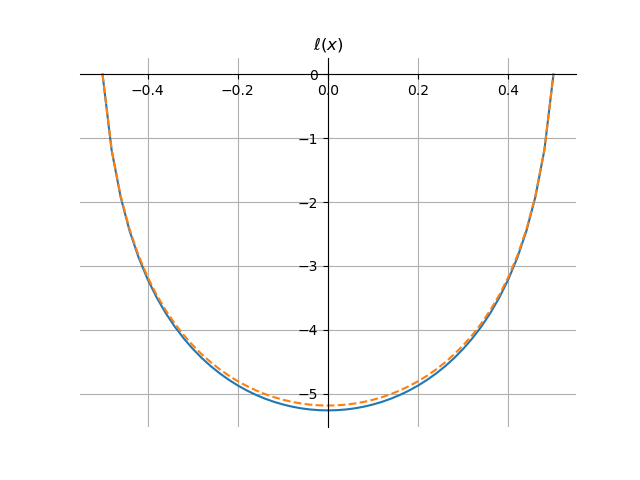}
        \subcaption{$\ell$ for $\kappa = 0.3$.}
    \end{subfigure}
    \qquad
    \begin{subfigure}[t]{0.4\textwidth}
        \centering
        \includegraphics[width=0.7\textwidth]{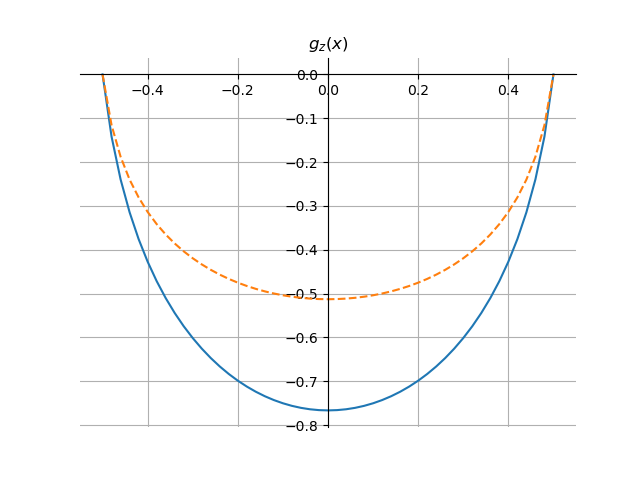}
        \subcaption{$g_z$ for $\kappa = 0.3$.}
    \end{subfigure}
    \caption{Comparison of the analytic approximations (orange), using $\ell^{(0)}$ and $g_z^{(1)}$, to the numerics (blue) at $\beta\bJ = 40$. The $x$-axis in the plots uses $x = \frac{\tau}{\beta} - \frac{1}{2} \in \left[-\frac{1}{2},\frac{1}{2}\right]$.}
    \label{fig:Chaotic_approx_compare_to_numerics}
\end{figure}

\paragraph{The chaotic phase}
The equations of motion, written in terms of the convenient variables for this case, take the form 
\begin{equation}
    \begin{aligned}
        \partial_{\tau}^{2}\ell &= 2\left(\bJ\nu\right)^{2}e^{\ell} + 2\left(\bJ\kappa\right)^{2}e^{\ell}e^{-g_{z}}\, ,\\
        \partial_{\tau}^{2}g_{z} &= 2\left(\bJ\kappa\right)^{2}e^{\ell}e^{-g_{z}} \,.
    \end{aligned}
\end{equation}
We would now like to solve the equations perturbatively in $\kappa$, which means that the solutions are continuously connected to those of the pure chaotic model at $\kappa = 0$. We do this by expanding the fields 
\begin{equation}
    \ell(\tau) = \ell^{\left(0\right)}\left(\tau\right) + \kappa^{2}\ell^{\left(1\right)}\left(\tau\right)\,, \qquad g_{z}(\tau) = g_z^{(0)}(\tau) + \kappa^{2}g_{z}^{\left(1\right)}(\tau) \,.
\end{equation}
At leading order we have 
\begin{equation}
    \ell^{\left(0\right)} = 2\log\left[\frac{\cos\left(\frac{\pi v}{2}\right)}{\cos\left[\pi v\left(\frac{1}{2} - \frac{\tau}{\beta}\right)\right]}\right]\,, \qquad g_{z}^{\left(0\right)}=0,\qquad\beta\bJ = \frac{\pi v}{\cos\left(\frac{\pi v}{2}\right)}\,,\qquad \tau\in\left[0,\beta\right] \,,
\end{equation}
and at the subleading order we can find\footnote{We thank Josef Seitz on discussions of this point.} 
\begin{equation}
    g_{z}^{\left(1\right)} = 2\log\left[\frac{\cos\left(\frac{\pi v}{2}\right)}{\cos\left[\pi v\left(\frac{1}{2} - \frac{\tau}{\beta}\right)\right]}\right] \,, \qquad \ell^{(1)} = 0 \,.
\end{equation}
These agree very well with the numerics in Figure~\ref{fig:Chaotic_approx_compare_to_numerics} for small $\kappa$ and reasonably well for moderate $\kappa$, up to expected $O(\kappa^2)$ corrections.
At leading order we have $g_z = 0$ and $\ell$ is simply the single chord solution with (dimensionless) temperature $\beta\bJ$, and therefore the action is
\begin{equation}\label{eq:SYK_phase_action_0th_orderGenTemp}
    S_{\text{chaotic phase}} = \frac{\pi^2v^2}{2} - 2\pi v \tan\left(\frac{\pi v}{2}\right) + O(\kappa^4) \,.
\end{equation}
At low temperatures $\beta\bJ\gg 1$ we can simplify this expression as $v \approx 1 - \frac{2}{\beta\bJ}$, and 
\begin{equation}
\label{eq:SYK_phase_action_0th_order}
    S_{\text{chaotic phase}} = -2\beta\bJ +O(\kappa^4)\,,
\end{equation}
where further subleading corrections are suppressed by $\kappa$ and by $\beta\bJ$.

\begin{figure}[t]
    \centering
    \begin{subfigure}[t]{0.4\textwidth}
        \centering
        \includegraphics[width=0.7\textwidth]{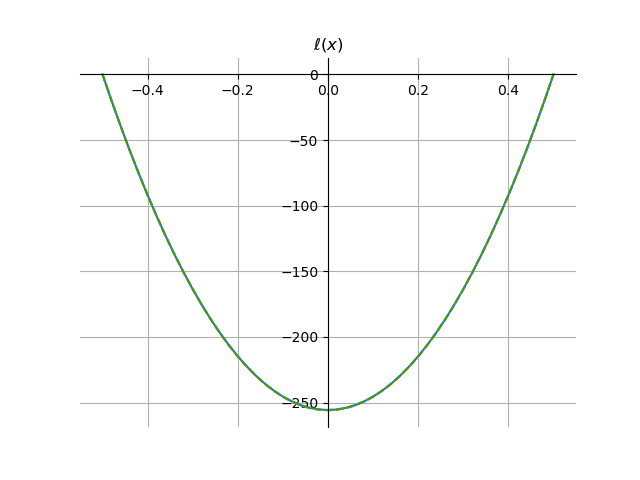}
        \subcaption{$\ell$ for $\kappa = 0.8$.}
    \end{subfigure}
    \qquad
    \begin{subfigure}[t]{0.4\textwidth}
        \centering
        \includegraphics[width=0.7\textwidth]{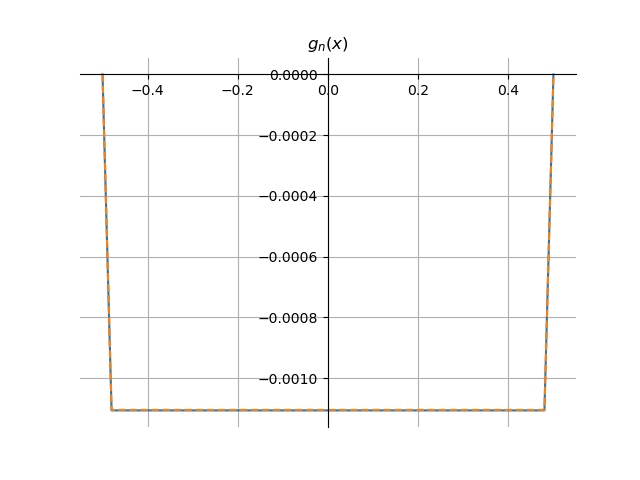}
        \subcaption{$g_n$ for $\kappa = 0.8$, $\frac{\nu^2}{\kappa^4(\beta\bJ)^2} \approx 5\cdot10^{-4}$.}
    \end{subfigure}
    
    \centering
    \begin{subfigure}[t]{0.4\textwidth}
        \centering
        \includegraphics[width=0.7\textwidth]{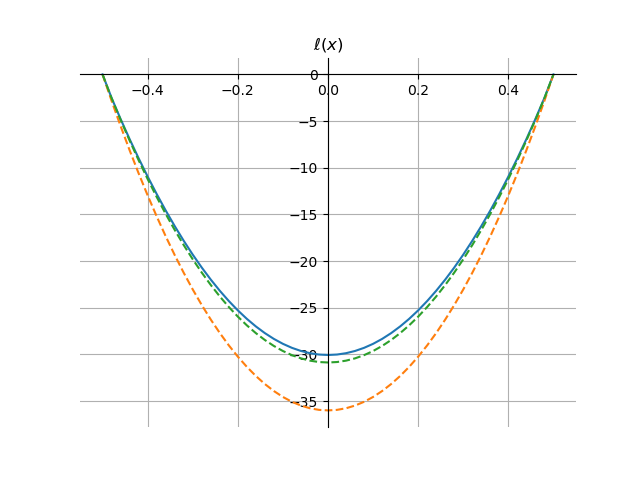}
        \subcaption{$\ell$ for $\kappa = 0.3$.}
    \end{subfigure}
    \qquad
    \begin{subfigure}[t]{0.4\textwidth}
        \centering
        \includegraphics[width=0.7\textwidth]{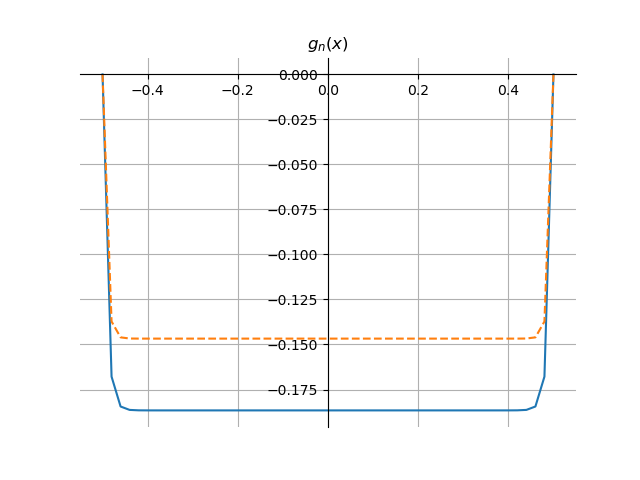}
        \subcaption{$g_n$ for $\kappa = 0.3$, $\frac{\nu^2}{\kappa^4(\beta\bJ)^2} \approx 0.07$.}
    \end{subfigure}
    \caption{Comparison of the leading nontrivial approximations ($\ell^{(0)}$ and $g_n^{(1)}$, orange) and the corrected solution up to the subleading order ($\ell^{(0)} + \nu^2 \ell^{(1)}$, green) to the numeric solutions (blue) at $\beta\bJ = 40$. In (a)---(b) they sit just on top of each other. The $x$-axis in the plots uses $x = \frac{\tau}{\beta} - \frac{1}{2} \in \left[-\frac{1}{2},\frac{1}{2}\right]$.}
    \label{fig:Integrable_approx_compare_to_numerics}
\end{figure}

\begin{figure}[t]
    \centering
    \includegraphics[scale=0.35]{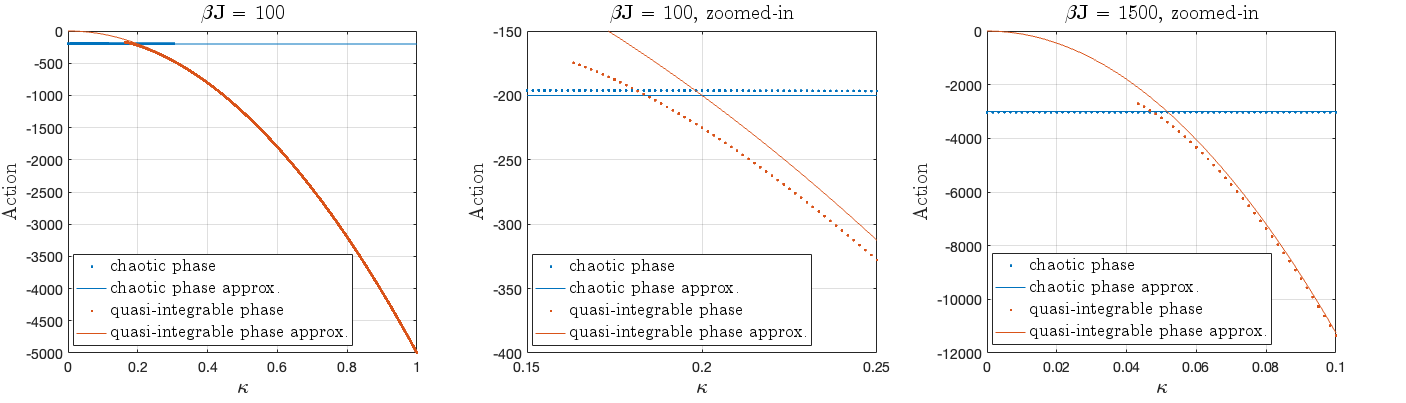}
    \caption{Comparison of the numerics vs leading order approximation for the action of the two phases. Away from the phase transition region, for temperature as low as $\beta \bJ =100$ the agreement with the zeroth-order approximations \eqref{eq:SYK_phase_action_0th_order} and \eqref{eq:integ_approx_0th_order} is already excellent (left figure).  Deviations are more significant near the phase transition region (middle figure), but even this deviation gets smaller and smaller when temperature is further lowered ($\beta \bJ =1500$, right figure).}
    \label{fig:actions_numeric_vs_0th_order}
\end{figure}

\begin{figure}[h]
    \centering
  \includegraphics[width=0.95\textwidth]{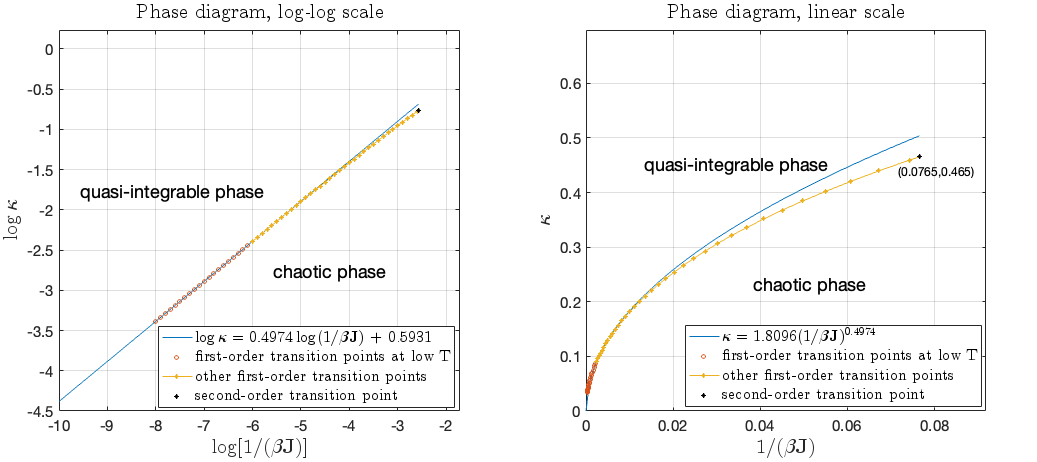}
    \caption{Phase diagrams in the $\kappa-1/\beta \mathbb{J}$ plane, obtained from solving equation \eqref{eq:eom 2 chords} numerically. Left: log-log scale. Right: linear scale.   The red and yellow dots are first-order phase transition points. The red dots are in the low temperature region, and are used to obtain the linear fit (blue curves) in log-log scale. Yellow dots are not used for fitting. The black dot is where the first-order transition line terminates.}
    \label{fig:phase-diag}
\end{figure}

\paragraph{The integrable phase}
Now let us solve the equations perturbatively when $\nu \ll 1$, i.e., for the case where the solution is continuously connected to the purely integrable model. At the moment we will treat $\kappa$ as independent of $\nu$. We would like to solve the equations
\begin{equation}
\label{eq:eom_dimensionless_ell_integ}
    \begin{aligned}
        \partial_{\tau}^{2}\ell &= 2\left(\bJ\nu\right)^{2}e^{\ell} + 2\left(\bJ\kappa\right)^{2}e^{g_{n}} \,, \\
        \partial_{\tau}^{2}g_{n} &= 2\left(\bJ\nu\right)^{2}e^{\ell} \,,
    \end{aligned}
\end{equation}
which we will do order by order in $\nu$ (we will see later that this is also a low temperature expansion, not just a perturbative expansion in $\nu$). We write
\begin{equation}     
\begin{aligned}
    \ell &= \ell^{\left(0\right)}+\nu^{2}\ell^{\left(1\right)} \,, \\
    g_{n} &= g_{n}^{\left(0\right)}+\nu^{2}g_{n}^{\left(1\right)} \,,
\end{aligned} 
\end{equation} 
and find that at leading order in $\nu$
\begin{equation}
\label{eq:integ_approx_0th_order}
g_{n}^{\left(0\right)} = 0\,,\qquad \ell^{\left(0\right)} = \left(\bJ\kappa\right)^{2}\tau\left(\tau - \beta\right)\,.
\end{equation} 
Note the presence of $\kappa^2=1-\nu^2$ in $\ell^{(0)}$ means we are absorbing part of subleading corrections into the leading order. It turns out this will make the analysis cleaner.
At the next order the equations are
\begin{equation}     
\begin{aligned}
\label{eq:integ_approx_1st_order}
    \partial_{\tau}^{2}\ell^{\left(1\right)} &= 2\bJ^{2}e^{\ell^{\left(0\right)}} + 2\left(\bJ\kappa\right)^{2}g_{n}^{\left(1\right)} \,, \\
    \partial_{\tau}^{2}g_{n}^{\left(1\right)} &= 2\bJ^{2}e^{\ell^{\left(0\right)}} \,.
\end{aligned} 
\end{equation}  
While an explicit solution is possible, it is quite cumbersome%
\footnote{Explicitly, the solution is given by
\begin{equation}     
\begin{aligned}         
g_{n}^{\left(1\right)}\left(\tau\right) &= \frac{1}{\kappa^{2}}\left(1-\beta\bJ\kappa F\left(\frac{\beta\bJ\kappa}{2}\right)+e^{\frac{1}{4}\left(\beta\bJ\kappa\right)^{2}\left(4x^{2}-1\right)}\left(2\left(\beta\bJ\kappa\right)x\,F\left(x\beta\bJ\kappa\right)-1\right)\right) \,, \\
\ell^{\left(1\right)}\left(\tau\right) &= \frac{e^{-\frac{1}{4}\left(\beta\bJ\kappa\right)^{2}}}{12\kappa^{2}}\Bigg[\sqrt{\pi}\beta\bJ\kappa\left(\text{erfi}\left(\frac{\beta\bJ\kappa}{2}\right)\left(\left(\beta\bJ\kappa\right)^{2}\left(1-6x^{2}\right)-3\right)+2x\left(2\left(\beta\bJ\kappa\right)^2x^{2}+3\right)\text{erfi}\left(x\beta\bJ\kappa\right)\right) \\
&\qquad\qquad\qquad-4e^{\left(\beta\bJ\kappa\right)^{2}x^{2}}\left(\left(\beta\bJ\kappa\right)^{2}x^{2}+2\right)+2e^{\frac{1}{4}\left(\beta\bJ\kappa\right)^{2}}\left(\left(\beta\bJ\kappa\right)^{2}\left(6x^{2}-1\right)+4\right)\Bigg] \,,
\end{aligned} 
\end{equation} 
where $x = \frac{\tau}{\beta} - \frac{1}{2}$, $F$ denotes Dawson's integral and erfi is the imaginary error function.}.
% end of footnote
At low temperatures, $\beta\bJ \gg 1$, the expressions simplify and scale like\footnote{Actually, our action assumes that the number of chords is large. Since $n(\tau_i,\tau_j) \sim \partial_\tau^2 g_n$, we see from \eqref{eq:integ_approx_1st_order} that the discrete number of chords at antipodal points is $
    n_{ij} = \frac{\beta^2}{\lambda s^2} n(\tau_i,\tau_j) \sim \frac{1}{\lambda} \left(\frac{\beta \bJ}{s}\right)^2 \nu^2 e^{-(\beta\bJ\kappa)^2/4} $.
This means that we must first take the $\lambda \to 0$ limit and only then any other limit, such as low temperature or $\nu \to 0$. If we do want to go to temperatures as low as $\beta\bJ \sim 1/\lambda$, we need to keep the discrete $n_{ij}$ in our action. Since $n_{ij}$ is not of order $1/\lambda$ anymore, it will not affect the saddle point and will only give corrections at subleading orders in $\lambda$.}
\begin{equation}
\label{eq:integrable phase gn scaling}
    g_{n}\left(\tau\right) = O\left(\frac{\nu^2}{\kappa^4(\beta\bJ)^2}\right)\,, \qquad \ell^{(1)} = O\left(\frac{(\beta\bJ)^0}{\kappa^2}\right) \,.
\end{equation}
Corrections to our leading order solution are therefore suppressed at low temperatures, beyond the naive expansion in $\nu$. We can expect the approximation to fail when $\kappa \sim \frac{1}{\sqrt{\beta\bJ}}$ for large $\beta\bJ$. We note that a phase transition might happen much earlier, and one must determine the action at the other phase to find the transition point. In Figure~\ref{fig:Integrable_approx_compare_to_numerics} we compare our analytic approximation to the numerical solution for two values of $\kappa$, and observe that the results agree very well even when $\nu \approx 0.95$, since $\nu^2/(\kappa^4(\beta\bJ)^2)$ is still small.
At leading order we have $g_z = \ell = \ell^{(0)}$, and the on-shell action simply becomes
\begin{equation}
\label{eq:p-spin_phase_action_0th_order}
    S_{\text{quasi-integrable phase}} = -\frac{1}{2}(\beta\bJ\kappa)^2 \,.
\end{equation}
What we have been doing is to compute the actions to $\nu^2$-order (the subleading order). Moreover, some of the $\nu^2$-order contributions are suppressed by temperature $1/\betaJ$.  Our final expression only keeps those which are unsuppressed by temperature.
\paragraph{The phase transition}
The phase transition happens when, as we increase $\kappa$, the action of the quasi-integrable phase \eqref{eq:p-spin_phase_action_0th_order} becomes more negative than that of the chaotic phase \eqref{eq:SYK_phase_action_0th_order}. The phase transition temperature is that in which the actions are equal,
\begin{equation}
    S_{\text{chaotic phase}} = S_{\text{quasi-integrable phase}} \,.
\end{equation}
At low temperatures $\beta\bJ \gg 1$, the transition temperature is
\begin{equation}
\label{eq:critical kappa low temp}
    \kappa_* \approx \frac{2}{\sqrt{\beta\bJ}} \qquad \text{(at low temperatures)} \,.
\end{equation}
In Figure~\ref{fig:Chaotic_approx_compare_to_numerics} and Figure~\ref{fig:Integrable_approx_compare_to_numerics} we see an example where the two numerical solutions overlap, and our low temperature approximations work quite well. In Figure~\ref{fig:actions_numeric_vs_0th_order} we present a comparison of the numeric and approximate values for the action and the transition point. In Figure~\ref{fig:phase-diag} we plot the phase diagram by numerically finding where one saddle becomes more dominant then the other. By fitting the transition point for low temperature, we verify that \eqref{eq:critical kappa low temp} gives a good approximation to the phase transition point. 

We note that the phase transition happens around where the integrable phase approximation breaks down. Corrections do not seem to be parametrically suppressed, as they are of order $O(\nu^2/(\beta\bJ\kappa^2)^2) = \frac{1}{16} + O(1/(\beta\bJ))$, but the numerical factor might be enough to explain the reasonable agreement with the numerics. We base the claim for the existence of phase transitions on the numerics, and treat the analytic approximation as an explanation for its features.

\begin{figure}[t]
    \begin{subfigure}{0.3\textwidth}
    \centering
    \includegraphics[width=1\textwidth]{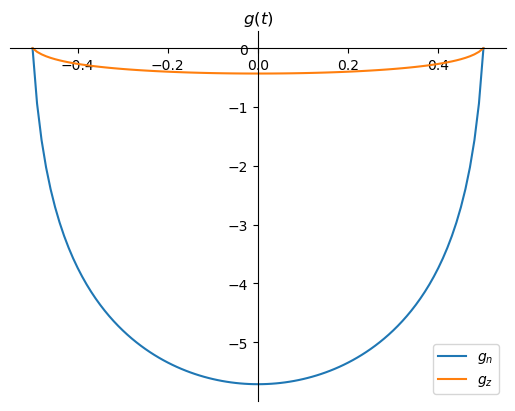}
    \subcaption{Chaotic phase.}
    \label{fig:g_numeric_SYK_like}
    \end{subfigure}
    \hfill
    \begin{subfigure}{0.3\textwidth}
    \includegraphics[width=1\textwidth]{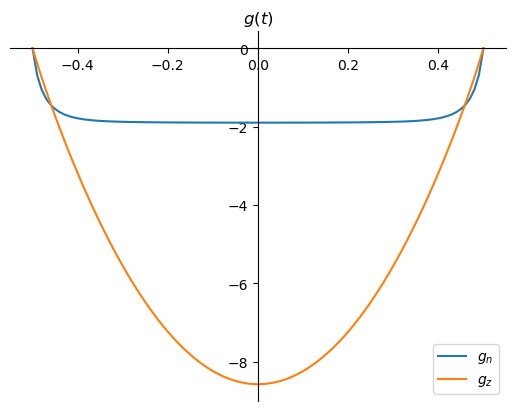}
    \subcaption{Subdominant saddle.}
    \label{fig:g_numeric_mixed}
    \end{subfigure}
    \hfill
    \begin{subfigure}{0.3\textwidth}
    \includegraphics[width=1\textwidth]{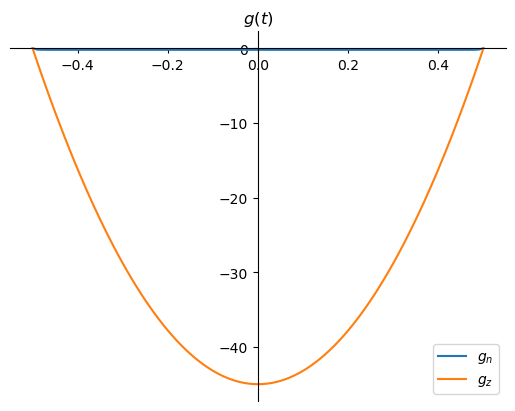}
    \subcaption{Integrable phase.}
    \label{fig:g_numeric_Z_like}
    \end{subfigure}
    \caption{The numerical solutions for $g_n(x)$ (blue) and $g_z(x)$ (orange) for $\beta \bJ = 65$ and $\kappa = 0.24$. We have shifted and rescaled the horizontal axis from $\tau \in [0, \beta]$ to $x\in [-0.5, 0.5]$ by $x = \tau/\beta -0.5$. }
    \label{fig:example}    
\end{figure}

\begin{figure}[t]
    \centering
    \includegraphics[width=\textwidth]{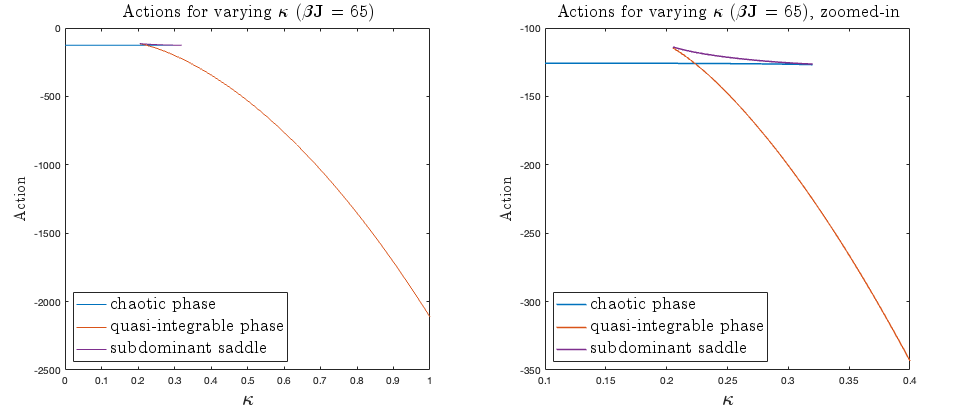}
    \caption{The action as a function of $\kappa$ when $\beta\bJ = 65$ for any $\kappa$ (left) and for the phase transition region (right). The transition is from the chaotic phase (blue) to the quasi-integrable phase (red). We also show the additional  saddle-point actions that are always subdominant (purple).}
    \label{fig:all_phases_beta_65}
    \end{figure}

\paragraph{Other saddles}
Do these two approximations capture all the different solutions to the equations of motion \eqref{eq:eom 2 chords}? The answer is negative. In Figure~\ref{fig:example} we numerically find for the same temperature and $\kappa$ three different solutions for the equations of motion -- one corresponds to the chaotic phase, another to the quasi-integrable phase, but there is also a third solution corresponds to neither. The results of more systematic numerics are shown in Figure~\ref{fig:all_phases_beta_65}, where the actions are plotted for the different phases as we vary $\kappa$. As advertised, one of the phases (in blue) is continuously connected to the purely chaotic system, another (in red) to the purely integrable system, and the third solution to the saddle point equations (in purple) is connected to neither and is always subdominant. It therefore does not play a role in computing the free energy of the system in the thermodynamic limit. 

\section{Interpolation between generic Hamiltonians}
\label{sec:generic 2 chords}

The method developed above allows us to study a more general class of models that interpolate between any two Hamiltonians that are amenable to expansion in chord diagrams,
\begin{equation}
\label{eq:General interp Hamiltonian}
    H = \nu H_1 + \kappa H_2 \,, \qquad \nu^2 + \kappa^2 = 1\,, \qquad \nu,\kappa \in [0,1] \,,
\end{equation}
with generic intersection weights $q_{ij} \equiv e^{-\lambda_{ij}}$ for the chords associated with these two Hamiltonians. Note that such Hamiltonians can be realized microscopically as as two different polarized operators operators (as defined in \eqref{eq:polarized_op_def}).

In this section we will shortly describe the application of the techniques developed earlier to such systems---we will write down their action and equations of motion in the semi-classical limit, and analyze the phase diagram of these systems in Section~\ref{sec:Phase diagram generic}. We will see that, depending on the weights of intersections, we might find the same type of phase transition as before, no phase transitions at all, or zero temperature phase transitions. 

Our methods apply in the semi-classical limit where all the $q_{ij}\to 1$ but perhaps at different relative rate (i.e., we could have generic $l_{ij} = \lambda_{ij}/\lambda_{nn}$). We again denote the two types of chords by $n$ and $z$. The expansion in chord diagrams then gives
\begin{equation}\label{eqn:genericChordRules}
    \left\langle\Tr\left(H^{2k}\right)\right\rangle = \sum_{\substack{\text{chord diagrams with} \\ \text{$n+z=k$ chords}}} \nu^{2n}\kappa^{2z} q_{nn}^{\text{\#$n$-$n$ intersections}}q_{nz}^{\text{\#$n$-$z$ intersections}} q_{zz}^{\text{\#$z$-$z$ intersections}} \,.
\end{equation}
These cases include RG-flows of the kind analyzed by \cite{jiang2019, Anninos:2022qgy,Anninos:2020cwo}, who considered the case where both Hamiltonians are SYK-like of different lengths, and we reproduce their results at the end of this section. The algebra for the chord creation and annihilation operators is \cite{Speicher:1993kt}
\begin{equation}
    [a_i, a_j^+]_{q_{ij}} = \delta_{ij} \,.
\end{equation}
Following a similar derivation to the one described above, the action one finds is 
\begin{multline}
\label{eq:2 chord continuum action generic}
    S =
    \frac{1}{4}\int_{0}^{\beta}d\tau_{1}\int_{0}^{\beta}d\tau_{2}\int_{\tau_{1}}^{\tau_{2}}d\tau_{3}\int_{\tau_{2}}^{\tau_{1}}d\tau_{4}\,\Bigg[n_{12} n_{34} + 2l_{nz} n_{12} z_{34}
    + l_{zz} z_{12} z_{34}\Bigg] \\ + \frac{1}{2} \int_{0}^{\beta}d\tau_1 \int_{0}^{\beta}d\tau_2\,\Bigg[n_{12}\left[\log\left(\frac{n_{12}}{\nu^2\bJ^2}\right) - 1\right]
    + z_{12}\left[\log\left(\frac{z_{12}}{\kappa^2\bJ^2}\right) - 1\right]\Bigg] \,,
\end{multline}
where we denoted $n(\tau_i,\tau_j) \equiv n_{ij}$ for brevity, even though we are using the continuous fields. The equations of motion again have a simple form when expressed in terms of the $g_{n,z}$ functions \eqref{eq:def gn gz},
\begin{equation} \label{eq:Generic_2_chords_EOM}
\begin{aligned}
    \partial_{\tau_1}\partial_{\tau_2}g_n(\tau_1,\tau_2) &= -2\bJ^2\nu^2 e^{g_n(\tau_1,\tau_2) + l_{nz} g_z(\tau_1,\tau_2)} \,, \\ 
    \partial_{\tau_1}\partial_{\tau_2}g_z(\tau_1,\tau_2) &= -2\bJ^2\kappa^2 e^{l_{nz} g_n(\tau_1,\tau_2) + l_{zz} g_z(\tau_1,\tau_2)} \,.
\end{aligned}
\end{equation}
The on-shell action has a neat expression when one uses the equations of motion and expresses everything through the $g$-variables,
\begin{multline}
\label{eq:generic 2 chords on-shell action}
    S_{\rm on-shell} = -\frac{1}{4}\int_{0}^{\beta}d\tau_{1}\int_{0}^{\beta}d\tau_{2}\,\Bigg[\frac{1}{2}g_{n}\partial_{1}\partial_{2}g_{n}+l_{nz}g_{n}\partial_{1}\partial_{2}g_{z}+\frac{1}{2}l_{zz}g_{z}\partial_{1}\partial_{2}g_{z} \\
    +2\left(\nu\bJ\right)^{2}e^{g_{n}+l_{nz}g_{z}}+2\left(\kappa\bJ\right)^{2}e^{l_{nz}g_{n}+l_{zz}g_{z}}\Bigg] \,.
\end{multline}

We also note that the two point function, $\frac{1}{Z}\braket{\Tr\left(e^{-\beta H} M(\tau_1) M(\tau_2)\right)}$, of random operators $M$ can be computed, see Appendix~\ref{app:CrshCrdDiag} for the definitions of these operators. All we need is to assume that this matter operator can be describe by a new type of chords, matter chord, whose intersections with $n$-chords are weighted by $e^{-\lambda \alpha_n}$ while its intersections with $z$-chords are weighted by $e^{-\lambda \alpha_z}$, in a way similar to that of the single chord case \cite{Berkooz:2018jqr}. Since $g_{n,z}(\tau_1,\tau_2)$ count the number of $n$- and $z$-chords that cross a chord that stretches between $\tau_1$ and $\tau_2$, the two-point function is given by the expectation value
\begin{equation}
    G_{\alpha_n,\alpha_z}(\tau_1,\tau_2) = \braket{e^{-\alpha_n g_n(\tau_1,\tau_2) - \alpha_z g(\tau_1,\tau_2)}} \,.
\end{equation} 
In the semi-classical limit, and assuming the $\alpha$'s are finite, we can simply evaluate this at the saddle point found above.

\paragraph{SYK RG flows} As a sanity check, we compare with the results of \cite{jiang2019, Anninos:2022qgy}, who studied a particular family of interpolating Hamiltonians---those that describe relevant deformations of one SYK-like Hamiltonian by another, and regarded the transition as an RG flow. They started with an SYK Hamiltonian of the form \eqref{eq:H_SYK_def} of length $p$, deformed it by a Hamiltonian of length $p/n$ for some integer $n$, and studied the system in the $p\to\infty$ limit. We expect the double scaling limit to reproduce the large $p$ limit results after taking the $\lambda\to 0$ limit.  Insertions of the former Hamiltonian will result in $n$-chords while insertions of the latter will result in $z$-chords. The weights of chord intersections for two operators of lengths $p_1$, $p_2$ is $e^{-2 p_1 p_2 / N}$ \cite{Berkooz:2018qkz}. In our case, denoting $q \equiv e^{-\lambda_{nn}} = e^{-2p^2/N}$ as before, these weights are $q^{-l_{ij}}$, where $l_{nz} = \lambda_{nz}/\lambda_{nn} = 1/n$ and $l_{zz} = \lambda_{zz}/\lambda_{nn} = 1/n^2$. The interpolation convention they used slightly differs from ours, and reads
\begin{equation}
    H_{\text{AGS}} = H^{(p)}_{\text{SYK}} + s_{\text{AGS}} H^{(p/n)}_{\text{SYK}} \,,
\end{equation}
where $s_{\text{AGS}} \in \bR$, such that $\Tr(H^2_{\text{AGS}}) = \frac{1}{\lambda}\cJ_{\text{AGS}}^2 \left(1 + s^2_{\text{AGS}} n^2\right)$. In order to match to our conventions, in which $\Tr(H^2) = \frac{1}{\lambda}\bJ^2$ throughout the interpolation, 
\begin{equation}
    H = \nu H^{(p)}_{\text{SYK}} + \kappa H^{(p/n)}_{\text{SYK}}  \,, \qquad \kappa^2 + \nu^2 = 1\,,\qquad \kappa,\nu\in[0,1] \,,
\end{equation}
we need to identify $s_{\text{AGS}} = \frac{\kappa}{n \nu}$ and $\cJ_{AGS} = \nu \bJ$. The two-point function of a random SYK-like operator of length $\tilde p$ is given by
\begin{equation}
    G_\Delta(\tau_1,\tau_2) = e^{-\Delta g(\tau_1,\tau_2)} \,, \qquad g(\tau_1,\tau_2) = g_n(\tau_1,\tau_2) + \frac{1}{n}g_z(\tau_1,\tau_2) \,, \qquad \Delta = \tilde p / p\,.
\end{equation}
By using \eqref{eq:Generic_2_chords_EOM} for the equation of motion for $g$, and assuming that the saddle only depends on $\tau = \tau_1-\tau_2$, we find
\begin{equation}
    \partial_{\tau}^2 g(\tau) = 2n{\cJ}_{\text{AGS}}^2 s_{\text{AGS}}^2 e^{g(\tau)/n} + 2{\cJ}_{\text{AGS}}^2 e^{g(\tau)} \,,
\end{equation}
exactly reproducing (3.5) of \cite{Anninos:2022qgy}, and the on-shell action \eqref{eq:generic 2 chords on-shell action} reproduces\footnote{Actually, it only reproduces the second term there. The origin of the first term is the normalization of the trace, $\Tr(\1) = 2^{N/2}$ instead of our $\Tr(\1) = 1$.} their (3.6).

\subsection{The phase diagram for generic Hamiltonians}
\label{sec:Phase diagram generic}
We can now address the phase diagram of the generic Hamiltonian \eqref{eq:General interp Hamiltonian}, and learn three lessons: first, we will see that the phase transition exists also when the interaction length of the integrable Hamiltonian differs from that of the chaotic one. This may be counter-intuitive, as usually we think of shorter operators as more relevant, yet here we find examples of an ostensibly irrelevant operator that controls the low temperature dynamics. Second, as in \cite{Anninos:2022qgy}, there are no phase transitions as we deform one SYK Hamiltonian by another of different length. Third, there are systems where one can find zero-temperature phase transitions.

\paragraph{Generalized chaos to integrability transition}
The first generalization we may consider is the one where we still interpolate between a chaotic and an integrable Hamiltonians, but when the intersections between the chaotic $n$-chords and the integrable $z$-chords have a different weight than the $n$-$n$ intersections. This amounts to picking $l_{nz} \neq 1$, $l_{zz} = 0$. Microscopically, this can be realized by choosing different lengths for the two Hamiltonians in Section~\ref{sec:model-defs}. 

\begin{figure}[t]
    \centering
    \includegraphics[width=\textwidth]{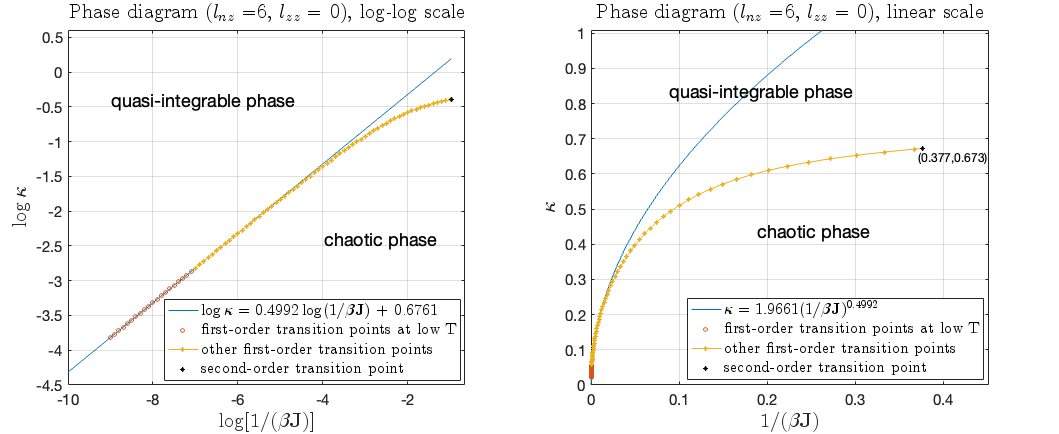}
    \caption{Phase diagrams obtained from solving  equation \eqref{eq:Generic_2_chords_EOM} numerically, with more general operator lengths ($l_{nz}=6, l_{zz}=0$). The red and yellow dots are first-order phase transition points. The red dots are in the low temperature region, and are used to obtain the linear fit (blue curves) in log-log scale. Yellow dots are not used for fitting. The black dot is where the first-order transition line terminates.}
    \label{fig:phaseMorel_nz}
\end{figure}

The approximations for the two phases in Section~\ref{sec:phase transition} still hold, where at leading order the only change is that for the integrable phase we should amend $\kappa \to \kappa \sqrt{l_{nz}}$ in the equation of motion \eqref{eq:eom_dimensionless_ell_integ}. This does not change the on-shell action, though, and \eqref{eq:critical kappa low temp} still holds at low temperatures. The subleading corrections do depend on $l_{nz}$. Numerical analysis confirms this, see Figure~\ref{fig:phaseMorel_nz}. The analytic approximation improves as we increase $l_{nz}$. As before, the analysis breaks down at high enough temperatures, and again the phase transition ends at some critical point, which happens at higher temperatures and larger $\kappa$ as we increase $l_{nz}$.

The final comment is that the first order phase transition happens when deforming by an integrable Hamiltonian of any length. At low enough temperatures we always find the integrable phase, which is surprising as usually in SYK longer operators are considered irrelevant. 

\paragraph{SYK RG flows}
The other extreme case is that in which we interpolate between two different SYK models of different lengths, as analysed in \cite{jiang2019,Anninos:2022qgy}, and corresponds to $l_{zz} = l_{nz}^2$. In this case there is no phase transition, but rather the action rather interpolates smoothly between the two cases, as seen by \cite{jiang2019, Anninos:2022qgy}.

\paragraph{General systems with a small $l_{zz}$} 
We saw that some systems have a phase transition and some do not. A natural question to ask is then what happens for a given  $l_{nz}$ and varying $l_{zz}$. It turns out that once $l_{zz}$ is nonzero (but small), we have 
\begin{itemize}
\item a zero-temperature phase transition, 
\item at low temperatures, the first-order line becomes linear in temperature,
\item the location of the 2nd order end of the line does not seem to change by a lot going from small $l_{zz}$ to zero $l_{zz}$.
\end{itemize}
The first two behaviors are qualitatively different from both the chaos-quasi-integrable case ($l_{nz} \neq 0, l_{zz}=0$) and the SYK RG flow case ($l_{zz} = l_{nz}^2$). As we keep on increasing $l_{zz}$ the phase transitions disappear, which happens before $l_{zz}=l_{nz}^2$ but we did not compute the exact point. Let us demonstrate the above points both numerically and analytically.

\begin{figure}[t!]
    \centering
    \includegraphics[width=\textwidth]{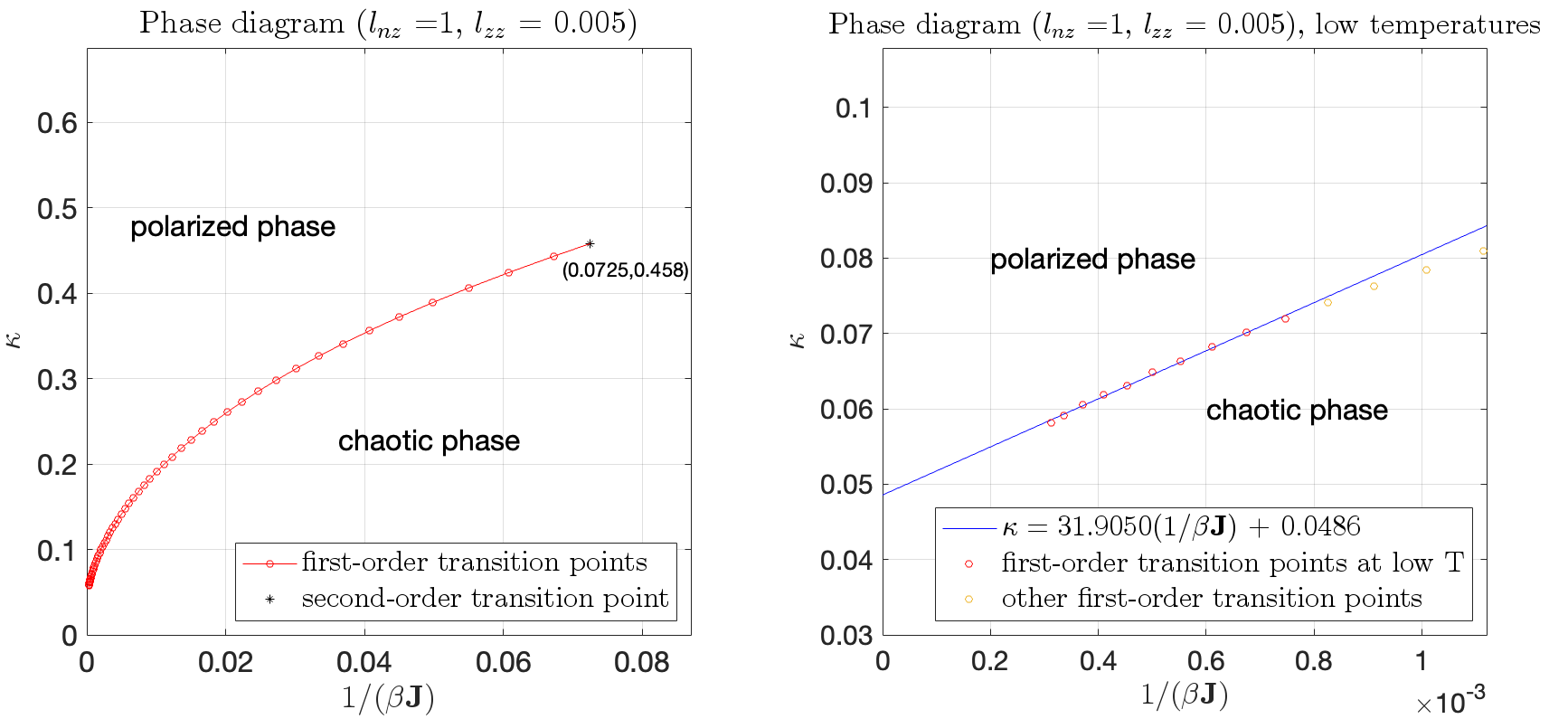}
    \caption{Phase diagram for $l_{nz}=1, l_{zz}=0.005$. Both are in linear scale. Left: the phase diagram up to the second order transition point (no fitting). Right: a zoom-in of the phase diagram into the low-temperature region, and the blue line is a linear fit using the lowest data points.}
    \label{fig:transitionLinelz=0p005}
\end{figure}

We will focus on the $l_{nz}=1$ scenario for simplicity. 
At low temperature $\beta \bJ \to \infty$ we repeat the analysis done in the $l_{zz}=0$ case to subleading order in $\kappa^2$ and $\nu^2$.  Near $\nu = 1$, $\kappa = 0$ we rewrite the on-shell action and the equation of motions in terms of the new variable $\ell = g_n +g_z$ and $g_z$.  It turns out $l_{zz}$ does not enter the analysis up to subleading order in $\kappa^2$, and we end up with exactly the same action as equation \eqref{eq:SYK_phase_action_0th_order} in the chaotic phase analysis. So we have that at low temperatures
\begin{equation}\label{eqn:smallLzzChaoticActionApprox}
    S_1 = -2 \beta \bJ + \frac{\pi^2}{2} + O((\betaJ)^{-1}).
\end{equation}
The first  correction in $\kappa$ comes at the order of $\kappa^4$ as before.  Notice this time we have expanded the action \eqref{eq:SYK_phase_action_0th_orderGenTemp} to subleading order in temperature, which turns out to be needed if one wants to see phase transitions at nonzero temperatures. 
Near $\nu = 0$, $\kappa = 1$ the analysis is quite different. In this phase we define the length operator to be 
\begin{equation}
    \ell =  g_n + l_{zz} g_z.
\end{equation}
In terms of their perturbative forms
\begin{equation}
\begin{aligned}
        \ell &= \ell^{(0)} + \nu^2 \ell^{(1)} \\
        g_n &=  g_n^{(0)} + \nu^2 g_n^{(1)},
\end{aligned}
\end{equation}
the leading order equations are
\begin{equation}
\begin{aligned}
    \partial_{\tau}^2 \ell^{(0)}  &= 2l_{zz}\kappa^2 \bJ^2 e^{\ell^{(0)}}\,, \\ 
    \partial_{\tau}^2 g_n^{(0)} &= 0\,.
\end{aligned}
\end{equation}
Note the appearance of $\kappa^2$ means we are absorbing some of the $\nu^2$ contribution to the leading order, which turns out to make our analysis cleaner.   The solutions are
\begin{equation}
    \ell^{\left(0\right)} = 2\log\left[\frac{\cos\left(\frac{\pi v}{2}\right)}{\cos\left[\pi v\left(\frac{1}{2} - \frac{\tau}{\beta}\right)\right]}\right]\,, \qquad g_{n}^{\left(0\right)}=0,\qquad  \sqrt{l_{zz}}\kappa\beta\bJ = \frac{\pi v}{\cos\left(\frac{\pi v}{2}\right)}\,,\qquad \tau\in\left[0,\beta\right] \,.
\end{equation}
The leading order action is given by 
\begin{equation}\label{eqn:polarizedActionLeading}
    S_2 =  -\frac{\betaJ^2 \kappa^2}{4}\int_{0}^{\beta}d\tau\, (2- l^{(0)}) e^{l^{(0)}}= -\frac{2\betaJ \kappa}{\sqrt{l_{zz}}}+  \frac{\pi^2}{2l_{zz}} + O((\kappa\sqrt{l_{zz}}\betaJ)^{-1}).
\end{equation}
Setting $S_1$ and $S_2$ to be equal, we get the following linear first-order transition line in the $\kappa -1/\betaJ$ plane:
\begin{equation}\label{eqn:smallLzzTransitionAnalytic}
    \kappa_* - \sqrt{l_{zz}} \approx \frac{1}{\beta \bJ} \frac{\pi^2}{4} (l_{zz}^{-\frac{1}{2}} - l_{zz}^{\frac{1}{2}}) \qquad \text{as $\beta \bJ \to \infty$}.
\end{equation}
 We will name the  phase continuously connected to $\kappa=1$ (described by the action $S_2$) the ``polarized phase'', because one way to microscopically realize it is to use randomly coupled Pauli matrices $\sigma^a (a=1,2,3)$  with different probability weights on the three possible Pauli matrices. See Appendix \ref{app:pauliChordRules}. In Figure~\ref{fig:actioncurveslz=0p005} we compare the numerical actions of the two phases to the numerics, while in Figure~\ref{fig:transitionLinelz=0p005} we plot the phase diagram.

Note since $S_2$ has a $\kappa$ dependence, we must complete the analysis at subleading order in $\nu^2$ to make sure equation \eqref{eqn:polarizedActionLeading}
is indeed the most dominant contribution to $S_2$.  At subleading $\nu^2$ order,  the action receives a contribution 
\begin{equation}
     - \frac{\nu^2 \betaJ^2}{4}  \int_0^\beta d\tau \left(\kappa^2  [ \ell^{(1)} - g_n^{(1)}-\ell^{(0)}\ell^{(1)}] e^{\ell^{(0)}}+ 2 e^{\ell^{(0)}/l_{zz}}\right).
\end{equation}
The equations of motion at subleading order in $\nu^2$ are
\begin{equation}
    \begin{aligned}
    \partial_{\tau}^2 \ell^{(1)}  &= 2l_{zz}\kappa^2\bJ^2 e^{\ell^{(0)}}\ell^{(1)} +2\bJ^2  e^{ \ell^{(0)}/{l_{zz}}}\,, \\ 
       \partial_{\tau}^2 g_n^{(1)} &= 2\bJ^2 e^{ \ell^{(0)}/l_{zz}} \,.
\end{aligned}
\end{equation} 
Combining this with the leading order equations allows us to simplify the subleading action to 
\begin{equation}
     - \frac{\nu^2\betaJ^2}{4}  \int_0^\beta d\tau \left(   \kappa^2 \ell^{(0)}\ell^{(1)} e^{\ell^{(0)}}+ 2 e^{\ell^{(0)}/l_{zz}}\right).
\end{equation}
Note the  $e^{\ell^{(0)}/l_{zz}}$ term is not suppressed by temperature upon integration. However, $e^{\ell^{(0)}/{l_{zz}}}$  decreases from 1 to 0 at a faster and faster rate as  $1/l_{zz}$ becomes larger, in an interval roughly of the size of $\beta/(\kappa \sqrt{l_{zz}}\betaJ)$.\footnote{ To be very careful, since the numerical experiment we are doing is lowering temperatures with $l_{zz}$ fixed, we should consider $\betaJ \to \infty$ limit as taken before taking  $l_{zz}^{-1}$ to be large.  Then we can state the following in terms of the rescaled coordinate $x=\tau/\beta$:
\begin{enumerate}
    \item for any fixed  (with respect to $\betaJ$) $x\in [0,1]$, $ e^{\ell^{(0)}/{l_{zz}} }$ quickly decays to zero as $1/(\sqrt{l_{zz}}\betaJ)^{2 / l_{zz}}$ at low temperatures.
    \item  $ e^{ \ell^{(0)}/l_{zz}}$ is of order (and smaller than) 1 in $\betaJ$ if $x\in [0,1/(\sqrt{l_{zz}}\betaJ)]$. So we can  approximate it as $(1- (\sqrt{l_{zz}}\betaJ)x)^{ \frac{l_{nz}}{l_{zz}}}$ 
\end{enumerate}
This means that for large $1 / l_{zz}$  and at low temperatures 
\[
 (\betaJ)^2\int_0^1 dx e^{\ell^{(0)}/l_{zz}}\propto \sqrt{l_{zz}}\betaJ.
\]}
The effect is that  its contribution to the action is suppressed by $l_{zz}$.   The same suppression is present for $\ell^{(1)}$ via the equation of motion, and due to the second derivative nature of the equation, its contribution to the action  is also suppressed by $1/(\kappa\sqrt{l_{zz}}\betaJ)$.  Therefore, at subleading order in $\nu^2$, we expect the low-temperature action to be corrected to 
\begin{equation}\label{eqn:smallLzzPolarizedActionApprox}
   S_2= -\frac{2\betaJ \kappa}{\sqrt{l_{zz}}} (1+ O(\nu^2 l_{zz}))+  \frac{\pi^2}{2l_{zz}} (1+ O(\nu^2 l_{zz})).
\end{equation}
Hence we have indeed accounted for the most dominant contributions to the action up to $\nu^2$ order when estimating for the phase transitions in equation \eqref{eqn:smallLzzTransitionAnalytic}.

Again, the above analysis is not  to be taken as evidence for the existence of a phase transition. 
Rather,  it is that when we are certain of a phase transition (by numerics), we may take equation \eqref{eqn:smallLzzTransitionAnalytic} as a first estimate of the location of transition. 
We will see from the numerics that it is not a bad estimate small $l_{zz}$.  In Figure \ref{fig:actioncurveslz=0p005}, we plot the actions for $l_{nz}=1$ and $l_{zz}=0.005$ as a function of $\kappa$ at low temperatures ($\beta \bJ =500, 1000$), both numerically and using the analytic approximation of equations \eqref{eqn:smallLzzChaoticActionApprox} and \eqref{eqn:smallLzzPolarizedActionApprox}. We can see the analytic approximation works almost perfectly away from the phase transition region.  There are small but visible deviations around the transition region,  but the deviations get smaller as we lower the temperatures. 
\begin{figure}[t!]
    \centering
    \includegraphics[scale=0.45]{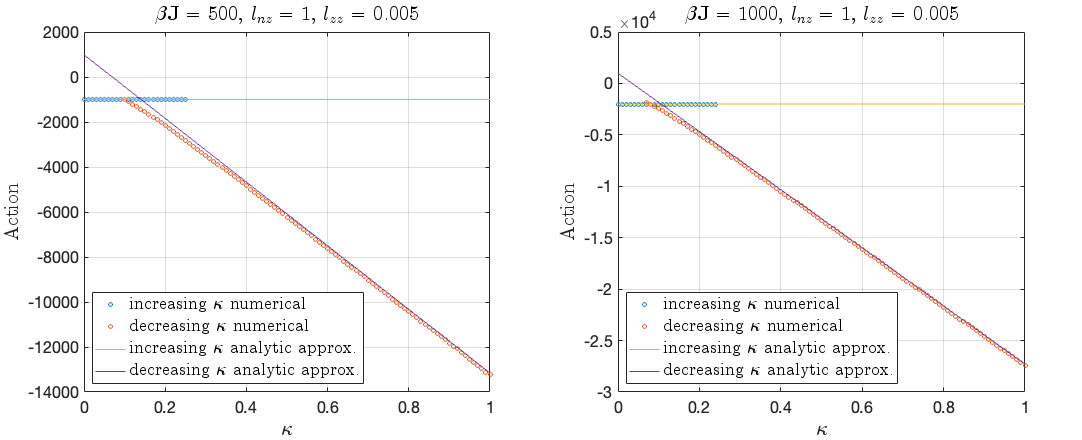}
    \caption{Action as a function of $\kappa$ for $l_{nz}=1, l_{zz}=0.005$. Circles are the numerical results and the sold lines are the analytic estimates. Left: $\beta \bJ =500$. Right: $\beta \bJ =1000$.}
    \label{fig:actioncurveslz=0p005}
\end{figure}
Therefore it seems we can indeed trust equation \eqref{eqn:smallLzzTransitionAnalytic} for small $l_{zz}$.  This implies we have a transition line that has a positive intercept $\kappa_0$ on the $\kappa$ axis, so that even at asymptotically low temperatures there is a chaotic phase as long as $\kappa<\kappa_0$, i.e., a zero temperature phase transition.  This is an interesting intermediate behavior between the $l_{zz}=0$ case (where there is no  chaotic phase at asymptotically low temperatures), and the $l_{zz} = l_{nz}^2$ case (where there is no first order transition line). Numerically this seems indeed the case.  In Figure \ref{fig:transitionLinelz=0p005}, we plot the phase diagram for the case of $l_{nz}=1$ and $l_{zz}=0.005$.  The left panel of Figure \ref{fig:transitionLinelz=0p005} shows the full phase transition curve; the right panel zooms into the low-temperature region and we take the lowest ten temperatures to perform a linear fit. The fit result is 
\begin{equation}
    \kappa_{*,\text{fit}} = 31.9 (\beta \bJ)^{-1} + 0.0486.
\end{equation}
The intercept 0.0486 may seem small but the fitting uncertainty is $4\times 10^{-4}$, so it is a significant fact that the intercept is nonzero. Meanwhile the analytic approximation of \eqref{eqn:smallLzzTransitionAnalytic} gives (for $l_{nz}=1$ and $l_{zz}=0.005$)
\begin{equation}
        \kappa_{*,\text{approx}} = 34.7 (\beta \bJ)^{-1} + 0.0707,
\end{equation}
which is in reasonable agreement with the numerical fit.

\paragraph{Acknowledgements}
We would like to thank Ahmed Almheiri, Dionysios Anninos, Andreas Blommaert, Ronny Frumkin, Damian Galante,  Antonio Garc\'{i}a-Garc\'{i}a, Vladimir Narovlansky, and Josef Seitz for useful discussions. YJ also thanks Chi-Ming Chang and Zhenbin Yang for their invitation to visit  Yau Mathematical Sciences Center (YMSC) and Institute for Advance Studies at Tsinghua University (IASTU), where part of this work is done.
 This work was supported in part by an Israel Science Foundation (ISF) center for excellence grant (grant number 2289/18), by ISF grant no. 2159/22, by the Minerva foundation with funding from the Federal German Ministry for Education and Research, by the German Research Foundation through a German-Israeli Project Cooperation (DIP) grant ``Holography and the Swampland''. YJ is also supported by the Koshland postdoctoral fellowship and by a research grant from Martin Eisenstein.
OM is supported by the ERC-COG grant NP-QFT No. 864583
``Non-perturbative dynamics of quantum fields: from new deconfined
phases of matter to quantum black holes'', by the MUR-FARE2020 grant
No. R20E8NR3HX ``The Emergence of Quantum Gravity from Strong Coupling Dynamics''. OM is also partially supported by
the INFN ``Iniziativa Specifica GAST''.

\appendix

\section{The integrable $p$-spin model}
\label{app:p-spin model}
The integrable $p$-spin model, as defined in the text around \eqref{eq:p-spin-Hamiltonian}, is a well studied model. In particular, at low energies it exhibits replica symmetry breaking and a spin-glass phase. In this appendix we show that in the double scaling limit and in our normalization, the critical temperature for the glassy transition is pushed to zero.

As given in the main text, the integrable $p$-spin model is defined by the following Hamiltonian acting on a system of $N$ qubits: 
\begin{equation}
    H_{p\text{-spin}} = \sum_{1\le b_1 < \dots < b_p \le N} B_{b_1\cdots b_p} \sigma^{z}_{b_1}\cdots \sigma^{z}_{b_p} \,,
\end{equation}
where the $B$'s are drawn from a random Gaussian distribution, 
\begin{equation}
    \braket{B_I} = 0 \,,\qquad \braket{B_I B_J} = \cJ^2 \binom{N}{p}^{-1} \delta_{IJ} \,.
\end{equation}
By a Jordan-Wigner transformation\footnote{The Jordan-Wigner form of the fermions is
\begin{equation}
\label{eq:JordanWigner}
\psi_{2k-1} = \overbrace{\sigma^{z} \otimes\cdots\otimes\sigma^{z}}^{k-1}  \otimes\sigma^{x}\otimes\overbrace{\mathbf{1}\otimes\cdots\otimes\mathbf{1}}^{N-k}\,, \qquad 
\psi_{2k} = \overbrace{\sigma^{z} \otimes\cdots\otimes\sigma^{z}}^{k-1}  \otimes\sigma^{y}\otimes\overbrace{\mathbf{1}\otimes\cdots\otimes\mathbf{1}}^{N-k}\,,
\end{equation}} the Pauli matrices can be written in terms of $2N$ Majornana fermions, where the model takes the form of commuting-SYK \cite{Gao:2023gta} Hamiltonian of the form \eqref{eq:commuting-SYK}.

  This Hamiltonian is integrable in the sense of level statistics: since all terms commute, each energy level is a sum of (weakly correlated) random numbers,  and the level spacing distribution must be Poisson. It is also integrable in the sense that it has as many conserved charges, $\{\sigma_i^z\}_{i=1}^N$, as there are degrees of freedom. The model exhibits spin glass behaviors at low temperatures \cite{gardener1985}, which will turn out to be irrelevant for our discussion. It is known that in the large $p$ limit (after the large $N$ limit is taken first), this model reduces to the random energy model (REM) \cite{derridaPRL,derridaPRB} which is much simpler.  
The REM is  defined by the following three properties:
\begin{enumerate}
    \item The system has $2^{N}$ energy levels $E_{i}$.

    \item  The one-level probability distribution of energy levels is Gaussian: $P\left(E\right)\propto e^{-\frac{E^{2}}{2 \cJ^{2}}}$.

    \item The energy levels are independent random variables. More precisely, $n$-level probability distributions all factorize when $n\ll N$.
\end{enumerate}
In the double-scaled limit, $p$ goes to infinity as well so we expect it to reduce to an REM too.  Indeed one can check for example the two-level distribution $P(E_1, E_2)$ factorizes into $P(E_1)P(E_2)$ in the double-scaled limit.
One of the main results of \cite{derridaPRL, derridaPRB} is that the phase transition temperature of the REM is given by
\begin{equation}
    T_c = \sqrt{\frac{\cJ^2}{2N\log 2}},
\end{equation}
below which the system exhibits glassy behavior, and above which the annealed averages can be trusted.
In the normalization of \cite{derridaPRL, derridaPRB}, $\cJ^2 \propto N$ and therefore the phase transition happens at an order-one temperature in their paper.   Our double-scaled normalization is
\begin{equation}
    \cJ^2 = \frac{1}{\lambda} \bJ^2, \qquad \lambda \equiv \frac{2p^2}{N},
\end{equation}
where $\bJ$ does not scale with $\lambda$ or $N$. Since the double-scaled limit also goes to REM,   we would have 
\begin{equation}
    T_c = \sqrt{\frac{\bJ^2}{2N \lambda \log 2}},
\end{equation}
and since the large $N$ limit is taken before the $\lambda \to 0$ limit, $T_c$ is zero in our normalization and therefore we need not worry about entering an ordered phase which would have invalidated the annealed computations.

\section{A crash course on chord diagrams} \label{app:CrshCrdDiag}

In this appendix we will derive the chord rules for three classes of models that give the same transfer matrix to leading order in $1/N$. These are  the double scaled limit of  SYK model (\ref{eq:H_SYK_def}), the double scaled limit of  the Pauli model (\ref{eq:def spin-SYK}) and  the Parisi hypercube model \cite{Jia:2020rfn,Berkooz:2023cqc,Berkooz:2023scv}. Then we will present the transfer matrix for a single chord species \cite{Berkooz:2018jqr,Berkooz:2018qkz}. In all these models we begin by presenting known results, which are chord rules for the chaotic Hamiltonian and for chaotic operators of general length. Then we move to define the integrable Hamiltonian and operators of general length from the same class. This allows us to derive the chord rules for the Integrable-to-Chaos Hamiltonian (\ref{eq:Cha_integ_Ham}). Lastly, we define a new class of operators which we call ``'polarized operators'', that give the usual intersection factor when crossing chaotic chords, but give a general factor when crossing each other. These have the nicer construction in the Pauli and Parisi models, but can also be constructed in the Majorana model.

\subsection{Fermionic model} \label{app:CrshCrdDiagMajo}

\subsubsection{Chord rules for DS-SYK}

\paragraph{Definition of the model:} Consider $N$ species of Majorana fermions $\psi_i$, $i=1,\cdots,N$ with the anti-commutation relations $\{\psi_i,\psi_j\}=2\delta_{ij}$, and let $p\in2\mathbb{N}$. Let us denote sets of $p$ distinct sites $1\leq i_1<\cdots<i_p\leq N$ by capital $I$ (with subscripts whenever we have several of those). Then we can write the model as
\begin{align}
H_{\text{SYK}} = i^{p/2}\sum_{|I|=p}J_I\psi_I \,,
\end{align}
where $\psi_I$ stands for the appropriate string $\psi_I=\psi_{i_1}\cdots\psi_{i_p}$ and the sum is over all ordered index sets of length $p$. The couplings $J$ are independent random Gaussian with 
\begin{align} 
\vev{J_I}=0,\qquad \vev{J_I J_{I'}} = {N \choose p}^{-1}\delta_{I,I'}\cJ^2 \,.   
\end{align}
We will set $\cJ=1$ for convenience, where $\langle\cdots\rangle$ stands for ensemble average. In this appendix we will not use $\bJ$, \eqref{eq:def bJ}. Had we used it, the moment $m_k$ would have had an additional factor of $\bJ^k\lambda^{-k/2}$.

\begin{figure}[t]
        \centering
        \includegraphics[width=0.25\textwidth]{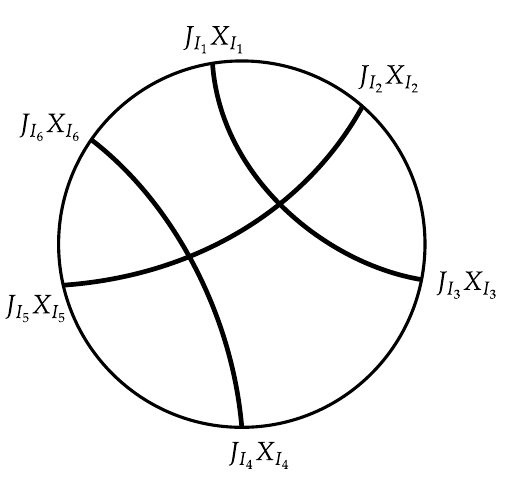}
    \caption{
    A chord diagram that contributes to $m_6$, representing the Wick contractions of $\langle J_{I_1} J_{I_3}\rangle \langle J_{I_2} J_{I_5}\rangle \langle J_{I_4} J_{I_6}\rangle \Tr(\psi_{I_1}\cdots \psi_{I_6})$. It contributes as $q^2$ to the sum as it has two chord intersections.}
    \label{fig:chord diagram apndx}
\end{figure}

\paragraph{Obtaining the chord picture:} To obtain the chord construction \cite{Berkooz:2018jqr,Berkooz:2018qkz} we can consider the partition function, which we expand using the moment expansion
\begin{align} \label{eq: def partition_moment}
    Z(\beta) = \vev{\Tr\left(e^{-\beta H}\right)} = \sum_{k=0}^{\infty}\frac{(-\beta)^k}{k!}m_k,\qquad m_k \equiv \vev{\Tr \left(H^k\right)} \,.
\end{align}
Here and in the main text we normalize the trace such that $\Tr(\1)=1$. We plug the Hamiltonian (\ref{eq:H_SYK_def}) into the definition of $m_k$ to obtain
\begin{align}
    m_{k}=i^{kp/2}\sum_{I_{1},\cdots I_{k}}\left\langle J_{I_{1}}\cdots J_{I_{k}}\right\rangle \text{Tr}\left(\psi_{I_{1}}\cdots\psi_{I_{k}}\right) \,.
\end{align}
Due to the Gaussian distribution (\ref{eq:cJ norm}), the expectation value over the coefficients is given by a sum over Wick contractions. This in turn means that the moment $m_k$ is given by all possible traces involving $k/2$ operator strings $\psi_I$, each of which appears twice in $m_k$, as
\begin{align} \label{eq:single_chord_moment}
    m_k = i^{kp/2}{\binom{N}{p}}^{-k/2}\sum_{\text{pairings}}\sum_{I_1,\cdots,I_{k/2}}\Tr(\psi_{I_1}\psi_{I_2}\cdots\psi_{I_1}\cdots) \,.
\end{align}
Each term in the sum over Wick contractions can be represented using a chord diagram: let each of the $k$ operator strings $\psi_I$ define a node on a circle. Each node is labelled by an index $j=1,\cdots,k$. We then connect the nodes in pairs, to designate which pairs have identical sets of species $I_i$. See Figure \ref{fig:chord diagram apndx} for an example of a chord diagram. 

Next we need to evaluate each diagram. Since the operators are composed of Majorana fermions, it is easy to see that two composite operators with the same index set ``annihilate'' each other,
\begin{equation}
    \psi_I \psi_I = i^p\,.
\end{equation}
Consider a specific diagram given by a specific pairing, 
%(still carrying the sum over $I_1,..,I_{k/2}$), i.e.,  
$
i^{kp/2}{N \choose p}^{-1}\sum_{I_{1},\cdots,I_{k/2}}\text{Tr}\left(\psi_{I_{1}}\psi_{I_{2}}\cdots\psi_{I_1}\cdots\right) \,.
$
To evaluate this expression we need to disentangle the diagram by exchanging nodes such that eventually the chords connect neighboring nodes, i.e., $\Tr(\psi_{I_1}\psi_{I_1}\psi_{I_2}\psi_{I_2}\cdots)$, and then annihilate the pairs. This corresponds to commuting the strings of $\psi$'s. Upon commuting $\psi_{I},\psi_{I'}$ we get a factor of $(-1)^{\left|I\cap I'\right|}$, where $\left|I\cap I'\right|$ is the number of indices that appear both in $I$ and in $I'$. Here we can use two major simplifications coming from the double scaled limit \cite{erdHos2014phase}:
\begin{enumerate}
    \item The number of overlapping site indices between any two index sets $\equiv|I\cap I'|$ is a Poisson distributed random variable with mean $p^2/N$.
    \item With probability $1$, the intersection of any three index sets vanishes, namely $|I_i\cap I_j \cap I_k|=0$, for $i\neq j\neq k$. This statement is summarized in lemma $(9)$ there, and subsequent discussion.
\end{enumerate}

The ${N \choose p}^{-k/2}$ prefactor  turns the counting of appearances of a certain type in the sum into probabilities of such events. Therefore, each intersection in the chord diagram gives a factor (summing over the possibilities for the number $m$ of sites in the intersection)
\begin{align}
    e^{-p^{2}/N}\sum_{m=0}^{\infty}\frac{\left(p^{2}/N\right)^{m}}{m!}\left(-1\right)^{m}=e^{-\lambda}=q,\qquad\lambda\equiv\frac{2p^{2}}{N}.
\end{align}
After commuting the terms, the pairs are neighboring, each giving just an $i^p$, such that all of these together cancel the $i^{kp/2}$ factor. 

We find that $m_{k}$ is given by a sum over chord diagrams, with each intersection of two chords simply assigned a factor of $q$. The moment \eqref{eq: def partition_moment} then takes the final form
\begin{align} \label{eq:ChordPF}
    m_k = \sum_{\text{CD}(k)}q^{ \text{No. of intersections}},
\end{align}
where $\text{CD}(k)$ are chord diagrams with $k$ nodes (i.e., $k/2$ chords).

\paragraph{Chaotic operators of general length}

Next we will consider chord rules for double scaled operators of general length $p_C\sim\sqrt{N}$. The operators that we will discuss in this subsection will be referred to as \textit{chaotic}, as they are in the same class as the chaotic SYK model. This is to be contrasted with \textit{integrable} operators which are operators of the same class as the integrable Hamiltonian, and \textit{polarized} operators, which are consisted of a product of fermions and fermion bi-linears.

We define a chaotic operator $M$ of length\footnote{We take here $p_C$ to be even as we will want to use the derived chord rules for the Hamiltonian. One could also define operators of odd length, where we get an additional minus sign upon commuting the Majoranas.} $p_C\in 2\mathbb{N}$  by
\begin{align}
    M_C=\sum_{|I|=p_C}\tilde{J}_{I}\psi_{I},
\end{align}
where $\tilde{J}$ is a new Gaussian random variable with
\begin{align}
\vev{\tilde{J}_I}=0,\qquad \vev{\tilde{J}_I \tilde{J}_{I'}} = {N \choose p_C}^{-1}\delta_{I,I'}.    
\end{align}
We wish to compute averaged traces involving some sequence 
 of $H_{\text{SYK}}$ and $M_C$, i.e.,
\begin{align} \label{eq:def_operator_sequence}
    \vev{\Tr\left(\prod_{i=1}^{k}\cO_i\right)},\ \ \  \cO_i\in\left\{H_{\text{SYK}},M_C \right\},\quad 1\leq i\leq k.
\end{align}
 These types of traces arise naturally when considering correlation functions \cite{Berkooz:2018jqr}, or moments in a theory where we deform $H_{\text{SYK}}$ by $M_C$, as we will consider later. As in the above case, we begin with Wick contractions, where we can only contract operators of the same type. Then we wish to disentangle the diagrams such that chords connect neighboring nodes, getting a factor of $(-1)^{|I\cap I'|}$ upon commutation of $\psi_I,\psi_I'$. Since $p,p_{C}\sim\sqrt{N}$, the analysis of \cite{erdHos2014phase} apply, and each specie appears at two different index sets/chords at most. The overlap of two index sets of length $p_{1},p_{2}$ is Poisson distributed with mean $\frac{p_{1}p_{2}}{N}$  \cite{Berkooz:2018qkz,Berkooz:2018jqr}, therefore,
\begin{itemize}
    \item for each $M_C-M_C$ crossing we have
    \begin{align}
        e^{-\frac{p_{C}^{2}}{N}}\sum_{m=0}^\infty \frac{\left(\frac{p_{C}^{2}}{N}\right)^{m}}{m!}\left(-1\right)^{m}=e^{-2\frac{p_{C}^{2}}{N}}\equiv q_C.
    \end{align}

    \item for each $H_{\text{SYK}}-M_C$ crossing we have
    \begin{align}
        e^{-\frac{p_{C}p}{N}}\sum_{m=0}^\infty \frac{\left(\frac{p_{C}p}{N}\right)^{m}}{m!}\left(-1\right)^{m}=e^{-2\frac{p_{C}p}{N}}\equiv q_{HC}.
    \end{align}
\end{itemize}
Then we find that 
\begin{align}
    \vev{\Tr\left(\prod_{i=1}^{k}\cO_i\right)} = \sum_{\text{CD}(\prod_{i}^{k}\cO_i)} q^{\# H-H \text{ intersections}} q_{C}^{\# M_C-M_C \text{ intersections}} q_{HC}^{\# H-M_C \text{ intersections}},
\end{align}
where $\text{CD}(\prod_{i}^{k}\cO_i)$ are all diagrams where the nodes are ordered according to the ordering of the operator product and only operators of the same kind are connected.

\subsubsection{Chord rules for integrable-to-chaos Hamiltonian} \label{sec:Chord_rules_int+cha}

Let us consider the Hamiltonian (\ref{eq:Cha_integ_Ham}), where we take (\ref{eq:H_SYK_def}) for the chaotic Hamiltonian and (\ref{eq:commuting-SYK}) as the integrable Hamiltonian. In fact, we will consider a generalized version in which the two Hamiltonians have different lengths, and specialize to the equal length case in the end. For $p_\text{Integ}\in 2\bN$ our Hamiltonian is then
\begin{align}
    H=\nu H_{\text{SYK}}+\kappa H_{\text{I-SYK}}=i^{p/2}\nu\sum_{\left|I\right|=p}J_{I}\psi_{I}+i^{p_{\text{Integ}}/2}\kappa\sum_{\left|L\right|=p_{\text{Integ}}}B_{L}\psi_{L},
\end{align}
where $L$ is a multi-index set
\begin{align} \label{eq:def:L_multi_index}
L=\left\{\left(2\ell_1-1,2\ell_1,\cdots,2\ell_{p_{\text{Integ}}/2}-1,2\ell_{p_{\text{Integ}}/2}\right) \;\bigg\vert\; 1 \le \ell_1 < \cdots < \ell_{p_{\text{Integ}}/2} \le N/2 \right\}.    
\end{align}
$B$ is an independent Gaussian variable with
\begin{align}
    \vev{B_L}=0,\qquad \vev{B_L B_{L'}}={N/2 \choose p_{\text{Integ}}/2}^{-1}\delta_{L,L'}.
\end{align}
Our main focus here, similar to the single chord case, is to compute $m_k\equiv \vev{\Tr \left(H^k\right)}$. By plugging in $H$ we see that $m_k$ is given by a sum of terms, 
\begin{align}
    m_{k}=\sum_{n+z=k}\text{Tr}\left\langle \left(\nu H_{\text{SYK}}\right)^{n_{1}}\left(\kappa H_{\text{I-SYK}}\right)^{z_{1}}\cdots\right\rangle ,\qquad n=\sum n_{i}, \quad z=\sum z_{i}\,.
\end{align}
Focus on a single term in this sum. Due to the Gaussian averaging we can write it as a sum over Wick contractions (we can only contract two Hamiltonians of the same type). Both Hamiltonians are of length $p\sim \sqrt{N}$, and so we have no triple or higher intersections \cite{erdHos2014phase}.

From the same arguments above, $I-I$ and $L-I$ overlaps (index sets associated with chaotic-chaotic and integrable-chaotic) are distributed as $\text{Pois}(p^2/N)$ and $\text{Pois}(p\cdot p_\text{Integ}/N)$ respectively. The $I-I$ case is the same as in the single chord case. For the $I-L$ overlap we use that the $I$ multi-index has a uniform distribution over the indices, and so the consecutive index constraint coming from $L$ is of no importance. Repeating the same arguments as above, the $L-I$ chord intersection factor is 
 \begin{equation}
 q_{nz}=e^{-2p\cdot p_\text{Integ}/N} \,.
 \end{equation}
There are also $L-L$ intersections  but since the operators commute it gives a factor of $1$ for the chord intersection.

\begin{figure}
    \centering
    \includegraphics[width=0.4\textwidth]{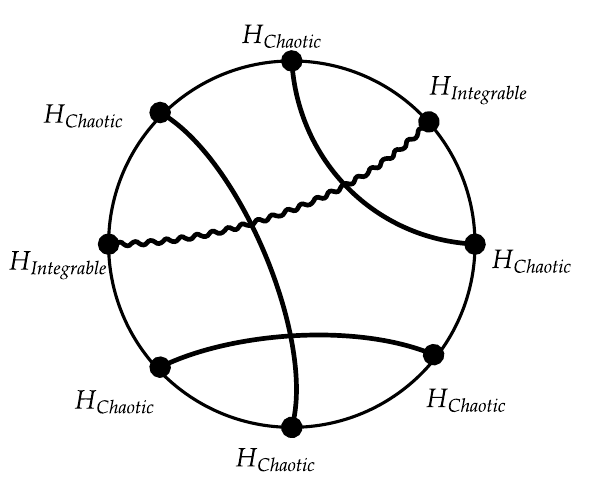}
    \caption{A chord diagram contributing to $m_8$ in (\ref{eq:Majorana_CPF})  with $n=6$ and $z=2$, coming from a specific Wick contraction of $\vev{\Tr(H_C^4H_{\text{Integ}}H_C^2H_{\text{Integ}})}$. It contributes $q\cdot q_{nz}^2$ to the sum over chord diagrams, and enters with a prefactor of $\kappa^6\nu^2$.}
    \label{fig:chord_diagram_2_chords}
\end{figure}

With each $H_{\text{SYK}}-H_{\text{SYK}}$ giving us an intersection factor of $q$ as before, we get that the Hamiltonian moment reads
\begin{align} \label{eq:Majorana_CPF}
    m_{k}=\left\langle \Tr \left(H^{k}\right)\right\rangle =\sum_{\text{\ensuremath{\substack{\text{chord diagrams with}\\
n+z = k\text{ nodes}
}
}}}\kappa^{n}\nu^{z}q^{\#n-n\text{ intersections}}q_{nz}^{\#n-z\text{ intersections}}.
\end{align}
To obtain the specific case of the Hamiltonian \eqref{eq:Interpolating_Hamiltonian_Majorana}, in which the chaotic and integrable Hamiltonians are of the same length, we set $p_\text{Integ}=p$. This sets $q_{nz}=q$. An example to a diagram contributing to $m_8$ is shown in Figure~\ref{fig:chord_diagram_2_chords}.

\subsubsection{Mixed operators}

We could also consider operators built from combinations of individual fermions and fermion-bilinears, allowing for generic intersection factors $q_{ij}$ between different types of chords. We will construct similar operators below using Pauli matrices, which is simpler for this purpose.

\subsection{The $p$-spin model} \label{app:pauliChordRules}

Another class of operators that gives rise to the same chord structure as DS-SYK is the Pauli spin model in the double scaled limit \cite{Berkooz:2018qkz, erdHos2014phase}, as seen in equation (\ref{eq:def spin-SYK}). Here we will present several types of operators and derive their chord intersection rules. In addition to the chaotic and integrable operators similar to the ones in the SYK model, it is easy to identify an intermediate class.

\subsubsection{Classes of operators}

Here we will present three different classes of operators---chaotic, integrable and polarized. While the microscopic operators are defined in terms of the allowed directions the Pauli operators can take, in the chord picture the difference comes about in the intersection factors: crossings of the chaotic and integrable chords (between themselves or of each other) are given purely in terms of the length of the operator, while for the polarized chords we have an additional degree of freedom that allows us to control the intersection factors.

Consider $N$ sites with spin $1/2$ degree of freedom on each. Denote the Pauli matrices acting on site $i=1,2,\cdots,N$ by $\sigma_{i}^{a}$, with $a=1,2,3$. Let $I$ be some ordered index set of some length $p$ such that $I\in\left\{ \left(i_{1},\cdots,i_{p}\right)\Big\vert1\leq i_{1}<\cdots<i_{p}\leq N\right\}$, and let $A$ be the corresponding directions vector $A\in\left\{ \left(a_{1},\cdots,a_{p}\right)\Big\vert1\leq a_{i}\leq 3\text{ for }1\leq i\leq p\right\}$. We will use multi indices to denote a product of Pauli matrices as $\sigma_{I}^{A}\equiv\sigma_{i_{1}}^{a_{1}}\cdots\sigma_{i_{p}}^{a_{p}}$. Then we define the following three classes of operators:

\begin{itemize}
    \item \textbf{Chaotic operators:} We define a chaotic operator of length $p_{C}$ as
    \begin{align}
        M_{C}=\sum_{\left|I\right|=\left|A\right|=p_{C}}J_{I}^{A}\sigma_{I}^{A},
    \end{align}
    where the sum is over all $I,A$ multi-index sets as defined above. The couplings $J$ are independent random Gaussians with
    \begin{align}
        \left\langle J_{I}^{A}\right\rangle =0,\qquad\left\langle J_{I}^{A}J_{I'}^{A'}\right\rangle =3^{-p_{C}}{N \choose p_{C}}^{-1}\delta_{I,I'}\delta_{A,A'}.
    \end{align}
    We note that the chaotic Hamiltonian (\ref{eq:def spin-SYK}) is nothing but a chaotic operator of length $p$.

    \item \textbf{Integrable operators:} We define an integrable operator of length $p_{\text{Integ}}$ as
    \begin{align}
        M_{\text{Integ}}=\sum_{\left|I\right|=p_{\text{Integ}}}B_{I}\sigma_{I}^{3},
    \end{align}
    where $\sigma_{I}^{3}\equiv\sigma_{i_{1}}^{3}\cdots\sigma_{i_{p}}^{3}$, and the couplings $B$ are independent random Gaussian variables with
    \begin{align}
        \left\langle B_{I}\right\rangle =0,\qquad\left\langle B_{I}B_{I'}\right\rangle ={N \choose p_{\text{Integ}}}^{-1}\delta_{I,I'}.
    \end{align}
    Note that the integrable Hamiltonian (\ref{eq:p-spin-Hamiltonian}) is a member of this operator class.

    \item \textbf{Polarized operators:} Consider the operator of length $p_P$
    \begin{align}
        \label{eq:polarized_op_def}
        M_{P}=\sum_{\left|I\right| =\left|A\right|=p_{P}}L_{I}^{A}\sigma_{I}^{A},
    \end{align}
    where now different directions in $A$ appear with different probabilities $0\leq\alpha_{1},\alpha_{2},\alpha_{3}\leq1$ and $\sum_{i}\alpha_{i}=1$:
    for example taking $\alpha_{i}=\frac{1}{3}$ for $i=1,2,3$ gives us a chaotic operator and $\alpha_{1}=\alpha_{2}=0$, $\alpha_{3}=1$ gives us an integrable operator. $L$ is a random Gaussian variable with
    \begin{align}
        \left\langle L_{I}^{A}\right\rangle =0,\qquad\left\langle L_{I}^{A}L_{I'}^{A'}\right\rangle ={N \choose p_{P}}^{-1}\delta_{I,I'}\delta_{A,A'}\prod_{j=1}^{3}\left(\alpha_j\right)^{\text{No. of } j \text{ in }A},
    \end{align}
    where ``No. of $j \text{ in }A$'' is the number of times the direction $j\in\{1,2,3\}$ appears in $A$.

\end{itemize}

\subsubsection{Chord rules}
As we did in the Majorana case in Section~\ref{app:CrshCrdDiagMajo}, we wish to know how to compute averaged traces\footnote{Note that we take the normalized trace $\text{Tr}\left(\1\right)=1$, which means that we define $\text{Tr}=2^{-N}\prod_{i=1}^{N}\text{tr}_{i}$, where $\text{tr}_{i}$ is the unnormalized trace in the $i^{\text{th}}$ site.} of the form
\begin{align}
    \vev{\Tr\left(\prod_{i=1}^{k}\cO_i\right)},\ \ \  \cO_i\in\left\{M_C, M_{\text{Integ}},M_P\right\} ,\quad 1\leq i\leq k.
\end{align}

The first step, as in the Majorana case, is to perform the average over the couplings. We get a sum over Wick contractions, where we can only contract two operators of the same type. As before this gives a chord diagram structure. Since we take the length of the operators $p_{C},p_{\text{Integ}},p_{P}\sim\sqrt{N}$ and $N\to\infty$, the results of \cite{erdHos2014phase,Berkooz:2018jqr} apply. We find that
\begin{enumerate}
    \item For $k\sim O\left(N^{0}\right)$, each site appears in two index-sets at most. 

    \item The overlap of two index sets of length $p_{1},p_{2}$ is Poisson distributed with parameter $p_{1}p_{2}/N$.
\end{enumerate}

Focus on a specific pairing in the sum over Wick contractions. We wish to evaluate the trace for each site $i=1,\cdots,N$. Due to property (1), a site can appear in either a single chord or two chords at most. If it appears in a single chord we get a factor of $1$. If it appears in two chords the result depends on the ordering of these two chords: remember that the trace is cyclic and so there are only two possibilities for an ordering of operators---non-intersecting (i.e., appear in the pairing as $\text{Tr}\left(\mathcal{O}_{1}\cdots\mathcal{O}_{1}\cdots\mathcal{O}_{2}\cdots\mathcal{O}_{2}\cdots\right)$), and intersecting (appear in the form $\text{Tr}\left(\mathcal{O}_{1}\cdots\mathcal{O}_{2}\cdots\mathcal{O}_{1}\cdots\mathcal{O}_{2}\cdots\right)$). 
We always find that non-intersecting chords, as well as the trace of a site appearing only in a single chord to give a factor of $1$, and a single-site trace for two intersecting chords gives
\begin{align}
    \frac{1}{2}\sum_{a,b=1,2,3}f_{a}g_{b}\text{tr}\left(\sigma^{a}\sigma^{b}\sigma^{a}\sigma^{b}\right),
\end{align}
where $f_{1}=f_{2}=f_{3}=\frac{1}{3}$ for a chaotic operator, $f_{1}=f_{2}=0$, $f_{3}=1$ for an integrable operator and $f_{1}=\alpha_{1}$, $f_{2}=\alpha_{2}$, $f_{3}=\alpha_{3}$ for a polarized operator (and the same for $g$). We summarize the results below (this factor is symmetric, and the part below the diagonal was omitted as to not overburden the reader)
\begin{align}
    \begin{matrix} & M_{C} & M_{\text{Integ}} & M_{P}\\
M_{C} & -1/3 & -1/3 & -1/3\\
M_{\text{Integ}} & \cdot & 1 & d_{\vec{\alpha}}\\
M_{P} & \cdot & \cdot & c_{\vec{\alpha}}
\end{matrix},\qquad\begin{matrix}c_{\vec{\alpha}} & \equiv & \alpha_{1}^{2}+\alpha_{2}^{2}+\alpha_{3}^{2}-2\left(\alpha_{1}\alpha_{2}+\alpha_{1}\alpha_{3}+\alpha_{2}\alpha_{3}\right)\\
d_{\vec{\alpha}} & \equiv & \alpha_{3}-\alpha_{1}-\alpha_{2}
\end{matrix}.
\end{align}

The total factor coming from an intersection of two chords is given by the amount of indices they share which is Poisson distributed as stated above. The general formula here is $e^{-\frac{p_{1}p_{2}}{N}}\sum_{m=0}^{\infty}\frac{\left(\frac{p_{1}p_{2}}{N}\right)^{m}}{m!}\left(\substack{\text{single site}\\ \text{intersection factor}}\right)^{m}$. The results are summarized in Table~\ref{tab:intersection factors}.
% \begin{align}
%     \begin{matrix} & M_{C} & M_{\text{Integ}} & M_{P}\\
% M_{C} & q_{C} & q_{CI} & q_{C\alpha}\\
% M_{\text{Integ}} & \cdot & 1 & q_{I\alpha}\\
% M_{P} & \cdot & \cdot & q_{P}
% \end{matrix},\qquad\begin{matrix}q_{C} & = & e^{-\frac{4p_{C}^{2}}{3N}}\\
% q_{CI} & = & e^{-\frac{4p_{C}p_{\text{Integ}}}{3N}}\\
% q_{C\alpha} & = & e^{-\frac{4p_{C}p_{P}}{3N}}\\
% q_{I\alpha} & = & e^{-\frac{p_{\text{Integ}}p_{P}}{N}\left(1-d_{\alpha}\right)}\\
% q_{P} & = & e^{-\frac{p_{P}^{2}}{N}\left(1-c_{\alpha}\right)}
% \end{matrix}.
% \end{align}
Then we find
\begin{align} \label{eq:General_Pauli_trace}
\begin{split}
    \vev{\Tr\left(\prod_{i=1}^{k}\cO_i\right)} =\sum_{\text{CD}\left(\prod_{i=1}^{k}\cO_i\right)}&q_{C}^{\#M_{C}-M_{C}\text{ intersections}}q_{P}^{\#M_{P}-M_{P}\text{ intersections}}q_{C\text{Integ}}^{\#M_{C}-M_{\text{Integ}}\text{ intersections}}\times
    \\
    & \qquad \times q_{CP}^{\#M_{C}-M_{P}\text{ intersections}}q_{\text{Integ}P}^{\#M_{\text{Integ}}-M_{P}\text{ intersections}},
\end{split}
\end{align}
where $\text{CD}\left(\prod_{i=1}^{k}\cO_i\right)$ are all chord diagrams where the nodes are ordered according to the operator product, and only two nodes of the same type can be contracted.

\begin{table}
    \centering
    % \begin{tabular}{| >{\centering\arraybackslash} m{1cm} || >{\centering\arraybackslash} m{3cm} | >{\centering\arraybackslash} m{3cm} | >{\centering\arraybackslash} m{4cm} |}
    \begin{tabular}{| c || c | c | c|}
        \hline
         & $M_{C}$ & $M_{\text{Integ}}$ & $M_{P}$\\
         \hline\hline
         \raisebox{-0.5\height}{$M_{C}$} & \raisebox{-0.5\height}{$q_C = e^{-\frac{4}{3}\frac{p_C^2}{N}}$} & \raisebox{-0.5\height}{$q_{C\text{Integ}} = e^{-\frac{4}{3} \frac{p_C p_{\text{Integ}}}{N}}$} & \raisebox{-0.5\height}{$q_{CP} = e^{-\frac{4}{3} \frac{p_C p_P}{N}}$} \\[1.5ex]
         \hline
         \raisebox{-0.5\height}{$M_{\text{Integ}}$} & \raisebox{-0.5\height}{$q_{C\text{Integ}}$} & \raisebox{-0.5\height}{$1$} & \raisebox{-.5\height}{$q_{\text{Integ}P} = e^{-\frac{p_{\text{Integ}} p_P}{N}(1-d_{\vec{\alpha}})}$} \\[1.5ex]
         \hline
         \raisebox{-0.5\height}{$M_{P}$} & \raisebox{-0.5\height}{$q_{CP}$} & \raisebox{-0.5\height}{$q_{\text{Integ}P}$} & \raisebox{-.5\height}{$q_P = e^{-\frac{p_P^2}{N}(1-c_{\vec{\alpha}})}$} \\[1.5ex]
         \hline
    \end{tabular}
    \caption{The intersection factors between chords.}
    \label{tab:intersection factors}
\end{table}

\subsubsection{Chord rules for the Hamiltonian (\ref{eq:def Pauli_combined_system})}
We can use the above (\ref{eq:General_Pauli_trace}) to find the chord rules for the Hamiltonian (\ref{eq:def Pauli_combined_system}) by taking $p_{C}=p_{\text{Integ}}=p$ without any $M_{P}$ operators, and by remembering that any $H_{\text{C-Spin}},H_{\text{I-Spin}}$ insertions come with a factor of $\kappa,\nu$ respectively. Then 
\begin{align}
    \left\langle \Tr\left(H^{k}\right)\right\rangle =\sum_{\substack{\text{chord diagrams with}\\
n+z=k\text{ nodes}
}
}\kappa^{n}\nu^{z}q^{\#n-n\text{ intersections}}q^{\#n-z\text{ intersections}}.
\end{align}

\subsection{The Parisi Hypercube model}\label{app:parisiChordRules}
It is observed that \cite{Jia:2020rfn,Berkooz:2023cqc,Berkooz:2023scv} the chord combinatorics of the DS-SYK model can be quite generally reproduced by starting from hopping operators that are highly fluxed in the Fock space. One can view this class of models as describing the dynamics of a many-body wavefunction in the presence of a large amount of random Berry curvatures. It was argued that the DS-SYK model and the DS Pauli spin model can be viewed as models in this class.  A particularly simple representative of this class is a hypercube model of Parisi \cite{Parisi:1994jg}. This model can be described on a Fock space of $N$ qubits, and the basic hopping operators are
 \begin{equation}
    T_i^+ = \sigma_i^+ e^{\frac{i}{4} \sum_{k, k\neq i}^N F_{ik} \sigma^3_k }, \quad    T_i^- =  (T_i^+)^\dagger,\quad \sigma^+ \equiv \frac{1}{2}(\sigma^1_i + i \sigma^2_i), \quad i = 1, \ldots,N, 
\end{equation}
where $\sigma^a_i$ is one of the Pauli matrices ($a=1,2,3$) acting on the $i^{\text{th}}$ qubit. 
The fluxes $F_{ij}$ are antisymmetric in $i$ and $j$. Moreover they are
quench disordered, and independently and identically distributed in distinct pairs of $[ij]$.  
The distribution function is even in $F_{ij}$, so that $\langle\sin F \rangle =0$ and $\langle \cos F \rangle$ is a tunable parameter. 
These fluxed hoppings satisfy the algebra 
 \begin{equation} \label{eqn:TopAlgebra}
 \begin{split}
        T_i^\pm  T_j^\pm =  T_j^\pm  T_i^\pm e^{i F_{ij}},& \quad 
     T_i^\pm  T_j^\mp =  T_j^\mp  T_i^\pm e^{-i F_{ij}}, \quad (i\neq j) \\
      ( T_i^\pm)^2=0,   & \quad T_i^\pm T_i^\mp = \sigma_i^\pm \sigma_i^\mp.
 \end{split}
\end{equation}
The Parisi's hypercube Hamiltonian is 
\begin{equation}
    H_\text{Parisi} = -\frac{1}{\sqrt{N}} \sum_{i=1}^N (T_i^+ +     T_i^-),
\end{equation}
which gives rise to the same chord combinatorics as those of the DS-SYK, with $q = \langle \cos F \rangle$. This Hamiltonian is chaotic in both the sense of the out-of-time-ordered correlation functions and the sense of random matrix level statistics.  

An interpolated Hamiltonian can be constructed as  
\begin{equation}
     H = -\frac{\nu}{\sqrt{N}} \sum_{i=1}^N (T_i^+ +     T_i^-) -\frac{\kappa}{\sqrt{N}} \sum_{i=1}^N (\tilde T_i^+ +     \tilde T_i^-)
\end{equation}
where $\tilde T_i^\pm$ are defined in the same way as $T_i^\pm$, but with a second random flux $\tilde F_{ij}$.  %This results in an additional set of algebraic relations 
% {\bf (MB Maybe we can drop this relation?)}
%   \begin{equation}\label{eqn:mixedFluxAlg}
%  T^\pm_i \tilde  T^\pm_j = e^{i \frac{F_{ij}+ \tilde F_{ij}}{2}}  \tilde  T^\pm_j  T^\pm_i , \quad  T^\pm_i \tilde  T^\mp_j = e^{-i \frac{F_{ij}+ \tilde F_{ij}}{2}}  \tilde  T^\mp_j  T^\pm_i.
%    \end{equation}
The chord rules are of the same form as equation \eqref{eqn:genericChordRules} with
\begin{equation}
    q_{nn}= \vev{\cos F}, \quad     q_{zz}= \vev{\cos \tilde F}, \quad  q_{nz}= \vev{\cos \frac{F+\tilde F}{2}},
\end{equation}
if we identify a $n$-chord with a $T-T$ contraction and a $z$-chord with a $\tilde T-\tilde T$ contraction.\footnote{$T$ and $\tilde T$ cannot contract, because they would give an exponentially small contribution that scales like $\langle \cos (F-\tilde F)/4 \rangle^N$.} If we set $\tilde F =0$,  we would get a chaos-integrable type of Hamiltonian because 
\begin{equation}
   \left. -\sum_{i=1}^N (\tilde T_i^+ +     \tilde T_i^-)\right|_{\tilde F= 0 } =  -\sum_{i=1}^N \sigma^1_i.
\end{equation}

\subsection{Transfer matrix for a single chord}

\begin{figure}
    \centering
    \includegraphics[width=0.5\textwidth,page=1]{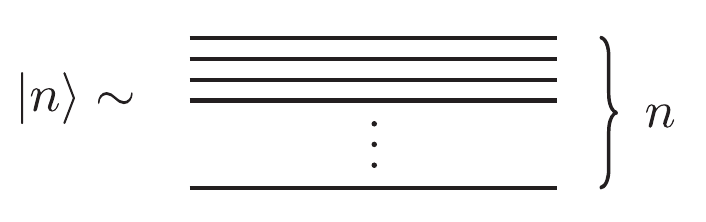}
    \caption{The vector $\ket{n}$ represents $n$ stacked chords.}
    \label{fig:StackedChords}
    \includegraphics[width=\textwidth,page=2]{Figures/TMatrixIntro.pdf}
    \caption{Acting with $T$}
    \label{fig:TMatrixExample}
\end{figure}

Let us use some linear algebra  to compute the weighted sum over all chord diagrams \eqref{eq:ChordPF}. Consider cutting the circle open at some point and going sequentially along the line. We define the Hilbert space $\mathcal{H}_{\text{aux}}$, which is spanned by a set of basis vectors $\{\ket{n}\}_{n=0}^{\infty}$. We take $\mathcal{H}_{\text{aux}}$ to be the Hilbert space of stacked chords, see Figure~\ref{fig:StackedChords}. We can think of $\left|n\right>$ as a state representing $n$ open chords, and a vector in this Hilbert space will be denoted by $\sum_{n\ge 0} v_n|n\rangle$. 

Let us define a \textit{transfer matrix} $T:\mathcal{H}_{\text{aux}}\to\mathcal{H}_{\text{aux}}$  on $\mathcal{H}_{\text{aux}}$. We think of $T$ as acting on a state $\ket{n}$ by opening a new chord or closing an existing one, see Figure~\ref{fig:TMatrixExample}. We can reproduce the sum \eqref{eq:ChordPF} if we decide that
\begin{enumerate}
    \item $T$ always opens a new chord below all existing chords. This means that chords cannot intersect when they open, i.e., $T\vert_{\text{raising}}\ket{n}=\ket{n+1}$
    \item Whenever a chord closes and intersects another chord, it does so with a factor of $q$. Hence $T\vert_{\text{lowering}}\ket{n}=\left(q^{n-1}+q^{n-2}+\cdots+1\right)\ket{n-1}$.
\end{enumerate}
This means that as we go over a node, the coefficients $v_n$ change by
\begin{align}
\begin{split}
    v_n(i+1) &=  v_{n-1}(i)+1\cdot v_{n+1}(i) +q\cdot  v_{n+1}(i) +\cdots+q^n\cdot v_{n+1}(i) \\
    & \qquad \qquad =v_{n-1}(i)+\frac{1-q^{n+1}}{1-q}v_{n+1}(i).
\end{split}
\end{align}
In this basis the matrix $T$ is given by 
\begin{align} \label{eq:TransferMatrix}
    T = \begin{pmatrix}
    0 & \frac{1-q}{1-q} & 0 & 0 & \cdots \\
    1 & 0 & \frac{1-q^2}{1-q} & 0 & \cdots \\
    0 & 1 & 0 & \frac{1-q^3}{1-q} & \cdots \\
    0 & 0 & 1 & 0 & \cdots \\
    \vdots & \vdots & \vdots & \vdots & \ddots
    \end{pmatrix}
\end{align}
Combining all of the above we see that in order to reproduce the sum appearing in (\ref{eq:ChordPF}) of all chord diagrams of length $k$, we need to consider the element 
\begin{align} \label{eq:TranMetMoment}
m_k = \braket{0|T^k|0}.
\end{align}

The task of finding the moment $m_k$ reduces to diagonalizing the matrix $T$ and taking its $k$th power. This is done in \cite{Berkooz:2018qkz}, and we will not repeat the derivation here, but merely cite the results. We have
\begin{align}
    m_k = \int_0^\pi d\theta \frac{(q;q)_{\infty}|(e^{2i\theta};q)_{\infty}|^2}{2\pi}\cdot\left(\frac{2\cos\theta}{\sqrt{1-q}}\right)^k, 
\end{align}
where $(a;q)_n$ is the $q$-Pochammer symbol, defined by
\begin{align}
    (a;q)_n\equiv\prod_{k=0}^{n-1}(1-aq^k),
\end{align}
and when $n=\infty$ we extend the product to an infinite product.
By resumming the $m_k$ into the thermal partition function, we get
\begin{align}
    \Tr[e^{-\beta H}]=\int_0^\pi \frac{d\theta}{2\pi}(q,e^{\pm 2i\theta};q)_{\infty} \exp\left[-\beta \frac{2\cos\theta}{\sqrt{1-q}}\right],
\end{align}
where $(a_1,a_2,\dots,a_k;q)_n\equiv \prod_{i=1}^k (a_k;q)_n$, and $(e^{\pm i\theta};q)\equiv (e^{+ i\theta};q)(e^{- i\theta};q)$.
We refer the reader to \cite{Berkooz:2018jqr} for the computation of two- and four-point functions.

\section{Special functions} 
\label{app:SpecialFunctions}
We define here some of the special functions used in the main text, and mention some useful facts about them.

\paragraph{The $q$-Pochhammer symbol}
Throughout we will use the $q$-Pochhammer symbol 
\begin{equation}
    \left(z;q\right)_{n}=\prod_{k=0}^{n-1}\left(1-zq^{k}\right) \,,
\end{equation}
and specifically its infinite version, $(z;q)_\infty$. The infinite Pochhammer symbol has both a power series expansion and a plethystic expansion,
\begin{equation}
\label{eq:Pochhammer plethystic}
\left(z;q\right)_{\infty} = \sum_{n=0}^{\infty}\frac{\left(-1\right)^{n}q^{\frac{n\left(n-1\right)}{2}}}{\left(q;q\right)_{n}}z^{n} = \exp\left[-\sum_{k=1}^{\infty}\frac{1}{k}\frac{z^{k}}{1-q^{k}}\right] \,.
\end{equation}
The plethystic expansion can be used to find a useful expansion when $q \to 1$, or rather $\lambda \to 0$ where $q = e^{-\lambda}$:
\begin{equation}
\begin{aligned}
\label{eq:pochhammer expansion in lambda}
    \log\left(z;q\right)_{\infty} &= -\sum_{k=1}^{\infty}\frac{1}{k}\frac{z^{k}}{1-q^{k}} = -\sum_{k=1}^{\infty}\frac{z^{k}}{k}\left[\frac{1}{k\lambda}-\sum_{\ell=0}^{\infty}\frac{\zeta\left(-\ell\right)}{\ell!}\left(\lambda k\right)^{\ell}\right] \\ 
    &=-\sum_{k=1}^{\infty}\frac{1}{\lambda}\frac{z^{k}}{k^{2}}+\sum_{\ell=0}^{\infty}\frac{\zeta\left(-\ell\right)\lambda^{\ell}}{\ell!}\sum_{k=1}^{\infty}\frac{z^{k}}{k^{1-\ell}} = -\frac{1}{\lambda}\Li_{2}\left(z\right)+\sum_{\ell=0}^{\infty}\lambda^{\ell}\frac{\zeta\left(-\ell\right)}{\ell!}\Li_{1-\ell}\left(z\right) \\
    &= -\frac{1}{\lambda}\Li_{2}\left(z\right)+\frac{1}{2}\log\left(1-z\right)-\frac{\lambda}{12}\frac{z}{1-z}+\frac{\lambda}{720}\frac{z\left(1+z\right)}{\left(1-z\right)^{3}}+\cdots \,,
\end{aligned}
\end{equation}
whose first few terms are reproduced in \cite{Zagier2007}. In the second equality we used the expansion $\frac{1}{1-e^{-x}} = \frac{1}{x} - \sum_{k=0}^{\infty}\frac{\zeta\left(-k\right)}{k!}x^{k}$.
The dilogarithm and the more general polylogarithm also appear in \eqref{eq:pochhammer expansion in lambda}, and are defined via
\begin{equation}
    \Li_2(z) = \sum_{k=1}^\infty \frac{z^k}{k^2} \,,\qquad \Li_s(z) = \sum_{k=1}^\infty \frac{z^k}{k^s} \,.
\end{equation}
The dilogarithm of an exponent 
%and its derivatives have 
has the convenient expansion \cite{Dilogarithm_expansions_wolfram}
\begin{equation}
\begin{aligned}
\label{eq:Dilogarithm expansions}
    \Li_{2}\left(e^{-x}\right) &= \frac{\pi^{2}}{6}+x\left(\log x-1\right)+\sum_{k=2}^{\infty}\frac{\zeta\left(2-k\right)}{k!}\left(-x\right)^{k} \,.
    % \log\left(1-e^{-x}\right) &= \frac{d}{dx}\Li_{2}\left(e^{-x}\right) = \log x-\sum_{k=1}^{\infty}\frac{\zeta\left(1-k\right)}{k!}\left(-x\right)^{k} \,, \\
    % \frac{e^{-x}}{1-e^{-x}} &= \frac{d^{2}}{dx^{2}}\Li_{2}\left(e^{-x}\right) = \frac{1}{x}+\sum_{k=0}^{\infty}\frac{\zeta\left(-k\right)}{k!}\left(-x\right)^{k} \,, \\
    % \Li_{1-\ell}\left(e^{-x}\right) &= \frac{d^{\ell+1}}{dx^{\ell+1}}\Li_{2}\left(e^{-x}\right) = \frac{\left(-1\right)^{\ell+1}\left(\ell-1\right)!}{x^{\ell}} - \left(-1\right)^{\ell}\sum_{k=0}^{\infty}\frac{\zeta\left(2-\ell-k\right)}{k!}\left(-x\right)^{k} \,, \qquad \ell \ge 1 \,.
\end{aligned}
\end{equation}
One can also find\footnote{A simple way of doing so is to relate it to the Dedekind eta function and use its modular properties.} an expansion for $(q;q)_\infty$,
% A simple way of doing so is using 
% \begin{equation}
%     \left(q;q\right)_{\infty} = \prod_{k=1}^{\infty}\left(1-q^{k}\right) = e^{-\frac{\lambda}{24}}\eta\left(\frac{i\lambda}{2\pi}\right) \,,
% \end{equation}
% where $\eta(\tau)$ is the Dedekind eta function, which has the modular property
% \begin{equation}
%     \eta\left(\tau\right) = \frac{\eta\left(-\frac{1}{\tau}\right)}{\sqrt{-i\tau}} \,,
% \end{equation}
% and so
% \begin{equation}
%     \left(q;q\right)_{\infty} = \sqrt{\frac{2\pi}{\lambda}}e^{-\frac{\lambda}{24}}\eta\left(\frac{2\pi i}{\lambda}\right)=\sqrt{\frac{2\pi}{\lambda}}e^{-\frac{\lambda}{24}-\frac{1}{\lambda}\frac{\pi^{2}}{6}}\left[1+\sum_{n=1}^{\infty}\left(-1\right)^{n}\left[e^{-\frac{2\pi^{2}}{\lambda}n\left(3n-1\right)}+e^{-\frac{2\pi^{2}}{\lambda}n\left(3n+1\right)}\right]\right] \,,
% \end{equation}
% or rather
\begin{equation}
    (q;q)_\infty = \sqrt{\frac{2\pi}{\lambda}}\exp\left[-\frac{1}{\lambda}\frac{\pi^{2}}{6} +\frac{\lambda}{24}\right] + O\left(e^{-4\pi^2/\lambda}\right) \,.
\end{equation}
To summarize, the two expansions which we will use heavily throughout the paper are
\begin{equation}
\begin{aligned}
\label{eq:q-Pochhammer approx}
    (z;q)_\infty &= \exp\left[-\frac{1}{\lambda} \Li_2(z) + \frac{1}{2}\log(1-z) - \frac{\lambda}{12}\frac{z}{1-z} + O(\lambda^2) \right] \,, \\ (q;q)_\infty &= \sqrt{\frac{2\pi}{\lambda}} \exp\left[-\frac{1}{\lambda}\frac{\pi^2}{6} + \frac{\lambda}{24}\right]\left(1 + O\left(e^{-4\pi^2/\lambda}\right)\right) \,.
\end{aligned}
\end{equation}

\paragraph{The $q$-factorial}
The $q$-Pochhammer symbol can also be used to define the $q$-factorial and $q$-Gamma function, 
\begin{equation}
\label{eq:q factorial def}
    \Gamma_{q}\left(n+1\right) \equiv \left[n\right]_{q}! \equiv \prod_{k=1}^n \left(1 + \cdots + q^{k-1}\right) = \frac{\left(q;q\right)_{\infty}}{\left(q^{n};q\right)_{\infty}}\frac{1-q^{n}}{\left(1-q\right)^{n}}=\frac{(q,q)_n}{(1-q)^n}\,.
\end{equation}
The $q$-factorial counts the number of permutations of a sequence of $n$ elements, giving weight $q$ to each inversion of two elements compared to the initial configuration, $\sum_{w\in S_n}q^{\text{No. of inv.}}=[n]_q!$. In our context, this is exactly the weight of all the possible configurations where $n$ chords stretch between two segments. One can also define the $q$-multinomial coefficients
\begin{equation}
\label{eq:q multinomial def}
    \binom{n}{a_1, \cdots, a_k}_q = \frac{[n]_q!}{[a_1]_q!\cdots[a_k]_q!} \,, \qquad \sum_{i=1}^k a_i = n \,,
\end{equation}
which also has a combinatorial meaning---suppose we have $n$ chords stretching between two segments, divided into $k$ sets of sizes $a_i$. The $q$-multinomial amounts to the weight of all diagrams where we allow the sets to intersect each other, without accounting for permutations within each set.

The $q\to1^-$ limit of these quantities are their usual combinatorial counterparts, and one can systematically analyze the higher order corrections. For example, for the $q$-Gamma function, following Gospar's proof in \cite{Andrews1986qseriesT}, 
\begin{multline}
\label{eq:semiclassical_limit_q_Gamma}
    \Gamma_{q}\left(x+1\right) = \left(\prod_{n=0}^{\infty}\frac{1-q^{n+1}}{1-q^{n+x+1}}\right)\left(1-q\right)^{-x} = \left(\prod_{n=1}^{\infty}\frac{1-q^{n}}{1-q^{n+x}}\right)\left(1-q\right)^{-x} = \prod_{n=1}^{\infty}\frac{\left(1-q^{n}\right)\left(1-q^{n+1}\right)^{x}}{\left(1-q^{n+x}\right)\left(1-q^{n}\right)^{x}} \\
    = \left(1 + \frac{x(1-x)\lambda^2}{24}  + O(\lambda^3)\right)\prod_{n=1}^{\infty}\frac{n}{n+x}\left(\frac{n+1}{n}\right)^{x} 
    % \\ = \left(1 + \frac{x(1-x)\lambda^2}{24}  + O(\lambda^3)\right)\prod_{n=1}^{\infty}\left(\frac{n}{n+x}\right)\left(1+\frac{1}{n}\right)^{x}
    = \left(1 + \frac{x(1-x)\lambda^2}{24} + O(\lambda^3)\right)\Gamma\left(x+1\right) \,.
\end{multline}

\paragraph{The $q$-exponential}
We also use here the $q$-exponential function,
\begin{equation}
    e_q(x) \equiv e_q^x \equiv \sum_{k=0}^\infty \frac{x^k}{[k]_q!} \,,
\end{equation}
which admits the Plethystic expansion
\begin{equation}
\label{eq:q exp plethystic}
    \log e_q^x = \sum_{n=1} \frac{1}{n}\frac{x^n (1-q)^n}{(1-q^n)} \,.
\end{equation}

\section{Matrix elements at high temperatures}
\label{app:matrix_element}
In Section~\ref{sec:path integral single chord} we explained our coarse graining procedure for double scaled SYK. The first step in this procedure required accounting for the weights of all sub-diagrams where $n$ chords leave a segment of length $\beta_i$, without accounting for the intersections of the outgoing chords. This weight is exactly the matrix element $\langle n|e^{-\beta_i T} | 0 \rangle$. In this appendix we give a different derivation for this weight, \eqref{eq:0-n amplitude high temp}, which can also be used for computing higher orders in $\lambda$ if needed.

The explicit form for the matrix element is given in \cite{Berkooz:2018jqr}, in our $\cJ = 1$ normalization \eqref{eq:cJ norm},
\begin{equation} \label{eq:Berkooz_zero_to_n_evol}
    \langle n|e^{-\beta_i T}|0\rangle = \left(1-q\right)^{n/2}\int_{0}^{\pi}\frac{d\theta}{2\pi}\left(q,e^{\pm2i\theta};q\right)_{\infty}e^{-\frac{2\beta_i\cos\theta}{\sqrt{1-q}}}\frac{H_{n}\left(\cos\theta|q\right)}{\left(q;q\right)_{n}} \,.
\end{equation}
where $(a;q)_n$ denotes the $q$-Pochhammer symbol, we use the shorthand $(a,b;q)_\infty \equiv (a;q)_\infty (b;q)_\infty$, and $H_{n}\left(x|q\right)$ are the $q$-Hermite polynomials. Their generating function,
$\sum_{n=0}^{\infty}\frac{H_{n}\left(x|q\right)}{\left(q;q\right)_{n}}t^{n}=\frac{1}{\left(te^{\pm i\theta};q\right)_{\infty}}$, allows us to rewrite\footnote{We want to pick the residue at zero, and that $\left(te^{\pm i\theta};q\right)_{\infty}$ has zeros when $t=e^{\mp i \theta}q^{-\ell}$ for any integer $\ell\ge0$, so the contour has $\left|t\right|<1$.} 
 \begin{equation}
    \frac{\left\langle n|e^{-\beta_i T}|0\right\rangle}{\left(1-q\right)^{n/2}} = \frac{1}{2\pi i}\int_{0}^{\pi}\frac{d\theta}{2\pi}\oint dt\frac{\left(q;q\right)_{\infty}\left(e^{\pm2i\theta};q\right)_{\infty}}{\left(te^{\pm i\theta};q\right)_{\infty}}e^{-\frac{2\beta_i\cos\theta}{\sqrt{1-q}}}t^{-n-1} \,.
\end{equation}
Let us now take the limit $q\to 1$ while keeping $\tilde\beta_i \equiv \sqrt\lambda \beta_i$, $\tilde n \equiv \lambda n$ fixed. We use the approximations \eqref{eq:q-Pochhammer approx} to find
\begin{equation}
    \begin{aligned}
        \frac{\left\langle n|e^{-\beta_i T}|0\right\rangle }{\left(1-q\right)^{n/2}} &= \frac{\left(q;q\right)_{\infty}e^{\frac{\pi^{2}}{6\lambda}}}{2\pi i}\int_{0}^{\pi}\frac{d\theta}{2\pi}\oint\frac{dt}{t}e^{-\frac{1}{\lambda} S} \\
        S &= 2\left(\frac{\pi}{2}-\theta\right)^{2}-\Li_{2}\left(te^{i\theta}\right)-\Li_{2}\left(te^{-i\theta}\right) + 2\tilde\beta_i\cos\theta + \tilde n\log t \,.
    \end{aligned}
\end{equation}
We will now solve the integral by a saddle point approximation, which requires solving the equations\footnote{We've used $\frac{d}{dz} \Li_2(z) = -\frac{\log(1-z)}{z} = \sum_{k=0}^\infty \frac{z^k}{k+1}$.}
\begin{equation}
    \begin{aligned}
        \tilde\beta_i \sin\theta &= 2\theta-\pi - \frac{i}{2}\log\left(1-te^{-i\theta}\right) + \frac{i}{2}\log\left(1-te^{i\theta}\right) \;,\\ 
        \tilde n &= -\log\left[\left(1-te^{i\theta}\right)\left(1-te^{-i\theta}\right)\right] \,.
    \end{aligned}
\end{equation}
Unfortunately, for general $\tilde\beta_i$, we do not know how to solve them. Luckily, in Section~\ref{sec:path integral single chord} we are interested in the case $\tilde \beta_i \ll 1$, where each segment is small. Since the segments are small, the number of chords emanating from them can also be assumed to be small, so we will assume $\tilde n \ll 1$. From the saddle point equations, we see that this is consistent with the assumption $|t| \ll 1$. We introduce a new variable, $v$, such that
\begin{equation}
    \theta = \frac{\pi}{2} + \frac{\pi v}{2} \,,
\end{equation}
and assume that $v \ll 1$ as well, which is also consistent with the full saddle point equations. In these limits the action simplifies to
\begin{equation}
    S = \frac{1}{2}\left(\pi v\right)^{2} + \left(t-\tilde{\beta}_i\right)\pi v + \frac{1}{2}t^{2} + \tilde{n}\log t\;,
\end{equation}
while the saddle point equations are
\begin{equation}
\pi v =\tilde{\beta}_i - t \,,\qquad
0 = t^{2} + \pi vt + \tilde{n} \,.
\end{equation}
They admit the solution
\begin{equation}
    \pi v 
 = \frac{\tilde{\beta}_i^{2}+\tilde{n}}{\tilde{\beta}_i} \,, \quad
t = -\frac{\tilde{n}}{\tilde{\beta}} \;,
\end{equation}
so the action at the saddle point is
\begin{equation}
S = -\frac{\tilde{\beta}_i^{2}}{2}+\tilde{n}\left(\log\frac{\tilde{n}}{\tilde{\beta}_i}-1\right) + \pi i\tilde{n} \;.
\end{equation}
Note that the last term is a pure phase, whose contribution to the partition function is equivalent to $(-1)^{n_i}$. After the saddle point approximation, the amplitude gives
\begin{equation}
\label{eq:0-n amplitude high temp saddle}
\left\langle n|e^{-\beta_i T}|0\right\rangle = (-1)^n \lambda^{n}e^{-\frac{1}{\lambda}\left[-\frac{\tilde{\beta}_i^{2}}{2}+\tilde{n}(\log\frac{\tilde{n}}{\tilde{\beta}_i} - 1) + O\left(\lambda\right)\right]} \;, \qquad \text{for }\tilde\beta \equiv \sqrt{\lambda}{\beta} \ll 1,\; \tilde n \equiv \lambda n \ll 1\,,
\end{equation}
This agrees with \eqref{eq:0-n amplitude high temp}, but allows for a systematic computation of higher orders in $\lambda$.

\printbibliography

@article{Lin:2022rbf,
    author = "Lin, Henry W.",
    title = "{The bulk Hilbert space of double scaled SYK}",
    eprint = "2208.07032",
    archivePrefix = "arXiv",
    primaryClass = "hep-th",
    doi = "10.1007/JHEP11(2022)060",
    journal = "JHEP",
    volume = "11",
    pages = "060",
    year = "2022"
}

@article{Berkooz:2018jqr,
    author = "Berkooz, Micha and Isachenkov, Mikhail and Narovlansky, Vladimir and Torrents, Genis",
    title = "{Towards a full solution of the large N double-scaled SYK model}",
    eprint = "1811.02584",
    archivePrefix = "arXiv",
    primaryClass = "hep-th",
    doi = "10.1007/JHEP03(2019)079",
    journal = "JHEP",
    volume = "03",
    pages = "079",
    year = "2019"
}

@article{Berkooz:2018qkz,
    author = "Berkooz, Micha and Narayan, Prithvi and Simon, Joan",
    title = "{Chord diagrams, exact correlators in spin glasses and black hole bulk reconstruction}",
    eprint = "1806.04380",
    archivePrefix = "arXiv",
    primaryClass = "hep-th",
    doi = "10.1007/JHEP08(2018)192",
    journal = "JHEP",
    volume = "08",
    pages = "192",
    year = "2018"
}

@article{Goel:2023svz,
    author = "Goel, Akash and Narovlansky, Vladimir and Verlinde, Herman",
    title = "{Semiclassical geometry in double-scaled SYK}",
    eprint = "2301.05732",
    archivePrefix = "arXiv",
    primaryClass = "hep-th",
    month = "1",
    year = "2023"
}

@article{Maldacena:2016hyu,
    author = "Maldacena, Juan and Stanford, Douglas",
    title = "{Remarks on the Sachdev-Ye-Kitaev model}",
    eprint = "1604.07818",
    archivePrefix = "arXiv",
    primaryClass = "hep-th",
    doi = "10.1103/PhysRevD.94.106002",
    journal = "Phys. Rev. D",
    volume = "94",
    number = "10",
    pages = "106002",
    year = "2016"
}

@inbook{Zagier2007,
	address = {Berlin, Heidelberg},
	author = {Zagier, Don},
	booktitle = {Frontiers in Number Theory, Physics, and Geometry II: On Conformal Field Theories, Discrete Groups and Renormalization},
	doi = {10.1007/978-3-540-30308-4_1},
	editor = {Cartier, Pierre and Moussa, Pierre and Julia, Bernard and Vanhove, Pierre},
	isbn = {978-3-540-30308-4},
	pages = {3--65},
	publisher = {Springer Berlin Heidelberg},
	title = {The Dilogarithm Function},
	url = {https://doi.org/10.1007/978-3-540-30308-4_1},
	year = {2007}}

@article{Mukhametzhanov:2023tcg,
    author = "Mukhametzhanov, Baur",
    title = "{Large p SYK from chord diagrams}",
    eprint = "2303.03474",
    archivePrefix = "arXiv",
    primaryClass = "hep-th",
    month = "3",
    year = "2023"
}

@article{Czech:2016xec,
    author = "Czech, Bartlomiej and Lamprou, Lampros and McCandlish, Samuel and Mosk, Benjamin and Sully, James",
    title = "{A Stereoscopic Look into the Bulk}",
    eprint = "1604.03110",
    archivePrefix = "arXiv",
    primaryClass = "hep-th",
    reportNumber = "SU-ITP-16-07",
    doi = "10.1007/JHEP07(2016)129",
    journal = "JHEP",
    volume = "07",
    pages = "129",
    year = "2016"
}

@article{erdHos2014phase,
    author = {Erd\H{o}s, L\'aszl\'o and Schr\"oder, Dominik},
    title = "{Phase Transition in the Density of States of Quantum Spin Glasses}",
    eprint = "1407.1552",
    archivePrefix = "arXiv",
    primaryClass = "math-ph",
    doi = "10.1007/s11040-014-9164-3",
    journal = "Math. Phys. Anal. Geom.",
    volume = "17",
    number = "3-4",
    pages = "441--464",
    year = "2014"
}

@article{derridaPRL,
  title = {Random-Energy Model: Limit of a Family of Disordered Models},
  author = {Derrida, B.},
  journal = {Phys. Rev. Lett.},
  volume = {45},
  issue = {2},
  pages = {79--82},
  numpages = {0},
  year = {1980},
  month = {Jul},
  publisher = {American Physical Society},
  doi = {10.1103/PhysRevLett.45.79},
  url = {https://link.aps.org/doi/10.1103/PhysRevLett.45.79}
}

@article{derridaPRB,
  title = {Random-energy model: An exactly solvable model of disordered systems},
  author = {Derrida, Bernard},
  journal = {Phys. Rev. B},
  volume = {24},
  issue = {5},
  pages = {2613--2626},
  numpages = {0},
  year = {1981},
  month = {Sep},
  publisher = {American Physical Society},
  doi = {10.1103/PhysRevB.24.2613},
  url = {https://link.aps.org/doi/10.1103/PhysRevB.24.2613}
}

@article{gross1984,
title = {The simplest spin glass},
journal = {Nuclear Physics B},
volume = {240},
number = {4},
pages = {431-452},
year = {1984},
issn = {0550-3213},
doi = {https://doi.org/10.1016/0550-3213(84)90237-2},
url = {https://www.sciencedirect.com/science/article/pii/0550321384902372},
author = {D.J. Gross and M. Mezard}
}

@article{Berkooz:2020xne,
    author = "Berkooz, Micha and Brukner, Nadav and Narovlansky, Vladimir and Raz, Amir",
    title = "{The double scaled limit of Super--Symmetric SYK models}",
    eprint = "2003.04405",
    archivePrefix = "arXiv",
    primaryClass = "hep-th",
    doi = "10.1007/JHEP12(2020)110",
    journal = "JHEP",
    volume = "12",
    pages = "110",
    year = "2020"
}

@article{pluma2022dynamical,
  title={A dynamical version of the SYK model and the q-Brownian motion},
  author={Pluma, Miguel and Speicher, Roland},
  journal={Random Matrices: Theory and Applications},
  volume={11},
  number={03},
  pages={2250031},
  year={2022},
  publisher={World Scientific}
}

@online{Dilogarithm_expansions_wolfram,
  author = {Wolfram Research Inc.},
  title = {Dilogarithm expansions, Wolfram},
  year = 1999,
  url = {https://functions.wolfram.com/ZetaFunctionsandPolylogarithms/PolyLog2/06/01/02/01/},
  urldate = {2023-06-12}
}

@article{katriel1996q, 
    title={The q-Zassenhaus formula}, 
    author={Katriel, Jacob and Rasetti, Mario and Solomon, Allan I}, 
    journal={Letters in Mathematical Physics}, 
    volume={37}, 
    pages={11--13}, 
    year={1996}, 
    publisher={Springer},
    doi={https://doi.org/10.1007/BF00400134}
}

@article{Anninos:2022qgy,
    author = "Anninos, Dionysios and Galante, Dami\'an A. and Sheorey, Sameer U.",
    title = "{Renormalisation Group Flows of the SYK Model}",
    eprint = "2212.04944",
    archivePrefix = "arXiv",
    primaryClass = "hep-th",
    month = "12",
    year = "2022"
}

@article{Cotler:2016fpe,
    author = "Cotler, Jordan S. and Gur-Ari, Guy and Hanada, Masanori and Polchinski, Joseph and Saad, Phil and Shenker, Stephen H. and Stanford, Douglas and Streicher, Alexandre and Tezuka, Masaki",
    title = "{Black Holes and Random Matrices}",
    eprint = "1611.04650",
    archivePrefix = "arXiv",
    primaryClass = "hep-th",
    reportNumber = "SU-ITP-16-19, SU-ITP-16/19, YITP-16-124",
    doi = "10.1007/JHEP05(2017)118",
    journal = "JHEP",
    volume = "05",
    pages = "118",
    year = "2017",
    note = "[Erratum: JHEP 09, 002 (2018)]"
}

@article{Gao:2023gta,
    author = "Gao, Ping",
    title = "{Commuting SYK: a pseudo-holographic model}",
    eprint = "2306.14988",
    archivePrefix = "arXiv",
    primaryClass = "hep-th",
    doi = "10.1007/JHEP01(2024)149",
    journal = "JHEP",
    volume = "01",
    pages = "149",
    year = "2024"
}

@article{gardener1985,
title = {Spin glasses with p-spin interactions},
journal = {Nuclear Physics B},
volume = {257},
pages = {747-765},
year = {1985},
issn = {0550-3213},
doi = {https://doi.org/10.1016/0550-3213(85)90374-8},
url = {https://www.sciencedirect.com/science/article/pii/0550321385903748},
author = {E. Gardner}
}

@article{Lin:2023trc,
    author = "Lin, Henry W. and Stanford, Douglas",
    title = "{A symmetry algebra in double-scaled SYK}",
    eprint = "2307.15725",
    archivePrefix = "arXiv",
    primaryClass = "hep-th",
    month = "7",
    year = "2023"
}

@inproceedings{Andrews1986qseriesT,
  title={q-series : their development and application in analysis, number theory, combinatorics, physics, and computer algebra},
  author={George E. Andrews},
  year={1986},
  url={https://api.semanticscholar.org/CorpusID:117872938}
}

@article{banerjee2017solvable,
  title={Solvable model for a dynamical quantum phase transition from fast to slow scrambling},
  author={Banerjee, Sumilan and Altman, Ehud},
  journal={Physical Review B},
  volume={95},
  number={13},
  pages={134302},
  year={2017},
  publisher={APS}
}

@article{Jian:2017unn,
    author = "Jian, Shao-Kai and Yao, Hong",
    title = "{Solvable Sachdev-Ye-Kitaev models in higher dimensions: from diffusion to many-body localization}",
    eprint = "1703.02051",
    archivePrefix = "arXiv",
    primaryClass = "cond-mat.str-el",
    doi = "10.1103/PhysRevLett.119.206602",
    journal = "Phys. Rev. Lett.",
    volume = "119",
    number = "20",
    pages = "206602",
    year = "2017"
}

@article{Garcia-Garcia:2017bkg,
    author = "Garc\'ia-Garc\'ia, Antonio M. and Loureiro, Bruno and Romero-Berm\'udez, Aurelio and Tezuka, Masaki",
    title = "{Chaotic-Integrable Transition in the Sachdev-Ye-Kitaev Model}",
    eprint = "1707.02197",
    archivePrefix = "arXiv",
    primaryClass = "hep-th",
    doi = "10.1103/PhysRevLett.120.241603",
    journal = "Phys. Rev. Lett.",
    volume = "120",
    number = "24",
    pages = "241603",
    year = "2018"
}

@article{Baldwin_2020,
   title={Quenched vs Annealed: Glassiness from SK to SYK},
   volume={10},
   ISSN={2160-3308},
   url={http://dx.doi.org/10.1103/PhysRevX.10.031026},
   DOI={10.1103/physrevx.10.031026},
   number={3},
   journal={Physical Review X},
   publisher={American Physical Society (APS)},
   author={Baldwin, C. L. and Swingle, B.},
   year={2020},
   month=aug }

@article{Swingle:2023nvv,
    author = "Swingle, Brian and Winer, Michael",
    title = "{A Bosonic Model of Quantum Holography}",
    eprint = "2311.01516",
    archivePrefix = "arXiv",
    primaryClass = "hep-th",
    month = "11",
    year = "2023"
}

@article{Anninos:2020cwo,
    author = "Anninos, Dionysios and Galante, Dami\'an A.",
    title = "{Constructing AdS$_{2}$ flow geometries}",
    eprint = "2011.01944",
    archivePrefix = "arXiv",
    primaryClass = "hep-th",
    doi = "10.1007/JHEP02(2021)045",
    journal = "JHEP",
    volume = "02",
    pages = "045",
    year = "2021"
}

@article{Okuyama:2023bch,
    author = "Okuyama, Kazumi and Suzuki, Kenta",
    title = "{Correlators of double scaled SYK at one-loop}",
    eprint = "2303.07552",
    archivePrefix = "arXiv",
    primaryClass = "hep-th",
    reportNumber = "RUP-23-5",
    doi = "10.1007/JHEP05(2023)117",
    journal = "JHEP",
    volume = "05",
    pages = "117",
    year = "2023"
}

@article{jian2017model,
  title={Model for continuous thermal metal to insulator transition},
  author={Jian, Chao-Ming and Bi, Zhen and Xu, Cenke},
  journal={Physical Review B},
  volume={96},
  number={11},
  pages={115122},
  year={2017},
  publisher={APS}
}

@misc{single-chord-future-paper,
    author = "Berkooz, Micha and Brukner, Nadav and Jia, Yiyang and Mamroud, Ohad",
    howpublished = "work in progress",
}

@article{Berkooz:2023scv,
    author = "Berkooz, Micha and Jia, Yiyang and Silberstein, Navot",
    title = "{Parisi's hypercube, Fock-space fluxes, and the microscopics of near-AdS$_2$/near-CFT$_1$ duality}",
    eprint = "2310.12335",
    archivePrefix = "arXiv",
    primaryClass = "hep-th",
    month = "10",
    year = "2023"
}

@article{Berkooz:2023cqc,
    author = "Berkooz, Micha and Jia, Yiyang and Silberstein, Navot",
    title = "{Parisi's hypercube, Fock-space frustration and near-AdS$_2$/near-CFT$_1$ holography}",
    eprint = "2303.18182",
    archivePrefix = "arXiv",
    primaryClass = "hep-th",
    month = "3",
    year = "2023"
}

@article{Jia:2020rfn,
    author = "Jia, Yiyang and Verbaarschot, Jacobus J. M.",
    title = "{Chaos on the hypercube}",
    eprint = "2005.13017",
    archivePrefix = "arXiv",
    primaryClass = "hep-th",
    doi = "10.1007/JHEP11(2020)154",
    journal = "JHEP",
    volume = "11",
    pages = "154",
    year = "2020"
}

@article{Saad:2018bqo,
    author = "Saad, Phil and Shenker, Stephen H. and Stanford, Douglas",
    title = "{A semiclassical ramp in SYK and in gravity}",
    eprint = "1806.06840",
    archivePrefix = "arXiv",
    primaryClass = "hep-th",
    month = "6",
    year = "2018"
}

@article{Polchinski:2016xgd,
    author = "Polchinski, Joseph and Rosenhaus, Vladimir",
    title = "{The Spectrum in the Sachdev-Ye-Kitaev Model}",
    eprint = "1601.06768",
    archivePrefix = "arXiv",
    primaryClass = "hep-th",
    doi = "10.1007/JHEP04(2016)001",
    journal = "JHEP",
    volume = "04",
    pages = "001",
    year = "2016"
}

@unpublished{kitaev2015simple,
	Author = {Alexander Kitaev},
	Note = {KITP strings seminar and Entanglement 2015 program, 12 February, 7 April and 27 May 2015, http://online.kitp.ucsb.edu/online/entangled15/},
	Title = {A simple model of quantum holography}}

@article{Maldacena:2015waa,
    author = "Maldacena, Juan and Shenker, Stephen H. and Stanford, Douglas",
    title = "{A bound on chaos}",
    eprint = "1503.01409",
    archivePrefix = "arXiv",
    primaryClass = "hep-th",
    doi = "10.1007/JHEP08(2016)106",
    journal = "JHEP",
    volume = "08",
    pages = "106",
    year = "2016"
}

@article{Maldacena:2016upp,
    author = "Maldacena, Juan and Stanford, Douglas and Yang, Zhenbin",
    title = "{Conformal symmetry and its breaking in two dimensional Nearly Anti-de-Sitter space}",
    eprint = "1606.01857",
    archivePrefix = "arXiv",
    primaryClass = "hep-th",
    doi = "10.1093/ptep/ptw124",
    journal = "PTEP",
    volume = "2016",
    number = "12",
    pages = "12C104",
    year = "2016"
}

@article{Berkooz:2022mfk,
    author = "Berkooz, Micha and Isachenkov, Misha and Isachenkov, Mikhail and Narayan, Prithvi and Narovlansky, Vladimir",
    title = "{Quantum groups, non-commutative AdS$_{2}$, and chords in the double-scaled SYK model}",
    eprint = "2212.13668",
    archivePrefix = "arXiv",
    primaryClass = "hep-th",
    doi = "10.1007/JHEP08(2023)076",
    journal = "JHEP",
    volume = "08",
    pages = "076",
    year = "2023"
}

@article{Blommaert:2023wad,
    author = "Blommaert, Andreas and Mertens, Thomas G. and Yao, Shunyu",
    title = "{The q-Schwarzian and Liouville gravity}",
    eprint = "2312.00871",
    archivePrefix = "arXiv",
    primaryClass = "hep-th",
    month = "12",
    year = "2023"
}

@article{Blommaert:2023opb,
    author = "Blommaert, Andreas and Mertens, Thomas G. and Yao, Shunyu",
    title = "{Dynamical actions and q-representation theory for double-scaled SYK}",
    eprint = "2306.00941",
    archivePrefix = "arXiv",
    primaryClass = "hep-th",
    month = "6",
    year = "2023"
}

@article{Speicher:1993kt,
    author = "Speicher, R.",
    title = "{Generalized statistics of macroscopic fields}",
    doi = "10.1007/BF00750677",
    journal = "Lett. Math. Phys.",
    volume = "27",
    pages = "97--104",
    year = "1993"
}

@article{Berkooz:2020uly,
    author = "Berkooz, Micha and Narovlansky, Vladimir and Raj, Himanshu",
    title = "{Complex Sachdev-Ye-Kitaev model in the double scaling limit}",
    eprint = "2006.13983",
    archivePrefix = "arXiv",
    primaryClass = "hep-th",
    doi = "10.1007/JHEP02(2021)113",
    journal = "JHEP",
    volume = "02",
    pages = "113",
    year = "2021"
}

@article{you2017sachdev,
    author = "You, Yi-Zhuang and Ludwig, Andreas W. W. and Xu, Cenke",
    title = "{Sachdev-Ye-Kitaev Model and Thermalization on the Boundary of Many-Body Localized Fermionic Symmetry Protected Topological States}",
    eprint = "1602.06964",
    archivePrefix = "arXiv",
    primaryClass = "cond-mat.str-el",
    doi = "10.1103/PhysRevB.95.115150",
    journal = "Phys. Rev. B",
    volume = "95",
    number = "11",
    pages = "115150",
    year = "2017"
}

@article{Sachdev:2015efa,
    author = "Sachdev, Subir",
    title = "{Bekenstein-Hawking Entropy and Strange Metals}",
    eprint = "1506.05111",
    archivePrefix = "arXiv",
    primaryClass = "hep-th",
    doi = "10.1103/PhysRevX.5.041025",
    journal = "Phys. Rev. X",
    volume = "5",
    number = "4",
    pages = "041025",
    year = "2015"
}

@article{Maldacena:2018lmt,
    author = "Maldacena, Juan and Qi, Xiao-Liang",
    title = "{Eternal traversable wormhole}",
    eprint = "1804.00491",
    archivePrefix = "arXiv",
    primaryClass = "hep-th",
    month = "4",
    year = "2018"
}

@article{Goel:2018ubv,
    author = "Goel, Akash and Lam, Ho Tat and Turiaci, Gustavo J. and Verlinde, Herman",
    title = "{Expanding the Black Hole Interior: Partially Entangled Thermal States in SYK}",
    eprint = "1807.03916",
    archivePrefix = "arXiv",
    primaryClass = "hep-th",
    doi = "10.1007/JHEP02(2019)156",
    journal = "JHEP",
    volume = "02",
    pages = "156",
    year = "2019"
}

@article{Jensen:2016pah,
    author = "Jensen, Kristan",
    title = "{Chaos in AdS$_2$ Holography}",
    eprint = "1605.06098",
    archivePrefix = "arXiv",
    primaryClass = "hep-th",
    doi = "10.1103/PhysRevLett.117.111601",
    journal = "Phys. Rev. Lett.",
    volume = "117",
    number = "11",
    pages = "111601",
    year = "2016"
}

@article{Streicher:2019wek,
    author = "Streicher, Alexandre",
    title = "{SYK Correlators for All Energies}",
    eprint = "1911.10171",
    archivePrefix = "arXiv",
    primaryClass = "hep-th",
    doi = "10.1007/JHEP02(2020)048",
    journal = "JHEP",
    volume = "02",
    pages = "048",
    year = "2020"
}

@article{Choi:2019bmd,
    author = "Choi, Changha and Mezei, M\'ark and S\'arosi, G\'abor",
    title = "{Exact four point function for large $q$ SYK from Regge theory}",
    eprint = "1912.00004",
    archivePrefix = "arXiv",
    primaryClass = "hep-th",
    reportNumber = "CERN-TH-2019-206",
    doi = "10.1007/JHEP05(2021)166",
    journal = "JHEP",
    volume = "05",
    pages = "166",
    year = "2021"
}

@article{garcia2016,
	Author = {Garc\'ia-Garc\'ia, Antonio M. and Verbaarschot, Jacobus J. M.},
	Doi = {10.1103/PhysRevD.94.126010},
	Issue = {12},
	Journal = {Phys. Rev. D},
	Month = {Dec},
	Numpages = {13},
	Pages = {126010},
	Publisher = {American Physical Society},
	Title = {Spectral and thermodynamic properties of the Sachdev-Ye-Kitaev model},
	Url = {https://link.aps.org/doi/10.1103/PhysRevD.94.126010},
	Volume = {94},
	Year = {2016}}

@article{garcia2017,
  title = {Analytical spectral density of the Sachdev-Ye-Kitaev model at finite $N$},
  author = {Garc\'ia-Garc\'ia, Antonio M. and Verbaarschot, Jacobus J. M.},
  journal = {Phys. Rev. D},
  volume = {96},
  issue = {6},
  pages = {066012},
  numpages = {10},
  year = {2017},
  month = {Sep},
  publisher = {American Physical Society},
  doi = {10.1103/PhysRevD.96.066012},
  url = {https://link.aps.org/doi/10.1103/PhysRevD.96.066012}
}

@article{sachdev1993,
	Author = {Sachdev, Subir and Ye, Jinwu},
	Doi = {10.1103/PhysRevLett.70.3339},
	Issue = {21},
	Journal = {Phys. Rev. Lett.},
	Month = {May},
	Numpages = {0},
	Pages = {3339--3342},
	Publisher = {American Physical Society},
	Title = {Gapless spin-fluid ground state in a random quantum Heisenberg magnet},
	Url = {http://link.aps.org/doi/10.1103/PhysRevLett.70.3339},
	Volume = {70},
	Year = {1993}}

@article{sachdev2010,
  title = {Holographic Metals and the Fractionalized Fermi Liquid},
  author = {Sachdev, Subir},
  journal = {Phys. Rev. Lett.},
  volume = {105},
  issue = {15},
  pages = {151602},
  numpages = {4},
  year = {2010},
  month = {Oct},
  publisher = {American Physical Society},
  doi = {10.1103/PhysRevLett.105.151602},
  url = {https://link.aps.org/doi/10.1103/PhysRevLett.105.151602}
}

@article{french1970,
	Author = {J.B. French and S.S.M. Wong},
	Doi = {http://dx.doi.org/10.1016/0370-2693(70)90213-3},
	Issn = {0370-2693},
	Journal = {Physics Letters B},
	Number = {7},
	Pages = {449 - 452},
	Title = {Validity of random matrix theories for many-particle systems},
	Url = {http://www.sciencedirect.com/science/article/pii/0370269370902133},
	Volume = {33},
	Year = {1970}}

@article{bohigas1971,
	Author = {O. Bohigas and J. Flores},
	Doi = {http://dx.doi.org/10.1016/0370-2693(71)90598-3},
	Issn = {0370-2693},
	Journal = {Physics Letters B},
	Number = {4},
	Pages = {261 - 263},
	Title = {Two-body random hamiltonian and level density},
	Url = {http://www.sciencedirect.com/science/article/pii/0370269371905983},
	Volume = {34},
	Year = {1971}}

@article{jiang2019,
   title={Thermodynamics and many body chaos for generalized large q SYK models},
   volume={2019},
   ISSN={1029-8479},
   url={http://dx.doi.org/10.1007/JHEP08(2019)019},
   DOI={10.1007/jhep08(2019)019},
   number={8},
   journal={Journal of High Energy Physics},
   publisher={Springer Science and Business Media LLC},
   author={Jiang, Jiaqi and Yang, Zhenbin},
   year={2019},
   month=aug }

@article{garcia2018c,
	title={Exact moments of the Sachdev-Ye-Kitaev model up to order $1/N^2$},
	author={Garc{\'i}a-Garc{\'i}a, Antonio M and Jia, Yiyang and Verbaarschot, Jacobus JM},
	journal={Journal of High Energy Physics},
	volume={2018},
	number={4},
	pages={146},
	year={2018},
	publisher={Springer},
doi={10.1007/JHEP04(2018)146}
}

@article{ismail1987combinatorics,
  title={The combinatorics of q-hermite polynomials and the askey—wilson integral},
  author={Ismail, Mourad EH and Stanton, Dennis and Viennot, G{\'e}rard},
  journal={European Journal of Combinatorics},
  volume={8},
  number={4},
  pages={379--392},
  year={1987},
  publisher={Elsevier}
}

@article{Jevicki:2016bwu,
    author = "Jevicki, Antal and Suzuki, Kenta and Yoon, Junggi",
    title = "{Bi-Local Holography in the SYK Model}",
    eprint = "1603.06246",
    archivePrefix = "arXiv",
    primaryClass = "hep-th",
    reportNumber = "BROWN-HET-1673",
    doi = "10.1007/JHEP07(2016)007",
    journal = "JHEP",
    volume = "07",
    pages = "007",
    year = "2016"
}

@online{Stanford-talk-kitp,
      title={Talk at KITP}, 
      author={Stanford, Douglas},
      year={2018},
      url={https://online.kitp.ucsb.edu/online/chord18/doublescale/rm/jwvideo.html}
}

@article{Jafferis:2022crx,
    author = "Jafferis, Daniel and Zlokapa, Alexander and Lykken, Joseph D. and Kolchmeyer, David K. and Davis, Samantha I. and Lauk, Nikolai and Neven, Hartmut and Spiropulu, Maria",
    title = "{Traversable wormhole dynamics on a quantum processor}",
    reportNumber = "FERMILAB-PUB-22-887-QIS",
    doi = "10.1038/s41586-022-05424-3",
    journal = "Nature",
    volume = "612",
    number = "7938",
    pages = "51--55",
    year = "2022"
}

@article{Pikulin:2017mhj,
    author = "Pikulin, D. I. and Franz, M.",
    title = "{Black Hole on a Chip: Proposal for a Physical Realization of the Sachdev-Ye-Kitaev model in a Solid-State System}",
    eprint = "1702.04426",
    archivePrefix = "arXiv",
    primaryClass = "cond-mat.dis-nn",
    doi = "10.1103/PhysRevX.7.031006",
    journal = "Phys. Rev. X",
    volume = "7",
    number = "3",
    pages = "031006",
    year = "2017"
}

@article{Parisi:1994jg,
    doi = {10.1088/0305-4470/27/23/007},
url = {https://dx.doi.org/10.1088/0305-4470/27/23/007},
year = {1994},
month = {dec},
publisher = {},
volume = {27},
number = {23},
pages = {7555},
author = {G Parisi},
title = {D-dimensional arrays of Josephson junctions, spin glasses and q-deformed harmonic oscillators},
journal = {Journal of Physics A: Mathematical and General}
}

@article{Berkooz:2024evs,
    author = "Berkooz, Micha and Brukner, Nadav and Jia, Yiyang and Mamroud, Ohad",
    title = "{From Chaos to Integrability in Double Scaled SYK}",
    eprint = "2403.01950",
    archivePrefix = "arXiv",
    primaryClass = "hep-th",
    month = "3",
    year = "2024"
}

@article{AlmehiriPaper,
    author = "Almheiri, Ahmed and Goel, Akash and Hu, Xu-Yao",
    title = "{Quantum gravity of the Heisenberg algebra}",
    eprint = "2403.18333",
    archivePrefix = "arXiv",
    primaryClass = "hep-th",
    month = "3",
    year = "2024"
}

@article{Peng:2017kro,
    author = "Peng, Cheng",
    title = "{Vector models and generalized SYK models}",
    eprint = "1704.04223",
    archivePrefix = "arXiv",
    primaryClass = "hep-th",
    doi = "10.1007/JHEP05(2017)129",
    journal = "JHEP",
    volume = "05",
    pages = "129",
    year = "2017"
}
\end{document}